\documentclass[aps,twocolumn,pra,superscriptaddress,amsmath,showpacs,tightenlines,pdflatex]{revtex4}

\usepackage{amsmath}

\usepackage{graphicx,times}
\usepackage{amstext}
\usepackage{amsmath}            
\usepackage{amssymb}            
\usepackage{graphicx}           
\usepackage{latexsym}
\usepackage{bm}
\usepackage{color}
 \def\extra#1{\emph{#1}}

\def \norm{{\cal N}}
\def \etal{\textit{et al.}}
\newcommand{\tr}{{\rm tr\,}}
\def\ket#1{{\,|#1\rangle}}
\def\bra#1{{\langle#1|\,}}
\newcommand{\real}{{\rm Re}\,}
\newcommand{\imag}{{\rm Im}\,}
\def\mean#1{{\langle #1\rangle}}

 \def\fig#1{{#1}}

\newcommand{\jpa}{J. Phys. A~}
\newcommand{\jpb}{J. Phys. B~}

\newcommand{\job}{J. Opt. B: Quantum Semicalssical Opt.}

\begin{document}

\title{State-dependent photon blockade via quantum-reservoir engineering}


\author{Adam Miranowicz}
\affiliation{Faculty of Physics, Adam Mickiewicz University,
PL-61-614 Pozna\'n, Poland} \affiliation{CEMS, RIKEN, Wako-shi,
Saitama 351-0198, Japan}

\author{Ji\v r\'\i\ Bajer}
\affiliation{Department of Optics, Palack\'{y} University, 772~00
Olomouc, Czech Republic}

\author{Ma\l{}gorzata Paprzycka}
\affiliation{Faculty of Physics, Adam Mickiewicz University,
PL-61-614 Pozna\'n, Poland}

\author{Yu-xi Liu}
\affiliation{Institute of Microelectronics, Tsinghua University,
Beijing 100084, China} \affiliation{Tsinghua National Laboratory
for Information Science and Technology (TNList),  Beijing 100084, China} \affiliation{CEMS, RIKEN,
Wako-shi, Saitama 351-0198, Japan}

\author{Alexandre M. Zagoskin}
\affiliation{CEMS, RIKEN, Wako-shi, Saitama 351-0198, Japan}
\affiliation{Department of Physics, Loughborough University,
Loughborough LE11 3TU, United Kingdom}

\author{Franco Nori}
\affiliation{CEMS, RIKEN, Wako-shi, Saitama 351-0198, Japan}
\affiliation{Physics Department, The University of Michigan, Ann
Arbor, Michigan 48109-1040, USA}

\date{\today}

\begin{abstract}
An arbitrary initial state of an optical or microwave field in a
lossy driven nonlinear cavity can be changed into a partially
incoherent superposition of only the vacuum and the single-photon
states. This effect is known as single-photon blockade, which is
usually analyzed for a Kerr-type nonlinear cavity parametrically
driven by a single-photon process assuming single-photon loss
mechanisms. We study photon blockade engineering via a nonlinear
reservoir, i.e., a quantum reservoir, where only two-photon
absorption is allowed. Namely, we analyze a lossy nonlinear cavity
parametrically driven by a two-photon process and allowing
two-photon loss mechanisms, as described by the master equation
derived for a two-photon absorbing reservoir. The nonlinear cavity
engineering can be realized by a linear cavity with a tunable
two-level system via the Jaynes-Cummings interaction in the
dispersive limit. We show that by tuning properly the frequencies
of the driving field and the two-level system, the steady state of
the cavity field can be the single-photon Fock state or a
partially incoherent superposition of several Fock states with
photon numbers, e.g., (0,2), (1,3), (0,1,2), or (0,2,4). At the
right (now fixed) frequencies, we observe that an arbitrary
initial coherent or incoherent superposition of Fock states with
an even (odd) number of photons is changed into a partially
incoherent superposition of a few Fock states of the same
photon-number parity. We find analytically approximate formulas
for these two kinds of solutions for several differently-tuned
systems. A general solution for an arbitrary initial state is a
weighted mixture of the above two solutions with even and odd
photon numbers, where the weights are given by the probabilities
of measuring the even and odd numbers of photons of the initial
cavity field, respectively. This can be interpreted as two
separate evolution-dissipation channels for even and odd-number
states. Thus, in contrast to the standard predictions of photon
blockade, we prove that the steady state of the cavity field, in
the engineered photon blockade, can depend on its initial state.
To make our results more explicit, we analyze photon blockades for
some initial infinite-dimensional quantum and classical states via
the Wigner and photon-number distributions.
\end{abstract}


\maketitle \pagenumbering{arabic}

\section{Introduction}

The progress in realizing macroscopic quantum coherent states in a
variety of systems (in particular, involving superconducting
devices~\cite{Xiang13}) makes many recently purely academic
problems very relevant for experimental research. Some such
problems are related to the interaction of photons in a cavity
with non-standard reservoirs (e.g., reservoirs with entanglement).
In this paper we consider the case of a two-photon absorbing
reservoir~\cite{Simaan78,Gilles93,Gilles94,
Guerra97,Dodonov97,Everitt13, Ispasoiu00,Klimov03, Buks06,
Yurke06, Karasik08, Boissonneault12,Voje13,Albert14} coupled to a
nonlinear cavity. Such a system can be realized, e.g., in the
microwave range, using a superconducting quantum interference
device (SQUID)~\cite{Kumar10,Everitt13}. A general framework of
two- and multi-photon dissipating models, within the Lindblad
master equations, was recently described in Ref.~\cite{Albert14}.
It is worth noting that in the years 2010s there has been a
renaissance of interest in quantum-reservoir engineering (also
known as dissipation engineering) (see, e.g., Refs.~\cite{Kumar10,
Boissonneault12, Voje13, Everitt13, Albert14,Gevorkyan10,
Mogilevtsev13, Leghtas13, Reiter13, Arenz13, Mirrahimi14, Lu14}),
which might be considered a new paradigm not only for quantum
state engineering but even for universal quantum
computation~\cite{Mirrahimi14}. Here we show how to realize
\emph{photon blockade} (PB) via a two-photon absorbing reservoir.

The term PB corresponds to the interpretation that a single photon
in a nonlinear cavity can block the transmission of a second
photon. Thus PB can be considered a photonic analog of solid-state
blockades including phonon blockade~\cite{Liu10} for quantum
oscillations of nanomechanical resonators, the celebrated Coulomb
blockade observed in single-electron tunneling experiments, or the
related Pauli spin blockade of electron transport due to spin
correlations. A detailed comparison showing the equivalence
between the photon and Coulomb blockades was given recently in
Ref.~\cite{Liu14}. We also note that, e.g., PB can be used to
demonstrate the occurrence of phonon blockade in optomechanical
systems in the microwave regime~\cite{Didier11}, where both
photon-photon and phonon-phonon interactions are induced by a
qubit (real or artificial two-level atom).

In the last two decades there has been considerable theoretical
and experimental interest in generating nonclassical light via
PB~\cite{Imamoglu97} in strongly coupled systems in cavity quantum
electrodynamics (QED)~\cite{Tian92,Werner99, Brecha99, Rebic99,
Kim99, Rebic02, Smolyaninov02}, and more recently also in circuit
QED~\cite{Hoffman11,Lang11,Didier11,Liu14} and quantum
optomechanics~\cite{Rabl11,Nunnenkamp11,Liao13,Liu14a}. PB was
demonstrated experimentally, e.g., in an optical cavity with a
single trapped atom~\cite{Birnbaum05}, in a photonic crystal
cavity with a quantum dot~\cite{Faraon08}, and in microwave
transmission-line resonators with a single superconducting
artificial atom~\cite{Hoffman11,Lang11}. Photon-induced tunneling,
experimentally demonstrated in Refs.~\cite{Smolyaninov02,
Faraon08,Majumdar12}, can also be explained in terms of PB.
Closely related experiments~\cite{Schuster08,Kubanek08}
demonstrated an observable optical nonlinearity (photon-photon
interaction) induced by a single atom in a cavity. Photon blockade
was also studied in the context of single-photon turnstile
devices~\cite{Imamoglu97}. For example, Ref.~\cite{Dayan08}
reports an experimental photon turnstile device dynamically
controlled by a single atom in a microscopic optical resonator.
The usual experimental realizations of single-photon turnstile
devices are based on Coulomb blockade in various semiconductor
systems~\cite{Kim99} (see Ref.~\cite{Shields07} for a review).
Finally, it is worth noting that typical optical nonlinearities
require strong light and macroscopic media.  The above-cited
impressive experiments, which can be considered as landmarks in
quantum and atom optics, showed the possibility to induce and
apply optical nonlinearities at the level of a single atom and one
or few photons.

Photon-photon interactions induced by a two-level system in a
linear cavity can be effectively described as a Kerr nonlinearity.
The occurrence of nonstationary PB in such Kerr nonlinear cavity
was predicted in Ref.~\cite{Leonski94}, and then studied in
various single-mode~\cite{Miran96,Leonski96,Miran14} and
two-mode~\cite{coupler,Kowalewska10} models. It should be stressed
that all these works discuss only the short-time evolution of
dissipation-free or sometimes dissipative nonlinear systems, so
the predicted effects can be referred to as nonstationary PB. This
is the opposite of the standard description of PB effects, which
are only considered in the steady-state limit. We also note that
this nonstationary Kerr-based PB is often referred to as a
nonlinear optical-state truncation or nonlinear quantum
scissors~(see reviews~\cite{Miran01,Leonski11}).  By contrast, the
effects of a linear optical-state truncation or linear quantum
scissors~\cite{Pegg98} are based on linear systems and conditional
measurements.

Nonclassical light generated via the standard single-photon
blockade~\cite{Imamoglu97,Leonski94} is a partially incoherent
superposition of the vacuum and single-photon states.  Recently,
the occurrence of two-photon blockades was predicted, where the
transmission of more than two photons can be effectively blocked
by single- and two-photon states~\cite{Miran13}. Thus, the
generated nonclassical light is a partially incoherent
superposition of the $n$-photon states for $n=0,1,2$. This
approach can be further generalized for multiphoton
blockades~\cite{Miran13,Hovsepyan14}.

In all these PB phenomena, the generated state of light was
independent of its initial state. Here we describe nonclassical
light, generated via a generalized PB, which can be sensitive to
its initial state, thus providing an additional method of its
(limited) measurement.

Namely, we predict here the occurrence of photon blockades in Kerr
nonlinear systems driven by a two-photon process and dissipating
by a two-photon absorption. We will show that there is no mixing
of number states of different parity during the evolution of such
Kerr nonlinear systems. Thus, this evolution can be described by
two solutions obtained for separate Hilbert spaces spanned either
by even- or odd-number states. By considering only initial states
of the same parity, the steady state does not depend on the
initial photon statistics. However, the general solution for an
initial state, which is a superposition of the even- and
odd-number Fock states, is a weighted mixture of the above two
solutions for different parities. The weights are determined by
the probabilities of measuring the even and odd photon numbers of
the initial field, respectively. Thus, even this simple analysis
reveals that the steady state can depend on the initial state,
although in this limited manner. We will discuss this problem in
detail in this work.

We will study photon-number statistics and a phase-space
description to compare various PB effects. In particular, we will
apply the standard Wigner function which, for a given state $\hat
\rho$, is defined by~\cite{VogelBook}:
\begin{equation}
W(\beta)\equiv W(q,p)=\frac{1}{\pi}\int\bra{q-x}\hat\rho \ket{q+x}
\exp\left(2ipx\right)dx, \label{Wigner}
\end{equation}
where $q=\real\beta$ and $p=\imag\beta$ are the canonical position
and momentum operators, respectively. The Wigner function for the
nonclassical states generated in PB can be experimentally
reconstructed by quantum state tomography~\cite{Lvovsky02} or even
directly measured by applying the method of
Ref.~\cite{Lutterbach97}. The power of the latter method was
demonstrated experimentally for the superpositions of a few
photons in cavity QED~\cite{Bertet02} and circuit
QED~\cite{Hofheinz09} systems.

The paper is organized as follows. Engineered photon blockade is
studied in the model described in Sec.~II. In particular, by
applying the Jaynes-Cummings model with a two-photon drive in the
dispersive limit, we derive an effective Hamiltonian describing a
driven Kerr-type nonlinearity. In Sec.~III, we present analytical
solutions describing nonstationary photon blockades and Rabi-type
oscillations for the model without dissipation. In Sec.~IV and
Appendix~A, we find and analyze steady-state solutions of a master
equation describing the two-photon loss mechanism. We discuss in
Sec.~V how photon blockade depends on specific initial fields. We
summarize our main results in the concluding section.

\section{Kerr nonlinearity with two-photon drive}

Here we derive an effective interaction model, describing a
Kerr-type nonlinearity driven by a two-photon process. We start
from the driven Jaynes-Cummings (JC) model in the dispersive
limit.

We analyze a two-level system (qubit), with a tunable transition
frequency $\omega_{\rm q}$, interacting with a cavity mode, with
frequency $\omega_{{\rm cav}}$, via the Jaynes-Cummings (JC)
model, described by the Hamiltonian $\hat H_{_{\rm JC}}$. We
assume that the cavity field is parametrically driven by a
two-photon process (with frequency $\omega_{\rm d}$), described by
the Hamiltonian $\hat H_{\rm d}$. Thus, the total Hamiltonian
$\hat H$ for our system, including the free Hamiltonian $\hat
H_{0}$ for the qubit and the cavity field, can be given as
follows:
\begin{align}
\hat H &=\hat H_{0}+\hat H_{_{\rm JC}}+\hat H_{\rm d},\\
\hat H_{0} & =\hbar\omega_{{\rm cav}}\hat a^{\dagger}\hat a
+\hbar\omega_{\rm q}\frac{\hat \sigma_{z}}{2},\\
\hat H_{_{\rm JC}} & =\hbar g(\hat a^{\dagger}\hat \sigma_{-}
+\hat a\hat \sigma_{+}),\\
\hat H_{\rm d} & =\hbar\epsilon_{0}[\hat a^{2}e^{i\omega_{\rm
d}t}+(\hat a^{\dagger})^{2}e^{-i\omega_{\rm d}t}]. \label{H1}
\end{align}
Here, $g$ is the qubit-field coupling strength, $\epsilon_0$ is a
driving field strength, for simplicity, assumed to be positive;
$\hat a$ ($\hat a^{\dagger}$) is the annihilation (creation)
operator of the cavity mode; the spin operators are $\hat
\sigma_{z}=|g\rangle\langle g|-|e\rangle\langle e|$, $\hat
\sigma_{+}=|e\rangle\langle g|$, and $\hat
\sigma_{-}=|g\rangle\langle e|$, where $|g\rangle$ $(|e\rangle)$
is the ground (excited) state of the qubit.

We analyze the Jaynes-Cummings interaction in the dispersive
limit, which occurs if the absolute value of the detuning
$\Delta=\omega_{\rm q}-\omega_{{\rm cav}}$ is much larger than the
qubit-field coupling $g$, i.e., we assume $|\lambda|\ll 1$ for the
parameter $\lambda=g/\Delta$.

\begin{figure}

 \fig{ \includegraphics[height=37mm]{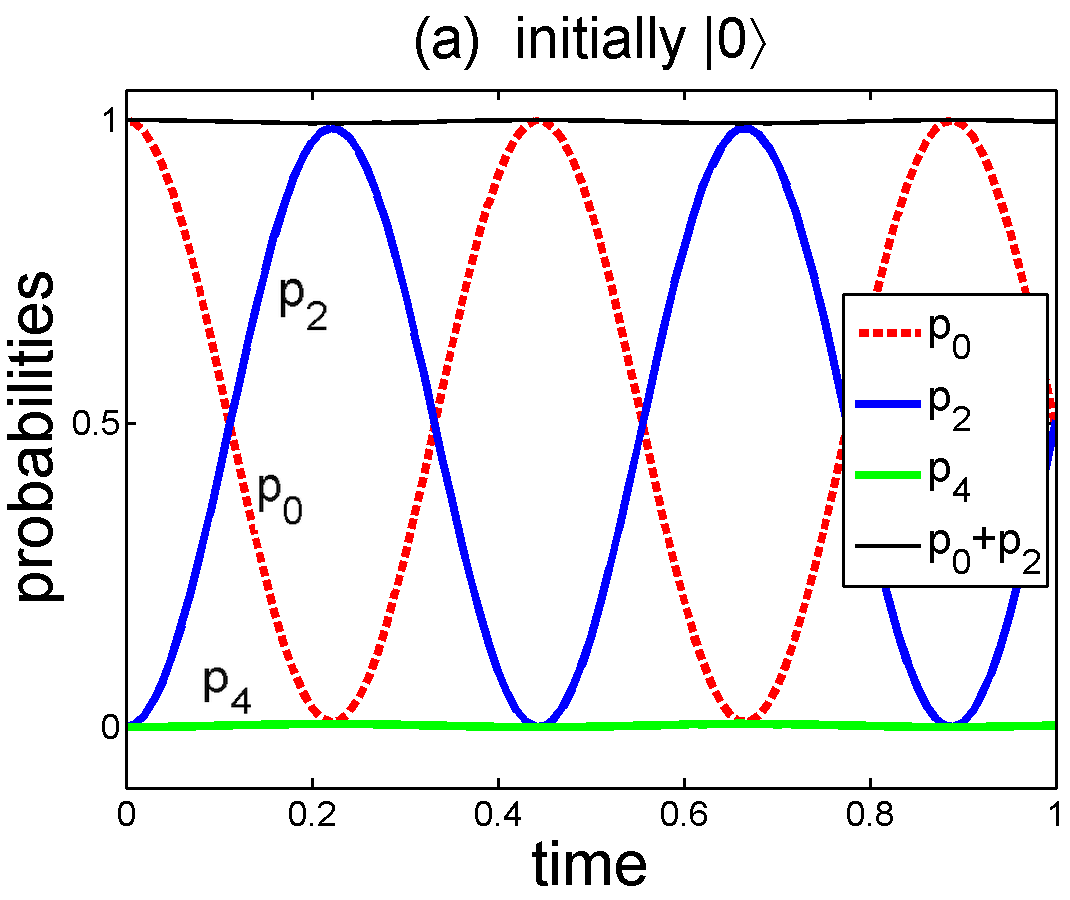}\includegraphics[height=37mm]{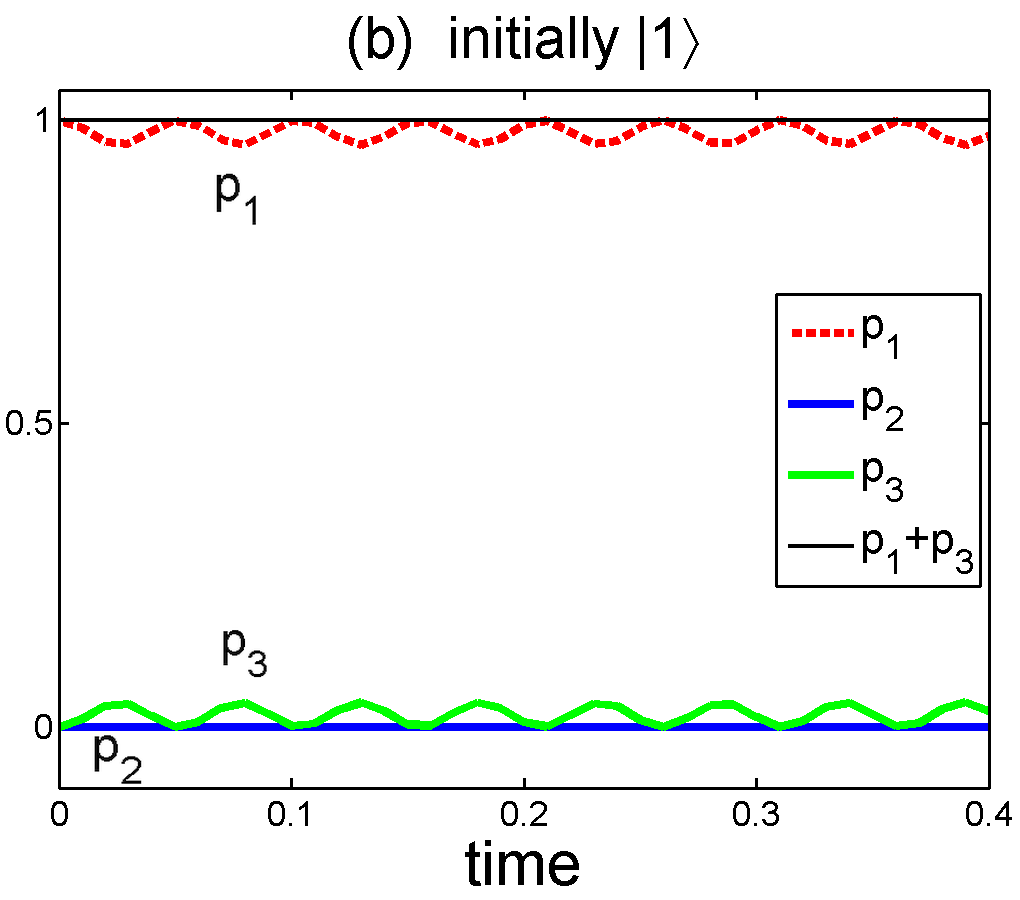}}

\vspace{1mm}

 \fig{ \includegraphics[height=37mm]{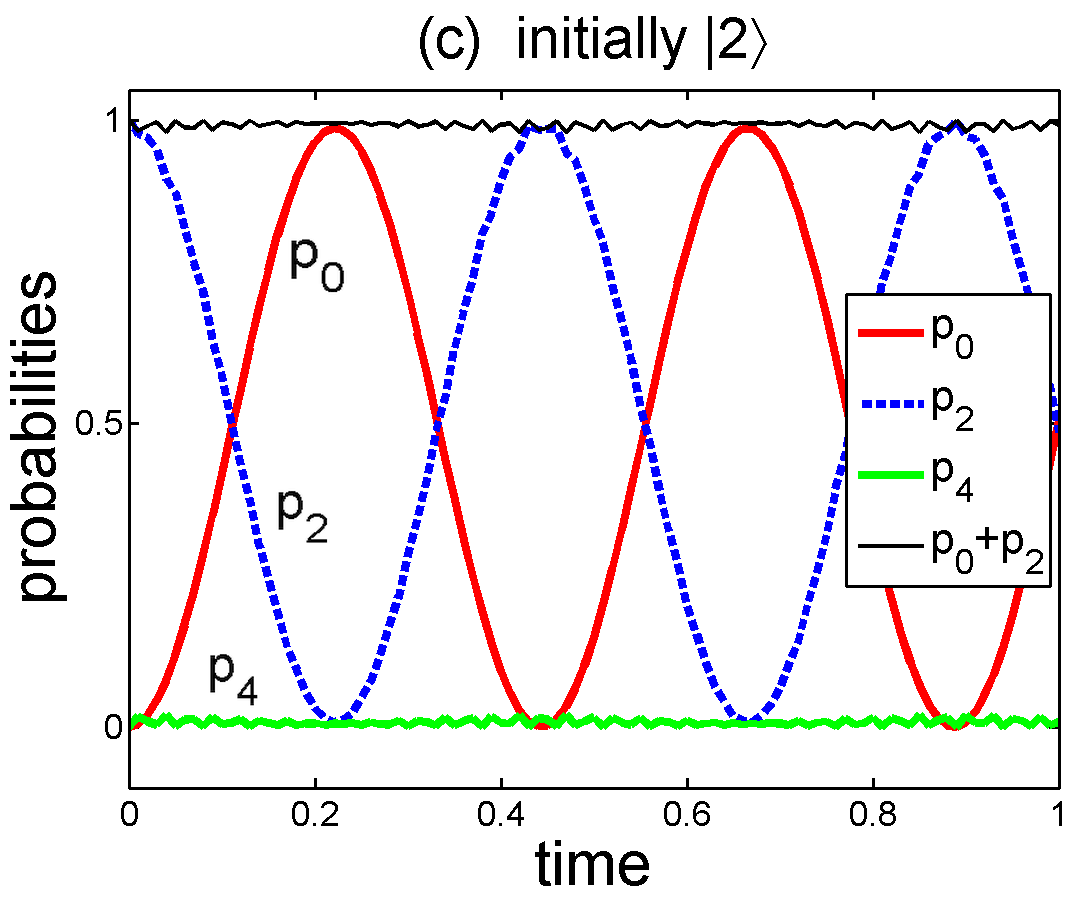}\includegraphics[height=37mm]{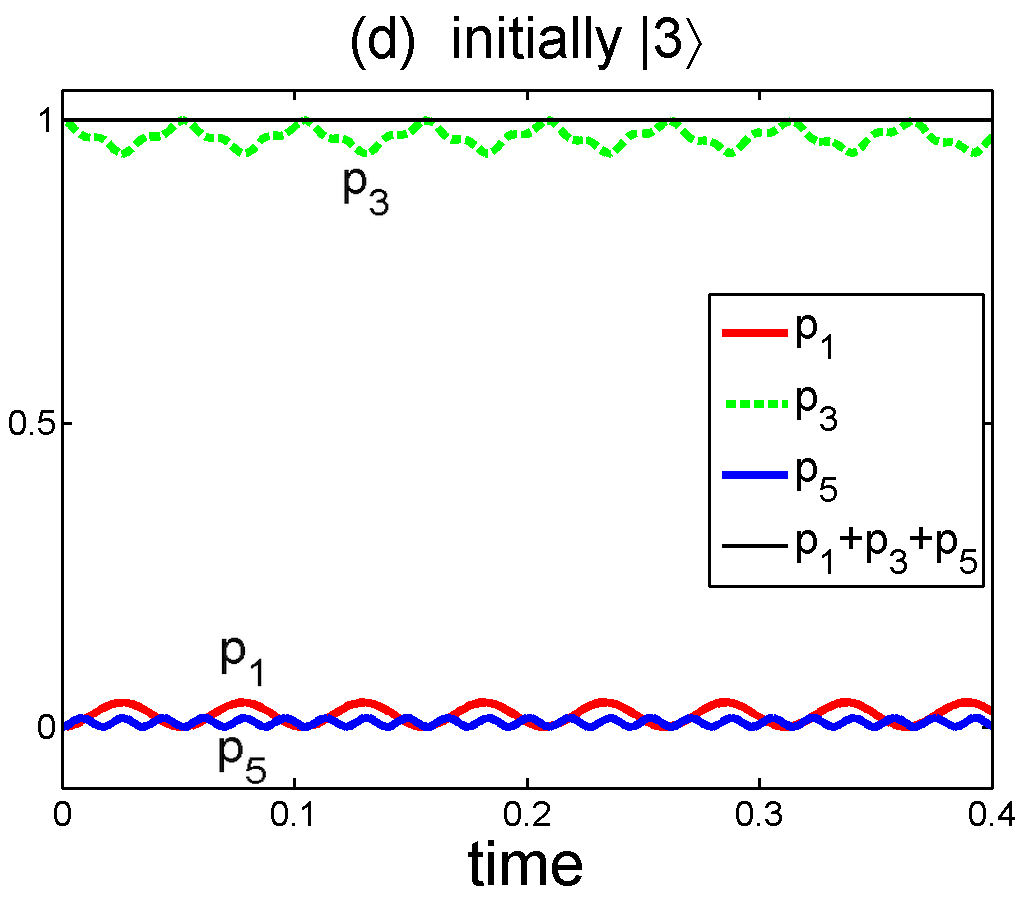}}

\caption{(Color online) Model~1: Dissipation-free evolution of the
photon-number probabilities $p_n(t)=|\langle n
\ket{\psi^{(m)}_{02}(t)}|^2$ and the photon-blockade fidelities
$F(t)=\sum_n p_n(t)$ for the Hamiltonian $\hat{H}_{02}$, given by
Eq.~(\ref{H02}), for several initial Fock states $\ket{m}$ (as
indicated in the panel titles). We set $\epsilon=\chi/6=5$. Panels
(a) and (c) show Rabi-type oscillations between the levels
$\ket{0}$ and $\ket{2}$, if at least one of them is initially
populated. Rabi-type oscillations are not observed if the other
levels are the only initially populated states, such as $\ket{1}$
and $\ket{3}$, as shown in panels (b) and (d), respectively. These
results have a simple physical explanation in terms of the
resonances shown in Fig.~\ref{fig02}.} \label{fig01}
\end{figure}

\begin{figure}

 \fig{ \includegraphics[height=50mm]{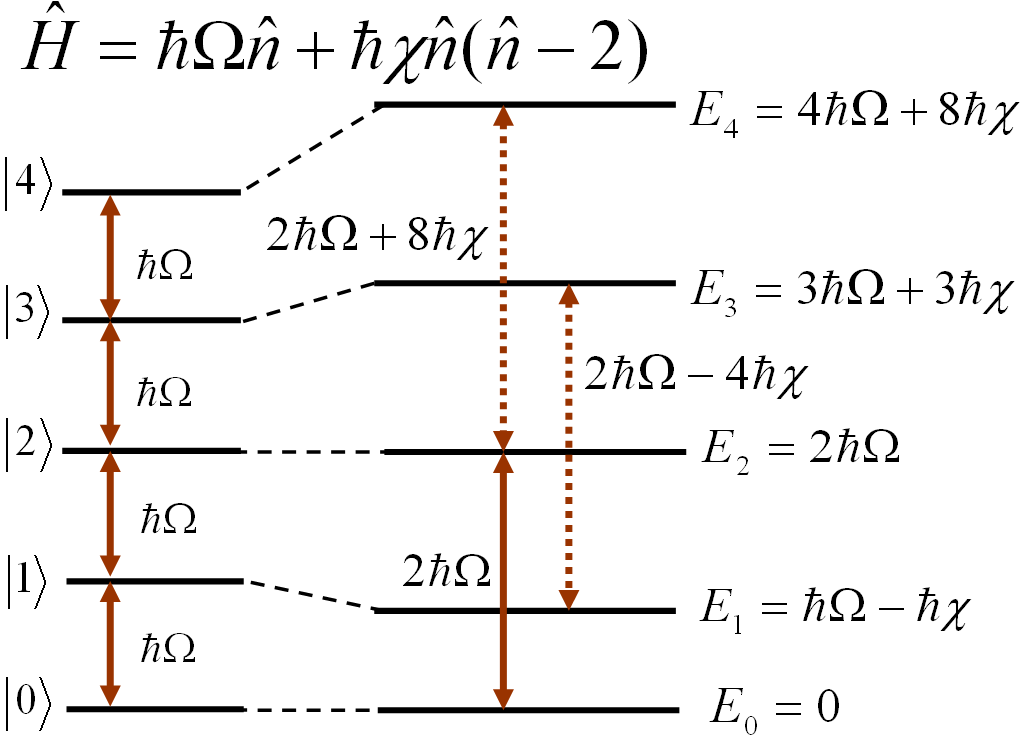}}

\caption{(Color online) Model~1: Explanation of the occurrence of
photon blockade via the energy levels of the Hamiltonian $\hat H$
corresponding to $\hat H_{\rm s}
(\Omega_{02}=\Omega,\Sigma_{02}=0)$, given by Eq.~(\ref{H02a}), in
the limit of a very small driving strength, $\epsilon\ll \chi$.
The Kerr-nonlinear term, proportional to $\chi$, changes the
harmonic spectrum (shown in the left spectrum) into an anharmonic
non-equidistant one (right side) with $E_{n+1}-E_{n}\neq\;$const,
where $E_n= n\hbar\Omega+n(n-2)\hbar\chi$ (with $n=0,1,...)$ are
the eigenvalues of the Hamiltonian $\hat H$. It is seen that the
two-photon transitions between the levels $|0\rangle$ and
$|2\rangle$ (shown by a solid double arrow in the right spectrum)
can be induced by a driving field with frequency $\omega_{\rm
d}=(E_{2}-E_{0})/\hbar=2\Omega$, which is the same as for the
harmonic system. The other transitions between the levels, e.g.,
$|1\rangle$ and $|3\rangle$, as well as $|2\rangle$ and
$|4\rangle$ (as shown by the dashed double arrows) are
off-resonance with $\Omega$ or its multiples.} \label{fig02}
\end{figure}

\begin{figure}

 \fig{ \includegraphics[height=37mm]{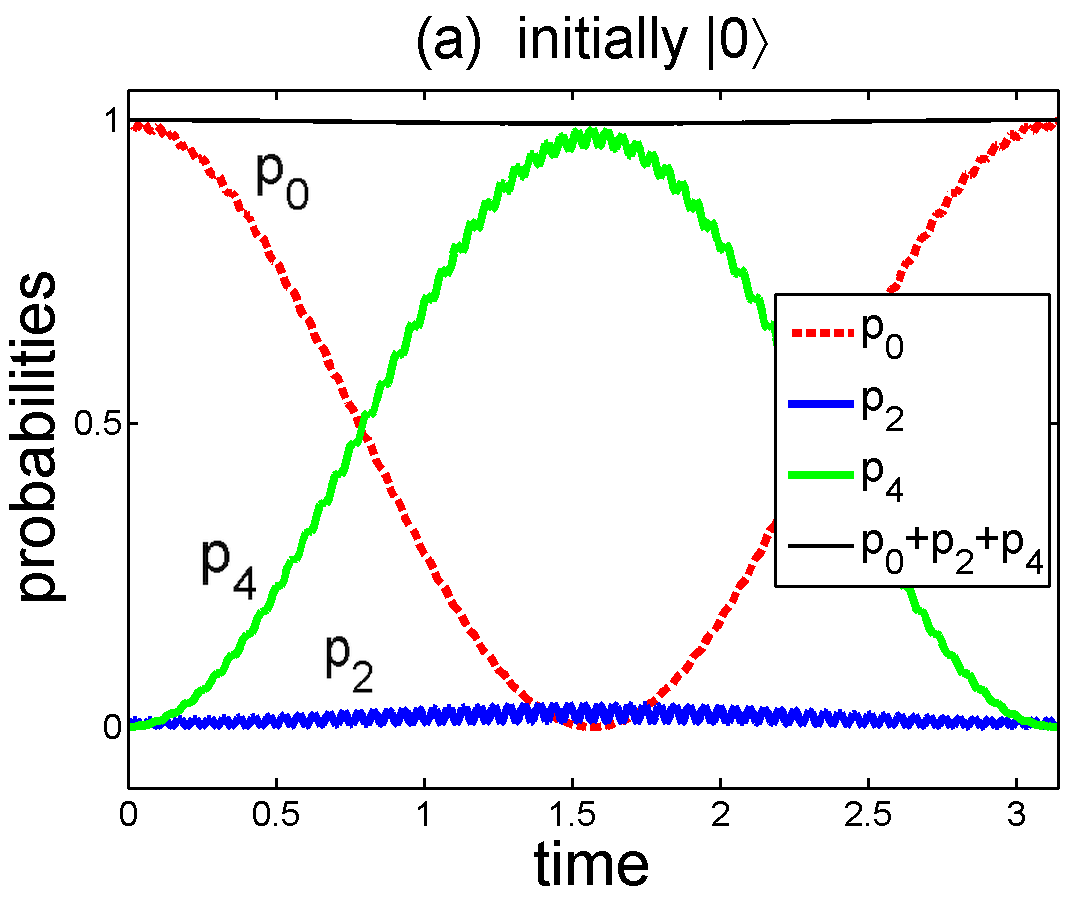}\includegraphics[height=37mm]{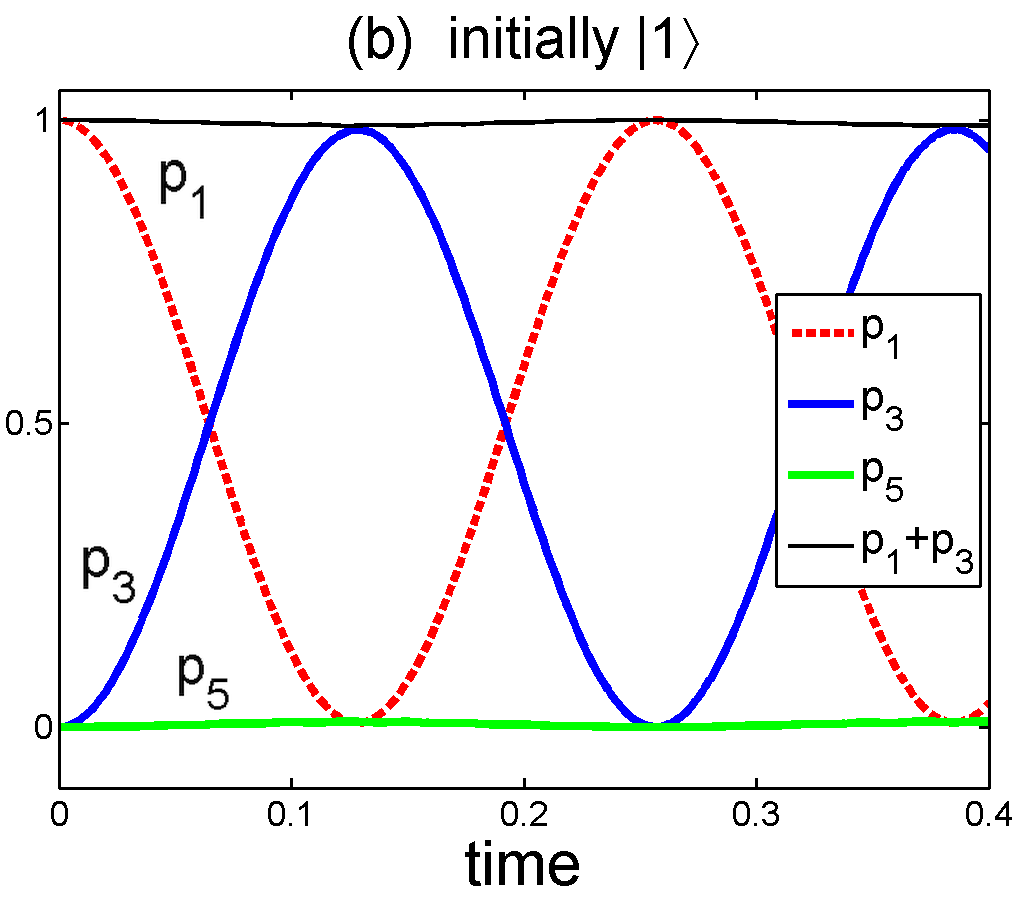}}

\vspace*{1mm}
 \fig{ \includegraphics[height=37mm]{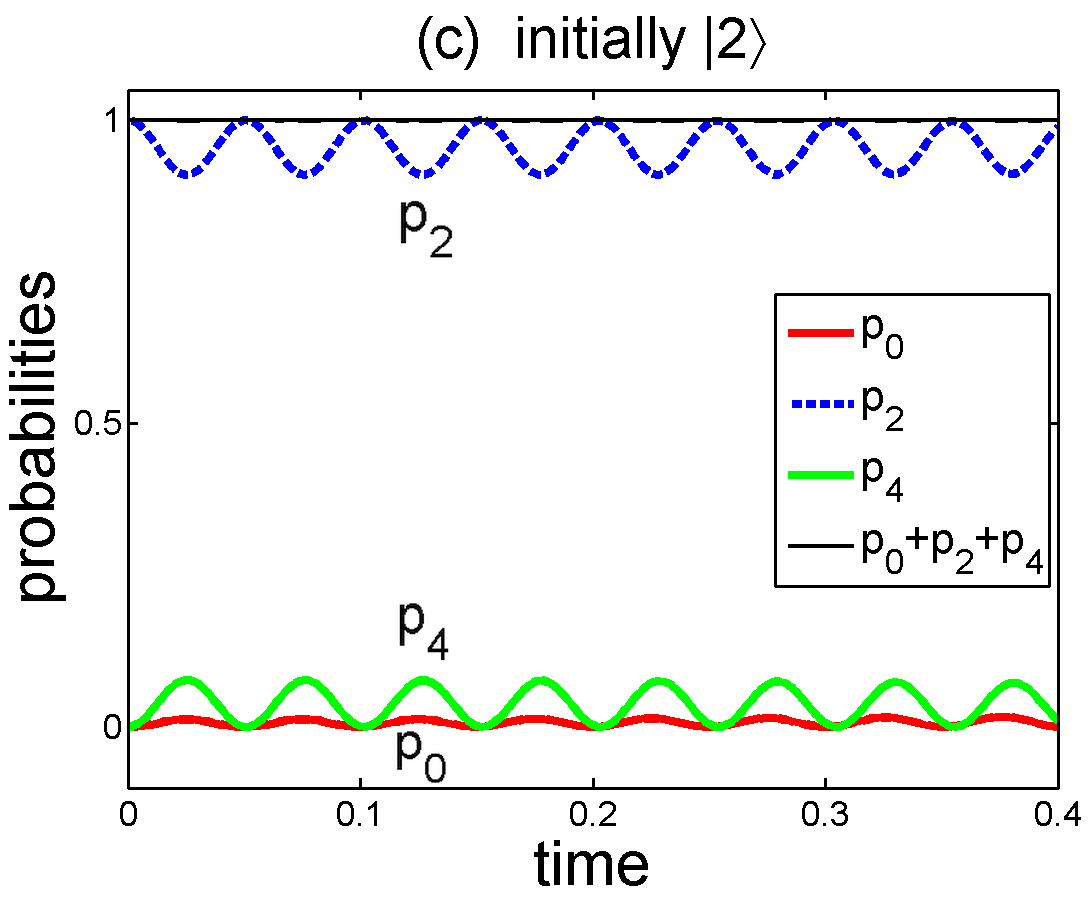}\includegraphics[height=37mm]{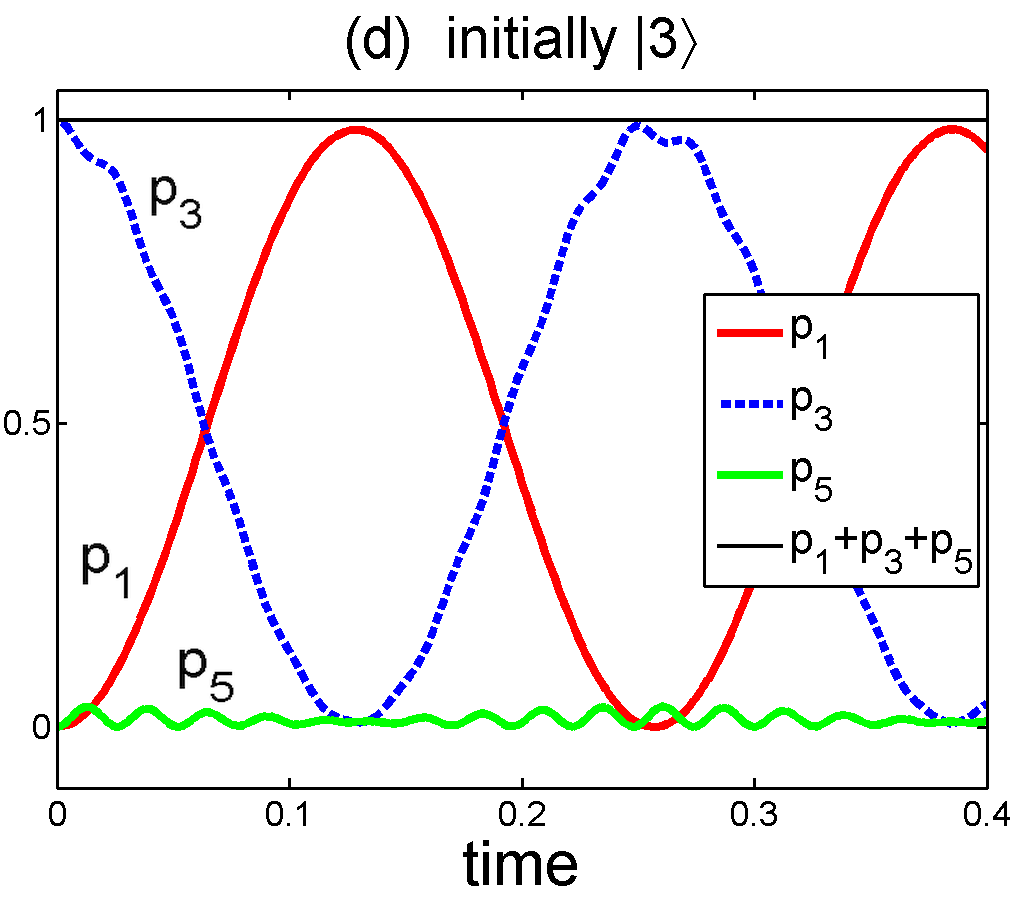}}

\caption{(Color online) Model~2: Same as in Fig.~\ref{fig01}, but
for the probabilities $p_n(t)=|\langle n
\ket{\psi^{(m)}_{13}(t)}|^2$ obtained for the Hamiltonian
$\hat{H}_{13}$, given by Eq.~(\ref{Hkl}) with $k=1,l=3$. Panel (a)
[(b) and (d)] show Rabi-type oscillations between the levels
$\ket{0}$ and $\ket{4}$ ($\ket{1}$ and $\ket{3}$), if at least one
of these levels is initially populated. Rabi-type oscillations are
not observed if the other levels are the only ones, which are
initially populated, such as $\ket{2}$, shown in panel (c). The
physical meaning of these results, analogously to those in
Fig.~\ref{fig01}, can be simply understood in terms of the
resonances shown in Fig.~\ref{fig04}. } \label{fig03}
\end{figure}

\begin{figure}

 \fig{ \includegraphics[height=50mm]{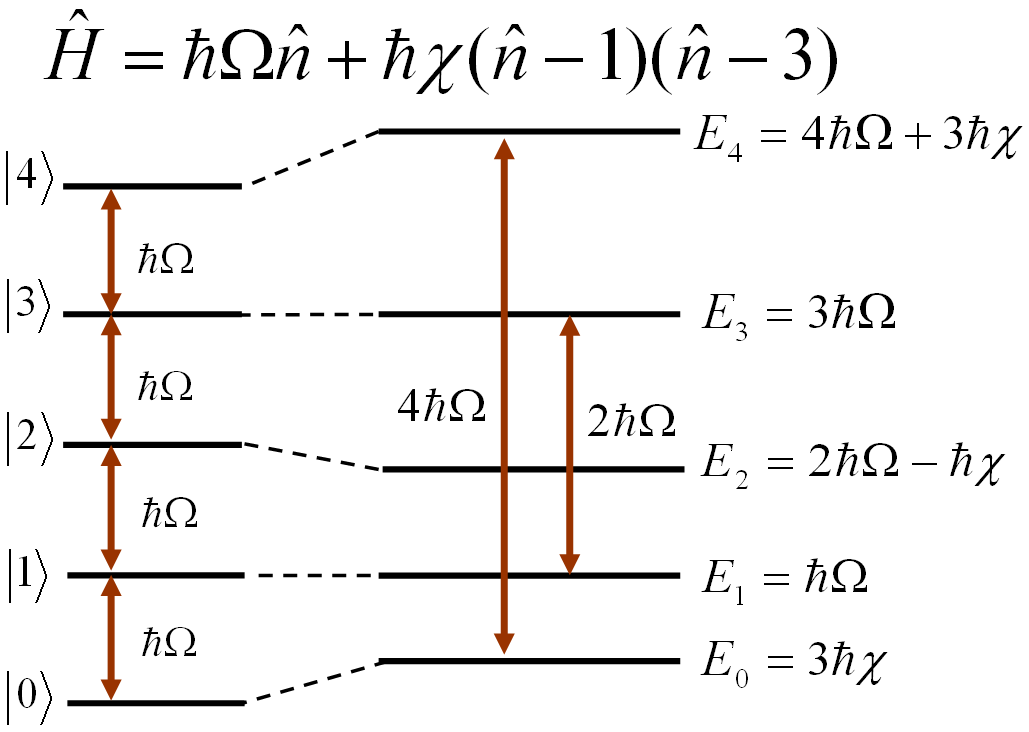}}

\vspace*{-4mm} \caption{(Color online) Model~2: Same as in
Fig.~\ref{fig02}, but for the Hamiltonian $\hat H$ corresponding
to $\hat H_{\rm s}(\Omega_{13}=\Omega,\Sigma_{13}=0)$, given by
Eq.~(\ref{Hkla}) with $k=1,l=3$ assuming $\epsilon\ll \chi$. Here
the two-photon (four-photon) transitions between the levels
$|1\rangle$ and $|3\rangle$ ($|0\rangle$ and $|4\rangle$), shown
by solid double arrows in the right spectrum, can be induced by a
driving field with frequency $\omega_{\rm
d}=(E_{3}-E_{1})/\hbar=2\Omega$ [$\omega_{\rm
d}=(E_{4}-E_{0})/\hbar=4\Omega$], which are the multiples of the
same frequency $\Omega$  of the harmonic system. } \label{fig04}
\end{figure}

\begin{figure}

 \fig{ \includegraphics[height=50mm]{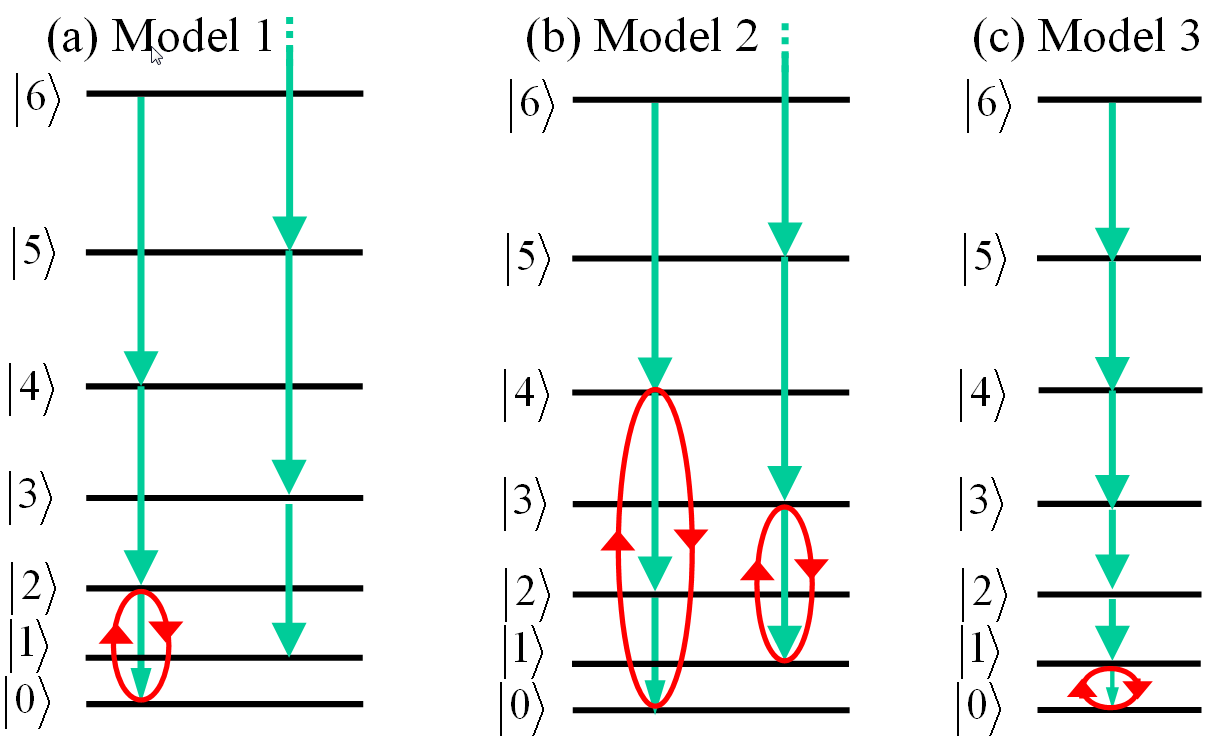}}

\vspace*{1mm} \caption{(Color online) An intuitive explanation of
the engineered photon blockades in Models~1 and 2 with two-photon
dissipation, and the standard photon blockade in Model~3 with
single-photon dissipation. The diagrams schematically show the
energy levels of three Kerr-type nonlinear systems driven by a
classical field with frequency $\omega_{\rm d}$ in resonance with
the desired transitions, as shown in Fig.~\ref{fig02} for Model~1
and Fig.~\ref{fig04} for Model~2. The red ellipses with arrows
describe these [(a),(b)] two-photon and (c) single-photon
drivings, together with the Rabi-type oscillations between the
corresponding levels. The systems are described by the
Hamiltonians: (a) $\hat H_{02}$, given by Eq.~(\ref{H02}), (b)
$\hat H_{13}$, given by Eq.~(\ref{Hkl}) for $k=1,l=3$, and $\hat
H_{\rm usual}$, given by Eq.~(\ref{H_usual}). The system
dissipation is governed by the master equations describing either
[(a), (b)] two-photon  or (c) single-photon absorption for
$\gamma\ll \epsilon\ll \chi$. The numerous green single arrows
pointing down describe these dissipations (absorptions). These
figures intuitively explain the occurrence of several kinds of the
engineered photon blockades, as well as two independent evolutions
of the initial Fock states with even and odd numbers of photons
for Models~1 and~2. This implies that the engineered PB effects in
panels (a,b) can depend on the initial state of a cavity, although
in a limited way, as they depend solely on the ratio of the
probabilities of measuring the photon numbers of different parity.
In contrast to this, the steady state generated in standard PB, as
shown in (c), is independent of the cavity initial state.}
\label{fig05}
\end{figure}

Following the approach of Ref.~\cite{Boissonneault09}, one can
apply the transformation $U=\exp[-f(\lambda)(\hat a^{\dagger}\hat
\sigma_{-}-\hat a\hat \sigma_{+})]$ to the Hamiltonian $\hat H$,
and expand the transformed Hamiltonian $\hat H'$ in power series
of $\lambda$, which results in
\begin{align}
\hat H' & =\hat U^{\dagger}\hat H\hat U =\hbar\omega'_{{\rm
cav}}\hat a^{\dagger}\hat a
+\hbar\hat \omega'_{\rm q}\frac{\hat \sigma_{z}}{2}\nonumber \\
 & \quad +\hbar\chi\hat  a^{\dagger}\hat a(\hat a^{\dagger}\hat a-2)
 \hat \sigma_{z}+\hat H'_{\rm d}+{\cal O}(\lambda^{4}).
 \label{H2}
\end{align}
Here,
$\omega'_{{\rm cav}} =\omega_{{\rm cav}}+\chi$, $\hat \omega'_{\rm
q} =\omega_{\rm q}-\eta+2(2\chi-\eta)\hat a^{\dagger}\hat a,$
 $\eta=-g\lambda(1-\lambda^{2})$, and $\chi=-g\lambda^{3}$ is
a Kerr-type nonlinearity coupling. Note that $\chi>0$ if
$\omega_{{\rm cav}}>\omega_{\rm q}$. Moreover, $f(\lambda)$ is
given explicitly in Ref.~\cite{Boissonneault09}, while $\hat
H_{\rm d}' =\hat U^{\dagger}\hat H_{\rm d}\hat U$ will be
specified below. By assuming that the qubit is in its ground
state, we have
\begin{align}
\langle g|\hat H'|g\rangle & =\hbar(\omega{}_{{\rm cav}}
+3\chi-\eta)\hat a^{\dagger}\hat a
+\hbar\chi\hat  a^{\dagger}\hat a(\hat a^{\dagger}\hat a-2)\nonumber \\
 &\quad +\hat H'_{\rm d}+\tfrac{1}{2}(\omega_{\rm q}-\eta)+{\cal
 O}(\lambda^{4}).
 \label{H3}
\end{align}
The annihilation operator transforms as~\cite{Boissonneault09}:
\begin{equation}
\hat a'=U^{\dagger}\hat a\hat U=\hat a\hat x+\lambda\hat  y\hat
\sigma_{-}+\lambda^{3}\hat a^{2}\hat \sigma_{+}+{\cal
O}(\lambda^{4}), \label{H4}
\end{equation}
where $\hat x=1+\lambda^{2}\hat \sigma_{z}/2$ and $\hat
y=1-3\lambda^{2}(\hat a^{\dagger}\hat a+1/2).$ By transforming the
driving interaction $\hat H_{\rm d}$ according to this expansion
of $\hat a'$, and assuming the qubit to be  in its ground state,
we find that
\begin{align}
\hat H'_{{\rm d}} & =\hbar\epsilon_{}[\hat a^{2}e^{i\omega_{\rm
d}t}+(\hat a^{\dagger})^{2}e^{-i\omega_{\rm d}t}]+{\cal
O}(\lambda^{4}), \label{H5}
\end{align}
where $\epsilon=(1+\lambda^{2})\epsilon_{0}$. We note that by
assuming that the qubit is in its excited state, then it would be
$\epsilon=(1-\lambda^{2})\epsilon_{0}$. Now we apply another
unitary operation $\hat{U}_{\rm rot}=\exp[-i(\omega_{\rm d}/2)
\hat{a}^{\dagger}\hat{a}t]$, which transforms the Hamiltonian
$\hat H'$ into
\begin{equation}
\hat{H}''= \hat{U}_{\rm rot}^{\dagger}\hat{H}'\hat{U}_{\rm rot}
-i\hbar\hat{U}_{\rm rot}^{\dagger}\frac{\partial}{\partial
t}\hat{U}_{\rm rot}.\label{Hrot}
\end{equation}
Here we also use the following operator-algebra
theorems~\cite{Louisell73}: $\hat a\hat f(\hat a^{\dagger}\hat
a)=\hat f(\hat a^{\dagger}\hat a+1)\hat a$ and $\hat f(\hat
a^{\dagger}\hat a)\hat a^{\dagger}=\hat a^{\dagger}\hat f(\hat
a^{\dagger}\hat a+1),$ which are valid for any function $\hat f$
of $\hat a^{\dagger}\hat a$. Then it is easy to observe that the
time-dependent Hamiltonian $\hat H'_{{\rm d}}$ is transformed by
$\hat{U}_{\rm rot}$ into the time-independent one $\hat H'_{{\rm
d}} \approx \hbar\epsilon(\hat a^{2}+\hat
a^{\dagger2})-\hbar(\omega_{\rm d}/2)\hat a^{\dagger}\hat a$.
Thus, in this rotating frame, we arrive at the following
time-independent effective Hamiltonian:
\begin{equation}
\langle g|\hat H''|g\rangle=\hat{U}^{\dagger}\langle g|\hat
H'|g\rangle\hat{U}=\hat H_{\rm s}+{\cal O}(\lambda^{4}) \label{H6}
\end{equation}
with
\begin{eqnarray}
\hat H_{\rm s}(\Omega_{02},\Sigma_{02})&=&\hbar\Omega_{02} \hat
a^{\dagger}\hat a+\hbar\chi \hat a^{\dagger}\hat a(\hat
a^{\dagger}\hat a-2) \nonumber \\&&+\hbar\epsilon[\hat a^{2}+(\hat
a^{\dagger})^{2}]+\hbar\Sigma_{02}, \label{H02a}
\end{eqnarray}
where
\begin{align}
\Omega_{02}=\;&\omega_{{\rm
cav}}+3\chi-\eta-\tfrac12\,\omega_{\rm d},\nonumber \\
\Sigma_{02}  =\;&\tfrac12 (\omega_{\rm q}-\eta). \label{Omega02}
\end{align}
These frequencies $\Omega_{02}$ and $\Sigma_{02}$ can be
simultaneously equal to zero by properly changing the detuning
$\Delta$  (i.e., the qubit transition frequency $\omega_{\rm q}$
or, equivalently, the cavity frequency $\omega_{{\rm cav}}$) and
the classical driving-field frequency $\omega_{\rm d}$. Thus,
under the above conditions, the effective Hamiltonian describing
our system, referred here to as Model~1, is given by
\begin{eqnarray}
\hat{H}_{02}&=&\hat H_{\rm s}(\Omega_{02}=0,\Sigma_{02}=0)\nonumber \\
&=& \hbar \chi
\hat a^{\dagger}\hat a(\hat a^{\dagger}\hat a-2) + \hbar
\epsilon(\hat{a}^{\dagger 2}+\hat{a}^{ 2}) \label{H02}
\end{eqnarray}
depending on the driving field strength $\epsilon$ and the Kerr
nonlinear coupling $\chi$. One can also rearrange terms in
Eq.~(\ref{H02a}) to obtain the following Hamiltonian
\begin{eqnarray}
\hat H_{\rm s}(\Omega_{kl},\Sigma_{kl})&=&\hbar\Omega_{kl} \hat
a^{\dagger}\hat a+\hbar\chi (\hat a^{\dagger}\hat a-k)(\hat
a^{\dagger}\hat a-l)
\nonumber \\
&&+\hbar\epsilon[\hat a^{2}+\hat a^{\dagger 2}] +\hbar\Sigma_{kl}, \label{Hkla}
\end{eqnarray}
where
\begin{align}
\Omega_{kl}=\;&\omega_{{\rm
cav}}+(k+l+1)\chi-\eta-\tfrac12\,\omega_{\rm d},\nonumber \\
\Sigma_{kl}  =\;&\tfrac12 (\omega_{\rm q}-2kl\chi-\eta).
\label{Omega13}
\end{align}
Analogously to the former case, one can avoid the contribution of
the terms proportional to the frequencies $\Omega_{kl}$ and
$\Sigma_{kl}$ by properly changing the detuning $\Delta$ and the
driving-field frequency $\omega_{\rm d}$. This results in the
following Hamiltonian
\begin{align}
\hat{H}_{kl}=&\hat H_{\rm
s}(\Omega_{kl}=0,\Sigma_{kl}=0)\nonumber\\=& \hbar \chi (\hat
a^{\dagger}\hat a-k)(\hat a^{\dagger}\hat a-l) + \hbar
\epsilon(\hat{a}^{2}+\hat{a}^{\dagger 2}). \label{Hkl}
\end{align}
Hereafter, we specify the Hamiltonian in Eq.~(\ref{Hkl}) to the
two special cases of $\hat{H}_{13}$ (referred to as Model~2) and
$\hat{H}_{02}$ (Model~1) in our analytical approaches and
numerical simulations shown in Figs.~\ref{fig01}--\ref{fig13}. For
clarity, we will usually explicitly denote by $\hat \rho^{kl}$,
the state generated by the action of the corresponding Hamiltonian
$\hat H_{kl}$.

This Kerr nonlinear oscillator driven by a two-photon (or
two-phonon) process is sometimes referred to as the Cassinian
oscillator, since its classical phase-space trajectories are the
ovals of Cassini~(see, e.g., Ref.~\cite{Meaney11} and references
therein). Various realizations of the Cassinian oscillator have
been proposed. In our context, the most promising implementations
seem to be those based on SQUIDs~\cite{Kumar10,Everitt13,Lin14}.

In particular, Ref.~\cite{Lin14} reports the experimental
realization of a parametric phase-locked oscillator (PO), also
referred to as a parametron. It is composed of a dc SQUID and a
superconducting coplanar waveguide linear resonator at a static
resonant frequency $\omega_0^{\rm PO}$. The SQUID, which is
formally equivalent to a qubit, introduces a Kerr-type
nonlinearity (as described by the nonlinearity parameter $\chi'$)
into the system. Thus, the PO can be described as an anharmonic
oscillator. The driving microwave field, at a frequency
$\omega_p$, is applied to a pump line being inductively coupled to
the SQUID. This driving field modulates the resonant frequency
around $\omega_0^{\rm PO}$. The static system Hamiltonian
reads~\cite{Lin14}:
\begin{equation}
\hat H_{{\rm sys}}(t)=\hbar\omega_{0}^{{\rm
PO}}\bigl[\hat a^{\dagger}\hat a+\bar\epsilon\cos(\omega_{{\rm
d}}t)(\hat a+\hat a^{\dagger})^{2}\bigr]+\hbar\chi'(\hat a+\hat a^{\dagger})^{4},
\end{equation}
where $\hat a$ is the annihilation operator of the resonator,
while $\omega_{\rm d}$ and $\bar\epsilon$ stand for the frequency
and strength of the parametric modulation, respectively. We
rewrite this Hamiltonian in normal order. We also transform it
into a rotating frame by applying the unitary operation
$\hat{U}_{{\rm rot}}=\exp[-i(\omega_{\rm
d}/2)\hat{a}^{\dagger}\hat{a}t]$, according to Eq.~(\ref{Hrot}),
and omit both the rapidly oscillating and constant terms. Thus,
one finally obtains the following approximate Hamiltonian
\begin{equation}
H'_{{\rm sys}}(t) =\hbar \Omega'
a^{\dagger}a+\hbar\epsilon'\left(a^{2}+a^{\dagger2}\right)+\hbar(6\chi'
)a^{\dagger}a( a^{\dagger}a-1),
\end{equation}
where $ \Omega'= \omega_{0}^{{\rm PO}}+12\chi' -\omega_{\rm d}/2$
and $\epsilon'=\omega_{0}^{{\rm PO}}\bar\epsilon/2$. At the
resonant condition $\Omega'=0,$ one obtains the Hamiltonian of the
Supplement of Ref.~\cite{Lin14} corresponding to our Hamiltonian
$H_{01}$, which is a special case of Eq.~(\ref{Hkl}) for
$\chi=6\chi'$ and $\epsilon=\epsilon'$. The general Hamiltonian
$H_{kl}$, given by Eq.~(\ref{Hkl}), with $k,l=0,1,2...,$ is
obtained by properly choosing $\omega_{\rm d}$ to satisfy the
condition $\Omega'+6(k+l-1)\chi'=0$.

For a comparison, it is worth noting that the standard predictions
of photon blockade were reported for systems described by the
following Hamiltonian~\cite{Imamoglu97,Leonski94}
\begin{align}
\hat{H}_{\rm usual}=& \hbar \chi \hat a^{\dagger}\hat a(\hat
a^{\dagger}\hat a-1) + \hbar \epsilon(\hat{a}+\hat{a}^\dagger),
\label{H_usual}
\end{align}
referred here to as Model~3, assuming a single-photon driving, as
described by the last term. Only for a brief comparison, we show
the solutions for  $\hat{H}_{01}$ and $\hat{H}_{\rm usual}$ in
Fig.~\ref{fig14}.

Let us also briefly consider the case when the frequency of the
single-photon driving field $\omega_{\rm d}$ is equal to the sum
of the Kerr nonlinearity $\chi$ and the cavity resonance frequency
$\omega_{{\rm cav}}$. Then, as shown in Ref.~\cite{Miran13},
Eq.~(\ref{H_usual}) can be replaced by
\begin{align}
\hat{H}'_{\rm usual}=& \hbar \chi \hat a^{\dagger}\hat a(\hat
a^{\dagger}\hat a-2) + \hbar \epsilon(\hat{a}+\hat{a}^\dagger),
\label{H_usual2}
\end{align}
referred here to as Model~4, which can lead to two-photon blockade
(two-photon state truncation) if $\epsilon\ll\chi$.

For the benefit of the reader, the various models defined here are
listed in Table~I.

\begin{widetext}

\begin{table}
\caption{Comparison of various kinds of photon blockades assuming
$m$ driving photons and $d$ dissipating photons (due to
absorption), with $d,m=1,2$. In particular, it is seen that the
steady states of these photon blockades can depend on the initial
states only for $d=m>1$. Our illustrations of the steady states
include their Wigner functions and photon-number probabilities.
Note that standard PB~\cite{Imamoglu97} is usually studied in
Model~3.}
\begin{tabular}{l l c c c c l c c l}
  \hline
      Model \hspace{0mm} &
      Hamiltonian  &
      Eq. &
      Kerr  \hspace{0mm} &
      $m$-photon  \hspace{0mm} &
      $d$-photon &
      initial state&
      populated  \hspace{2mm} &
      state  &
      examples
       \\
       &
       &
      &
      nonlinearity \hspace{0mm} &
      driving \hspace{0mm} &
      dissipation &
      &
      Fock states\footnote{The steady states generated via PB are partially incoherent superpositions of these Fock number states.} \hspace{2mm} &
      dependence &

       \\
\hline
   1 & $\hat H_{02}$ & (\ref{H02}) & $\hat{n}(\hat{n}-2)$  & $m=2$ & $d=2$  & even-number state & $\ket{0}$, $\ket{2}$ & no & Figs.~\ref{fig06}(a,b)
\\
    &  &   &&  &   & odd-number state & $\ket{1}$ & no & Figs.~\ref{fig06}(c,d)
\\
    & &   &&  &   & mixed-parity state\footnote{This can be a superposition or mixture of the even- and odd-number Fock states.} & $\ket{0}$, $\ket{1}$, $\ket{2}$ & yes & Fig.~\ref{fig10}
\\
     2 & $\hat H_{13}$ & (\ref{Hkl}) & $(\hat{n}-1)(\hat{n}-3)$  & 2 & 2  & even-number state & $\ket{0}$, $\ket{2}$,$\ket{4}$ & no & Figs.~7(a,b)
\\
    & &  & & &   & odd-number state & $\ket{1},$ $\ket{3}$ & no & Figs.~\ref{fig07}(c,d)
\\
    & &  & & &   & mixed-parity state$^b$ & $\ket{0},\ket{1},\ket{2},\ket{3},\ket{4}$ & yes & Fig.~\ref{fig11}
\\
   3 & $\hat{H}_{\rm usual}$ & (\ref{H_usual}) & $\hat{n}(\hat{n}-1)$  & $1$ & 1  & any & $\ket{0}$, $\ket{1}$ & no & Figs.~\ref{fig14}(a,b)
\\
   3' & $\hat{H}_{\rm usual}$ & (\ref{H_usual}) & $\hat{n}(\hat{n}-1)$  & 1 & 2  & any & $\ket{0}$, $\ket{1}$ & no & Figs.~\ref{fig14}(a,b)
\\
   4 & $\hat{H}'_{\rm usual}$ & (\ref{H_usual2})& $\hat{n}(\hat{n}-2)$  & 1 & 1  & any & $\ket{0}$, $\ket{1}$, $\ket{2}$ & no & Ref.~\cite{Miran13}
\\
   5 & $H_{01}$ & (\ref{Hkl}) & $\hat{n}(\hat{n}-1)$  & 2 & 1  & any & $\ket{0}$ & no & Figs.~\ref{fig14}(c,d)
\\
 \hline
\end{tabular}
\end{table}

\end{widetext}


\section{Nonstationary photon blockades and Rabi-type oscillations}

Here we briefly describe the evolution of the systems described by
Models~1 and~2 for some initial Fock states assuming no
dissipation. These evolutions lead to time-dependent PB (or
nonstationary PB), which can also be interpreted as an
optical-state truncation.

Assuming that the driving field strength $\epsilon$ is much weaker
than the Kerr nonlinearity $\chi$, one can find that the
pure-state evolution of the system, described by the Hamiltonian
$\hat{H}_{02}$ (Model~1), from the initial Fock states $\ket{0}$
and $\ket{2}$ can be approximately given as follows
\begin{eqnarray}
  \ket{\psi^{(0)}_{02}(t)} &\approx& \cos(\sqrt{2}\epsilon t)\ket{0}-i \sin(\sqrt{2}\epsilon
  t)\ket{2},
  \nonumber \\
  \ket{\psi^{(2)}_{02}(t)} &\approx& -i \sin(\sqrt{2}\epsilon t)\ket{0}+\cos(\sqrt{2}\epsilon
  t)\ket{2},
\label{free02}
\end{eqnarray}
respectively. These solutions are in a very good agreement with
the precise numerical solutions plotted in Figs.~\ref{fig01}(a)
and~\ref{fig01}(c). In the derivation of Eq.~(\ref{free02}), we
have ignored the contribution of $(\epsilon/\chi)^2$. The solution
$\ket{\psi^{(0)}_{02}(t)}$ can be referred to as a
three-dimensional squeezed vacuum~\cite{Miran96} or qutrit
squeezed vacuum.

The solutions in Eq.~(\ref{free02}) can be interpreted as
Rabi-type oscillations between the states $\ket{0}$ and $\ket{2}$
in an artificial two-level system dynamically truncated (or
generated) from the infinite-dimensional anharmonic system
described by the Hamiltonian $\hat{H}_{02}$ for $\epsilon\ll
\chi$. Thus, this phenomenon corresponds to a two-photon blockade,
where the excitation of more than two photons is
prohibited~\cite{Miran13}. The evolutions shown in
Figs.~\ref{fig01}(b) and~\ref{fig01}(d) are practically
negligible. This photon blockade can be physically understood via
the energy spectrum and resonances shown in Fig.~\ref{fig02}. Note
that our model, which leads to two-photon blockade induced by
two-photon driving, differs from that in Ref.~\cite{Miran13},
where a single-photon driving was assumed. We also mention that
the state $\ket{2}$ in the solution $\ket{\psi^{(0)}_{13}(t)}$ is
not populated, which is in contrast to the dissipative evolution
analyzed in the next section (see Table~I for comparison).

The dissipation-free system, given by the Hamiltonian
$\hat{H}_{13}$ (Model~2), evolves from the initial Fock states
$\ket{m}$ (for $m=0,1,3,4)$ as follows:
\begin{eqnarray}
  \ket{\psi^{(0)}_{13}(t)} &\approx& \cos(\tfrac15\epsilon t)\ket{0}-i \sin(\tfrac15\epsilon
  t)\ket{4},
  \nonumber \\
  \ket{\psi^{(1)}_{13}(t)} &\approx& \cos(\sqrt{6}\epsilon t)\ket{1}-i \sin(\sqrt{6}\epsilon
  t)\ket{3},
  \nonumber \\
  \ket{\psi^{(3)}_{13}(t)} &\approx& -i \sin(\sqrt{6}\epsilon t)\ket{1}+\cos(\sqrt{6}\epsilon
  t)\ket{3},
  \nonumber \\
  \ket{\psi^{(4)}_{13}(t)} &\approx& -i \sin(\tfrac15\epsilon t)\ket{0}+\cos(\tfrac15\epsilon
  t)\ket{4},
\label{free13}
\end{eqnarray}
respectively. These are relatively good approximations of our
precise numerical solutions plotted in
Figs.~\ref{fig03}(a),~\ref{fig03}(b), and~\ref{fig03}(d). In the
derivations of these approximate solutions for
$\ket{\psi^{(n)}_{13}(t)}$, the same as for
$\ket{\psi^{(n)}_{02}(t)}$, the contribution of
$(\epsilon/\chi)^2$ was omitted.

We interpret the solutions in Eqs.~(\ref{free13}) analogously to
those in Eqs.~(\ref{free02}), i.e., as generalized two-photon
(four-photon) blockades and Rabi-type oscillations between the
states $\ket{1}$ and $\ket{3}$ ($\ket{0}$ and $\ket{4}$) in an
artificial two-level system dynamically truncated from the
infinite-dimensional system of the Hamiltonian $\hat{H}_{13}$ if
$\epsilon\ll \chi$. The contributions of other Fock states are
practically negligible, as seen in Figs.~\ref{fig03}. These
phenomena can be easily understood by analyzing the energy spectra
and resonances shown in Fig.~\ref{fig04}.

For a comparison, we also recall the well-known approximate
solutions for the pure-state evolutions, under the interaction
described by the Hamiltonian $\hat{H}_{\rm
usual}$~\cite{Leonski94}:
\begin{eqnarray}
  \ket{\psi^{(0)}_{\rm usual}(t)} &\approx& \cos(\epsilon t)\ket{0}-i \sin(\epsilon
  t)\ket{1},
  \nonumber \\
  \ket{\psi^{(1)}_{\rm usual}(t)} &\approx& -i \sin(\epsilon t)\ket{0}+\cos(\epsilon
  t)\ket{1},
\label{free01}
\end{eqnarray}
assuming the initial vacuum and single-photon states,
respectively. These solutions can be interpreted as single-photon
blockade in the dissipation-free regime and two-dimensional (or
qubit) coherent states~\cite{Miran94}.

\section{Steady-state photon blockades via two-photon dissipation}

Here we explain in detail the occurrence of various kinds of
steady-state engineered PB effects, when the systems described by
the Hamiltonians $\hat{H}_{02}$ and $\hat{H}_{13}$ are affected by
two-photon loss mechanisms, as schematically shown in
Fig.~\ref{fig05}.

\subsection{Master equation describing two-photon absorption}


We assume that the system (s), described by the Hamiltonian
$\hat{H}_{kl}$, is coupled to an engineered reservoir (r) via
two-photon processes (see, e.g.,
Refs.~\cite{Simaan78,Gilles93,Gilles94, Guerra97,Dodonov97,
Everitt13, Ispasoiu00,Klimov03, Buks06, Yurke06, Karasik08,
Boissonneault12, Voje13,Albert14}) as described by $\hat H=\hat
H_{\rm s}+\hat H_{\rm r}+\hat H_{\rm sr},$ where
\begin{equation}
\hat H_{\rm sr}=\hbar g_{\rm sr}[\hat a^{2}\hat
\Gamma^{\dagger}+(\hat a^{\dagger})^{2}\hat \Gamma],
\label{H_Gamma}
\end{equation}
and $\hat H_{\rm r}$ can be given, depending on the physical
realization,  by, e.g., $\hbar\sum_{n}\omega_{n}\hat
\sigma_{z}^{(n)}$ or $\hbar\sum_{n}\omega_{n}\hat
a_{n}^{\dagger}\hat a_{n}$, while the collective reservoir
annihilation operator $\hat \Gamma$ is given by $\sum_{n}\hat
\sigma_{-}^{(n)}$ or $\sum_{n}\hat a_{n}$, respectively. Moreover,
$g_{\rm sr}$ is the system-reservoir coupling strength;
$\omega_{n}$ is the frequency of the $n$th mode of the reservoir,
$\hat \sigma_{z}^{(n)}$ and $\hat \sigma_{-}^{(n)}$ are the spin
operators for the $n$th qubit, defined analogously to those below
Eq.~(\ref{H1}), and $\hat a_{n}$($\hat a_{n}^{\dagger}$) is the
annihilation (creation) operator of the $n$th mode of the
reservoir. Thus, the evolution, under the Markov approximation, of
the reduced density matrix for the system can be given by the
following two-photon-absorption master equation in the Lindblad
form assuming zero temperature of the
reservoir~\cite{Simaan78,Gilles93,Guerra97,Karasik08},
\begin{equation}
\frac{d\hat\rho}{dt} = {\cal L}\hat\rho\equiv\ -\frac{i}{\hbar}[\hat
H_{\rm s},\hat \rho]+\gamma{\cal D}[\hat a^2]\hat \rho, \label{ME}
\end{equation}
where the superoperator ${\cal D}$ is defined by ${\cal D}[\hat
L]\hat\rho=\hat L\hat \rho \hat L^{\dag}-\frac{1}{2} (\hat L^\dag
\hat L \hat \rho + \hat \rho \hat L^{\dag}\hat L)$, and ${\cal L}$
is sometimes referred to as the Liouvillian (or Lindbladian)
superoperator. Moreover, $\gamma=\gamma_2$ is the two-photon
damping constant (two-photon decay rate).

It is worth noting that a single-mode squeezed state can be
generated by this two-photon absorption process via pure
dissipation~\cite{Gilles93}. This can be readily concluded by
noting that the Hamiltonian in Eq.~(\ref{H_Gamma}) corresponds to
a prototype squeezing Hamiltonian in the parametric approximation,
when the collective reservoir operator $\hat \Gamma$ is treated
classically. Thus, Eq.~(\ref{ME}) can be considered a master
equation obtained for a squeezing-generating reservoir. Note that
this equation is completely different from the standard master
equation for an amplifier whose reservoir consists of squeezed
white noise (squeezed-vacuum reservoir)~\cite{ScullyBook}.

A more general form of the master equation can
read~\cite{Everitt13}
\begin{equation}
\frac{d\hat\rho}{dt} = {\cal L'}\hat\rho=-\frac{i}{\hbar}[\hat
H_{\rm s},\hat\rho]+\gamma_\perp{\cal D}[\hat a^\dagger \hat
a]\hat \rho+\gamma_1{\cal D}[\hat a]\hat \rho+\gamma_{2}[\hat
a^2]\hat \rho, \label{ME2}
\end{equation}
to include also single-photon absorption with its decay rate
$\gamma_1$, and pure dephasing with its rate $\gamma_\perp$, in
addition to two-photon absorption. Note that it is still assumed
in this equation that the reservoir is at zero temperature, so
there is no transfer of reservoir fluctuations into the system.

Our former studies~\cite{Liu10,Miran13} showed that the standard
realizations of photon blockade can be very sensitive to these
thermal fluctuations. Nevertheless, for simplicity, we apply here
the zero-temperature master equations, given by Eq.~(\ref{ME}) or,
equivalently, by Eq.~(\ref{ME2}) assuming that
\begin{equation}
 0 \approx \gamma_\perp \approx \gamma_1 \ll \gamma_2 \ll \epsilon \ll
 \chi.
\label{N3}
\end{equation}
In order to visualize the steady-state solutions of the
two-photon-loss master equation, given by Eqs.~(\ref{ME}), we plot
their Wigner functions and photon-number probabilities
$p_n=\langle n |\hat \rho_{{\rm ss}}| n\rangle$ in
Figs.~\ref{fig06}--\ref{fig13}. Moreover, Fig.~\ref{fig14} shows
analogous solutions of the single-photon-loss master equation,
given in Eq.~(\ref{ME2}) assuming that the single-photon decay
rate $\gamma_1$ is dominantly larger than the two-photon decay
rate $\gamma_2$ and the dephasing rate $\gamma_\perp$.

\subsection{Steady-state solutions of the master equation}

Now we present our precise numerical and approximate analytical
steady-state solutions of the two-photon absorption master
equation, given by Eq.~(\ref{ME}), to show explicitly how photon
blockade in the discussed engineered reservoir depends on initial
states.

\begin{figure}

\fig{\includegraphics[height=35mm]{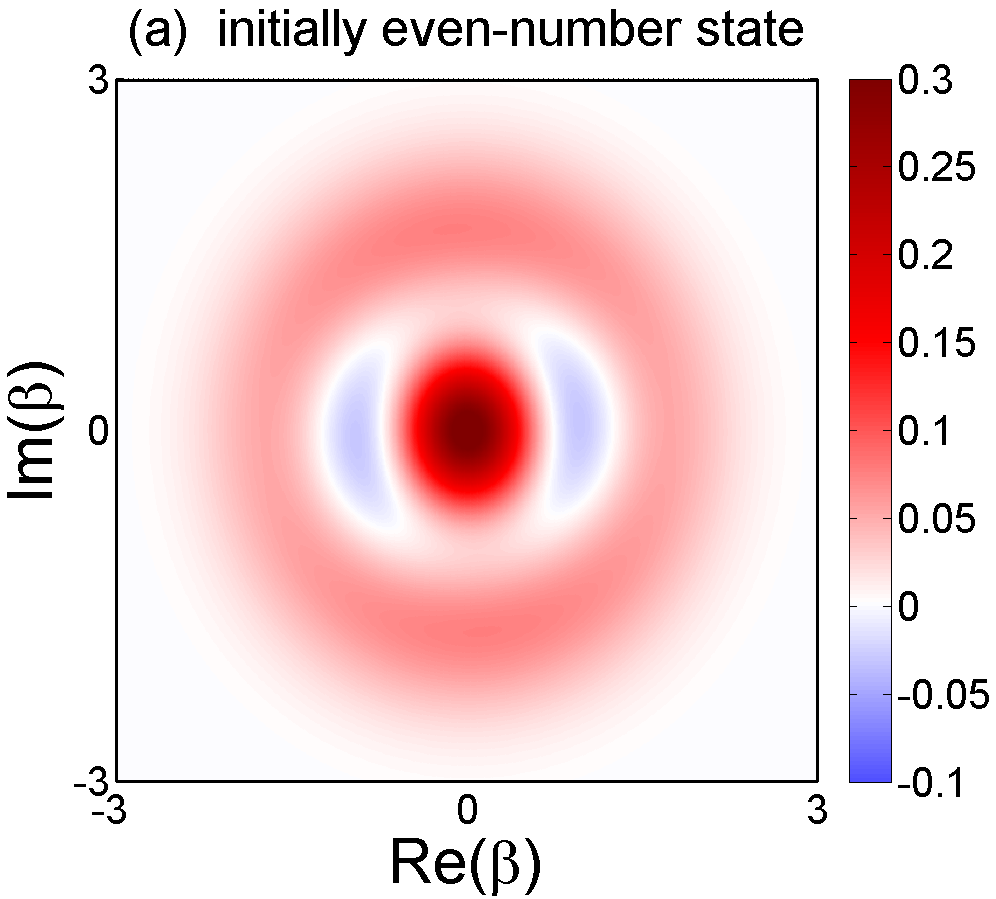}}
\fig{\includegraphics[height=35mm]{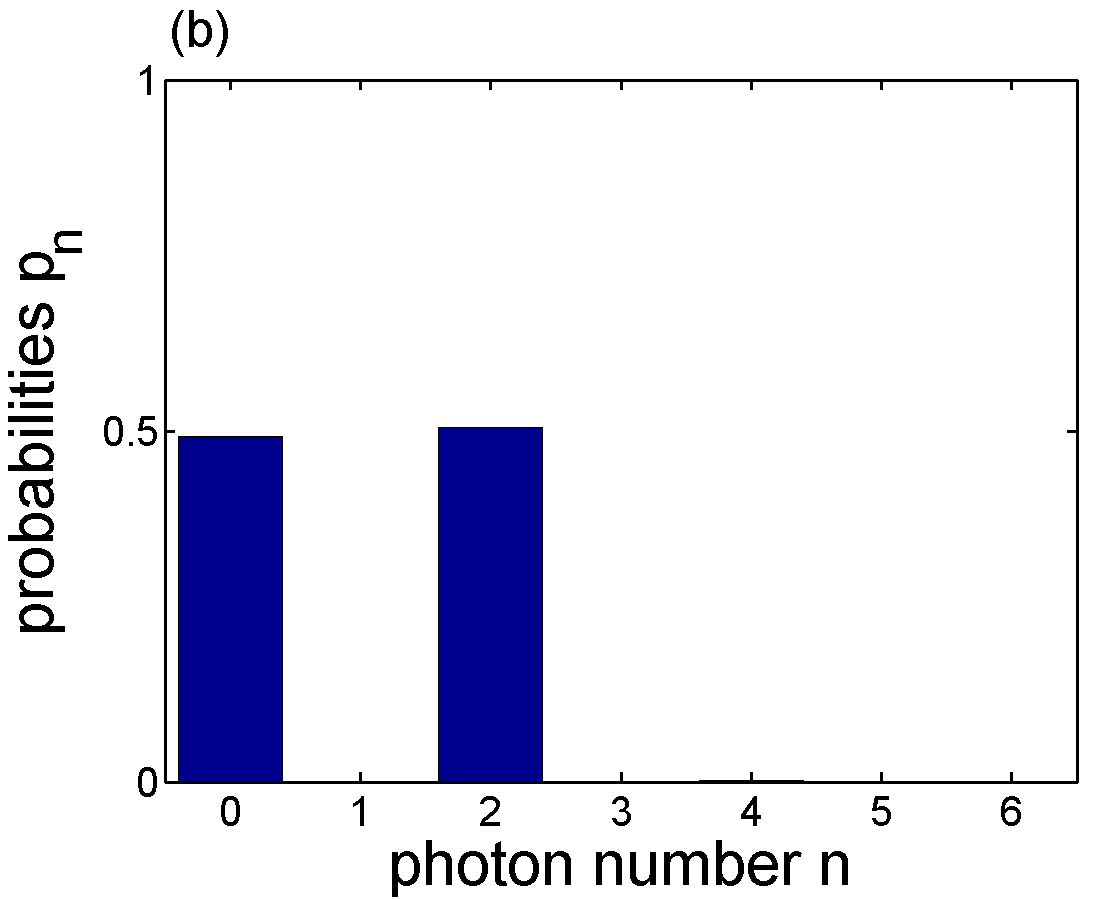}}

\vspace*{1mm}
\fig{\includegraphics[height=35mm]{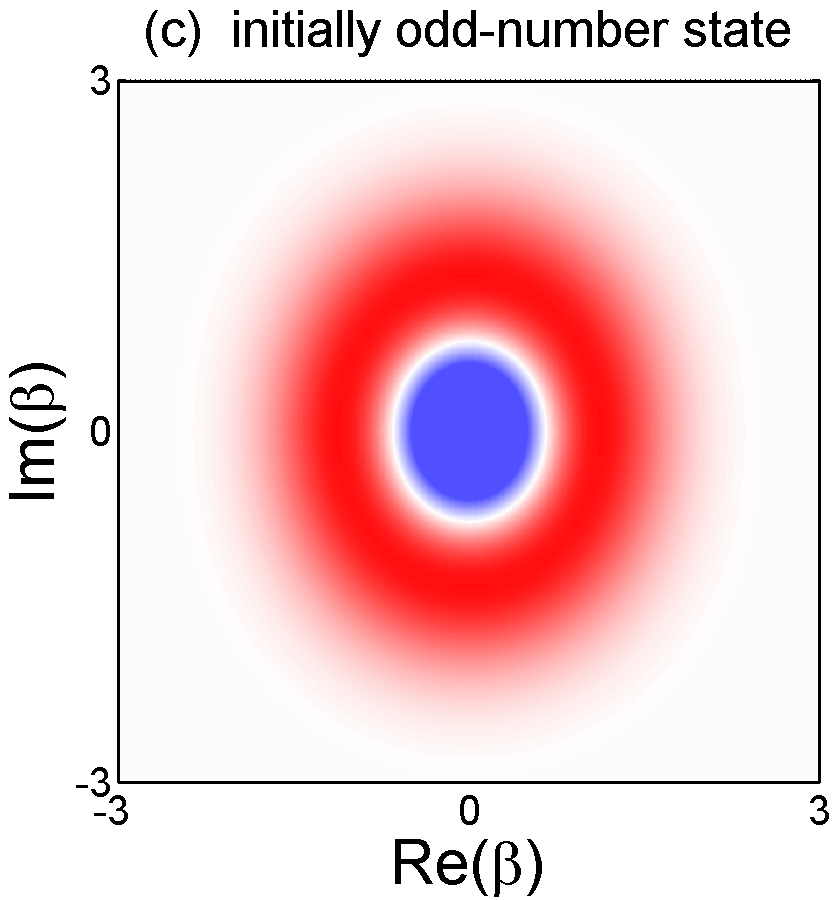}}\hspace{5mm}
\fig{\includegraphics[height=35mm]{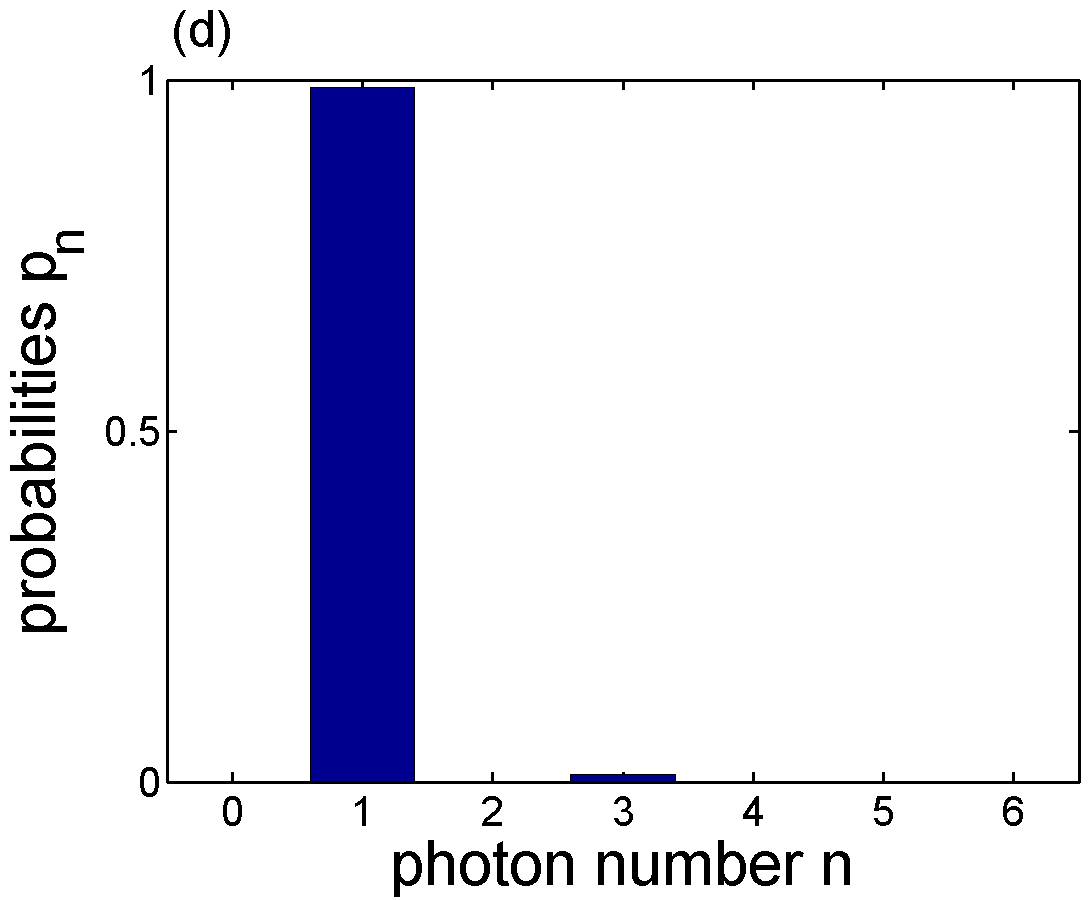}}

\caption{(Color online) Model~1 with two-photon dissipation: (a,c)
The Wigner functions $W(\beta)$ and (b,d) the photon-number
probabilities $p_n$ for the steady-state solutions $\hat
\rho^{02}_{{\rm ss}}$ of the master equation~(\ref{ME}) with the
Hamiltonian $\hat H_{02}$, given by Eq.~(\ref{H02}), assuming the
cavity field to be initially in an arbitrary (a,b) even- or (c,d)
odd-number state. We set the ratios of the driving field strength
$\epsilon$ and the Kerr nonlinear coupling $\chi$, and of the
damping constant $\gamma$ and $\epsilon$ to be small and equal to
$\delta=\epsilon/\chi=1/6$ and $\delta'=\gamma/\epsilon=1/25$. The
color codes in panel (c) (and all other figures of the Wigner
functions) are the same as in panel (a). Note that the negative
regions of the Wigner functions are marked in blue.} \label{fig06}
\end{figure}

\begin{figure}

\fig{\includegraphics[height=38mm]{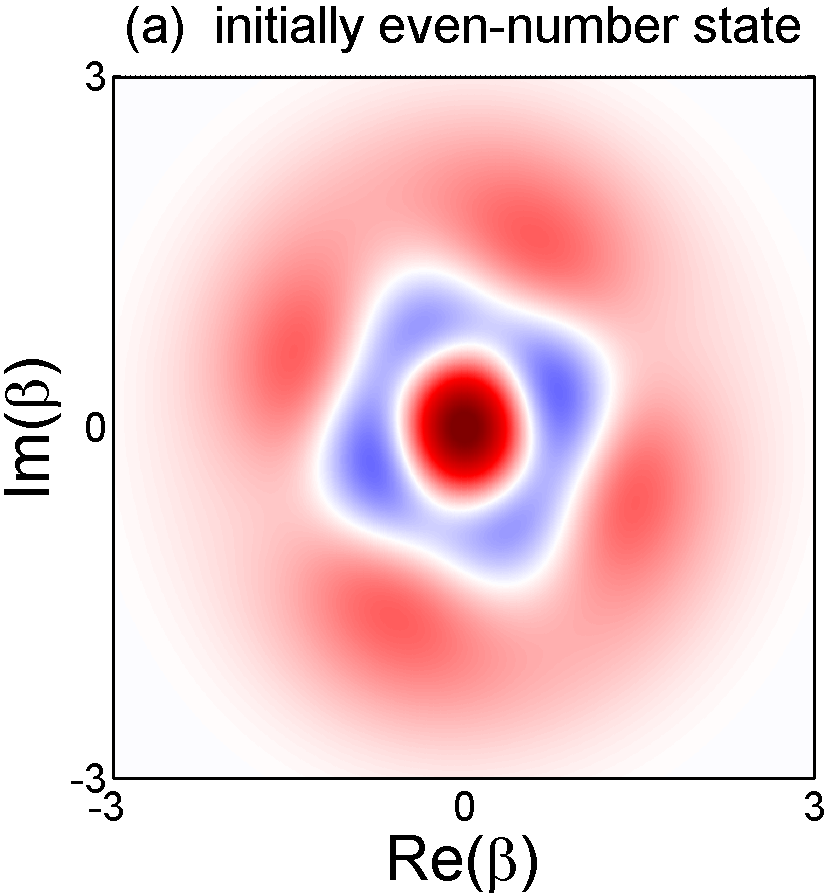}}
\fig{\includegraphics[height=38mm]{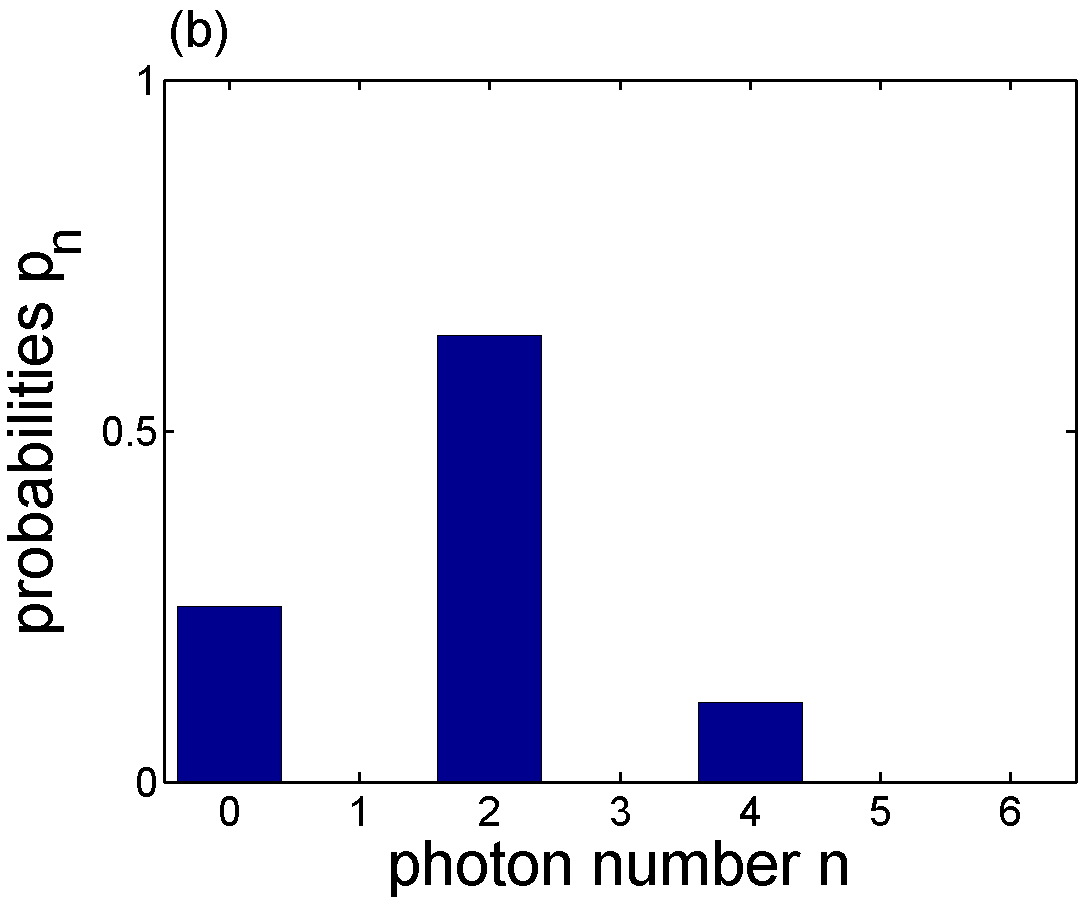}}

\fig{\includegraphics[height=38mm]{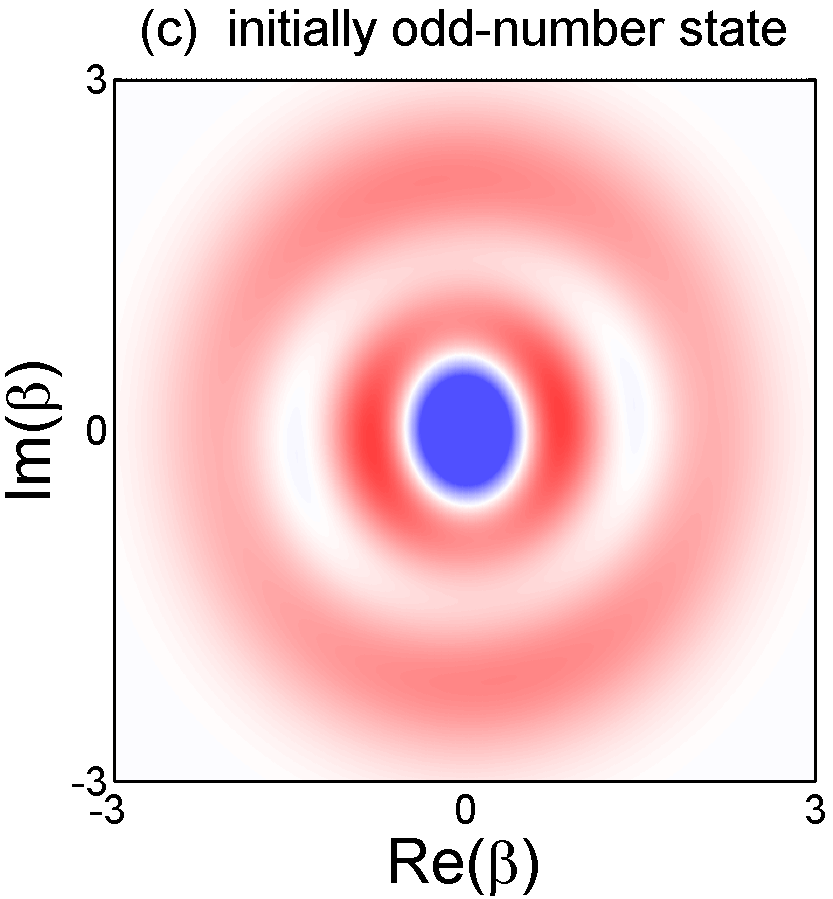}}
\fig{\includegraphics[height=38mm]{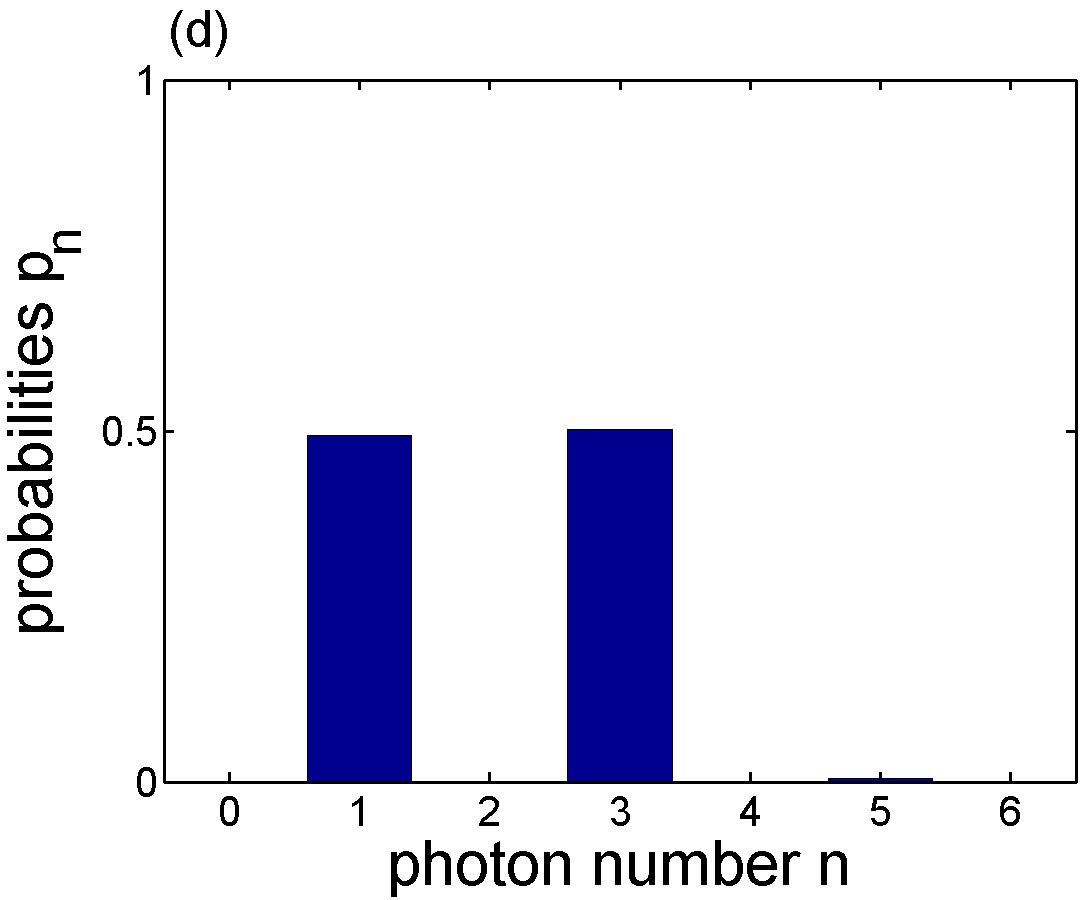}}

\vspace*{-4mm} \caption{(Color online) Model~2 with two-photon
dissipation: Same as in figure~\ref{fig06}, but for the
steady-state solutions $\hat \rho^{13}_{{\rm ss}}$ of the master
equation~(\ref{ME}) with the Hamiltonian $\hat H_{13}$, given by
Eq.~(\ref{Hkl}) for $k=1,l=3$.} \label{fig07}
\end{figure}

\begin{figure}

\fig{\includegraphics[height=50mm]{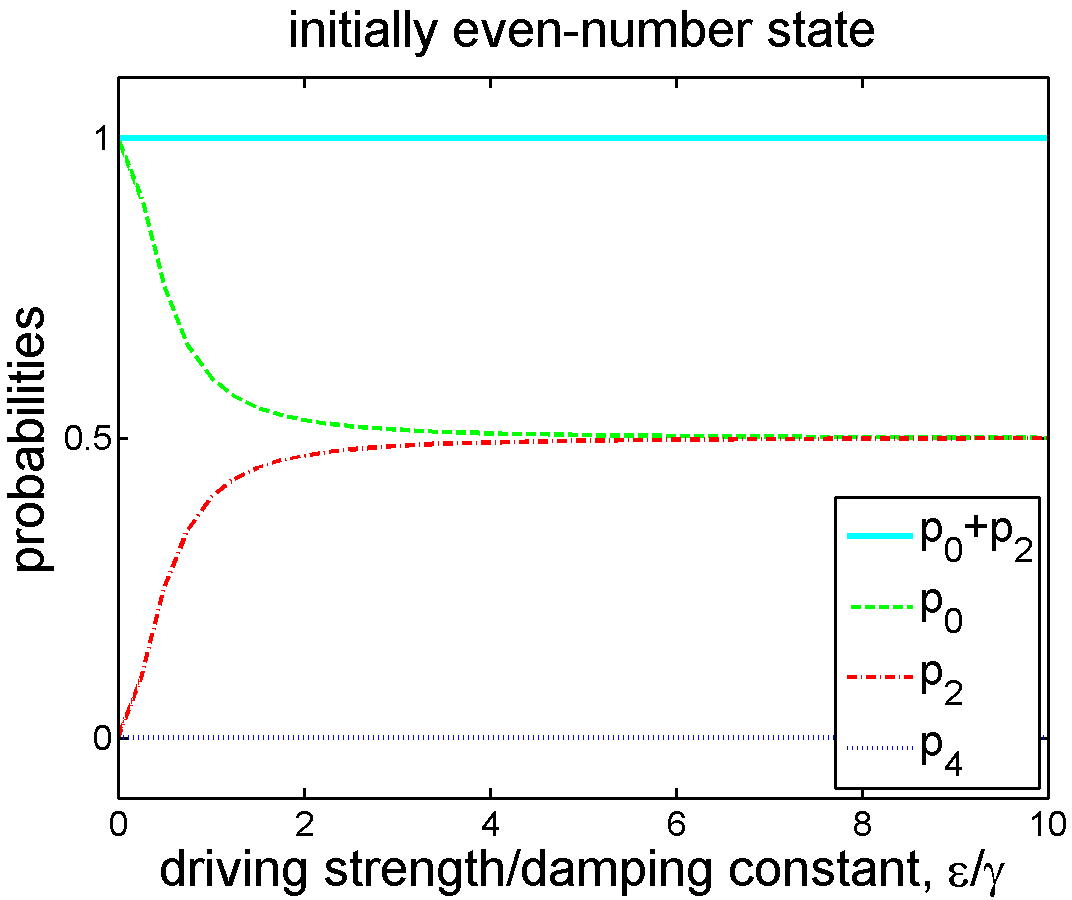}}

\vspace*{-4mm} \caption{(Color online) Model~1 with two-photon
dissipation: The photon-number probabilities $p_n=\bra{n}\hat
\rho^{02}_{{\rm ss}}\ket{n}$ and the fidelity $F=p_0+p_2$ of the
photon blockade versus the driving field strength $\epsilon$, in
units of the damping constant $\gamma$, assuming the initial state
to have an even number of photons and $\gamma/\chi=1/150$. The
analogous figure for an initial odd-number state is omitted since
$p_1\approx 1$ [see Fig.~\ref{fig06}(d)], at least, for
$\epsilon/\gamma\in[0,10]$. For brevity, analogous plots for
Model~2 are not presented here either.} \label{fig08}
\end{figure}

\begin{figure}

\fig{\includegraphics[height=45mm]{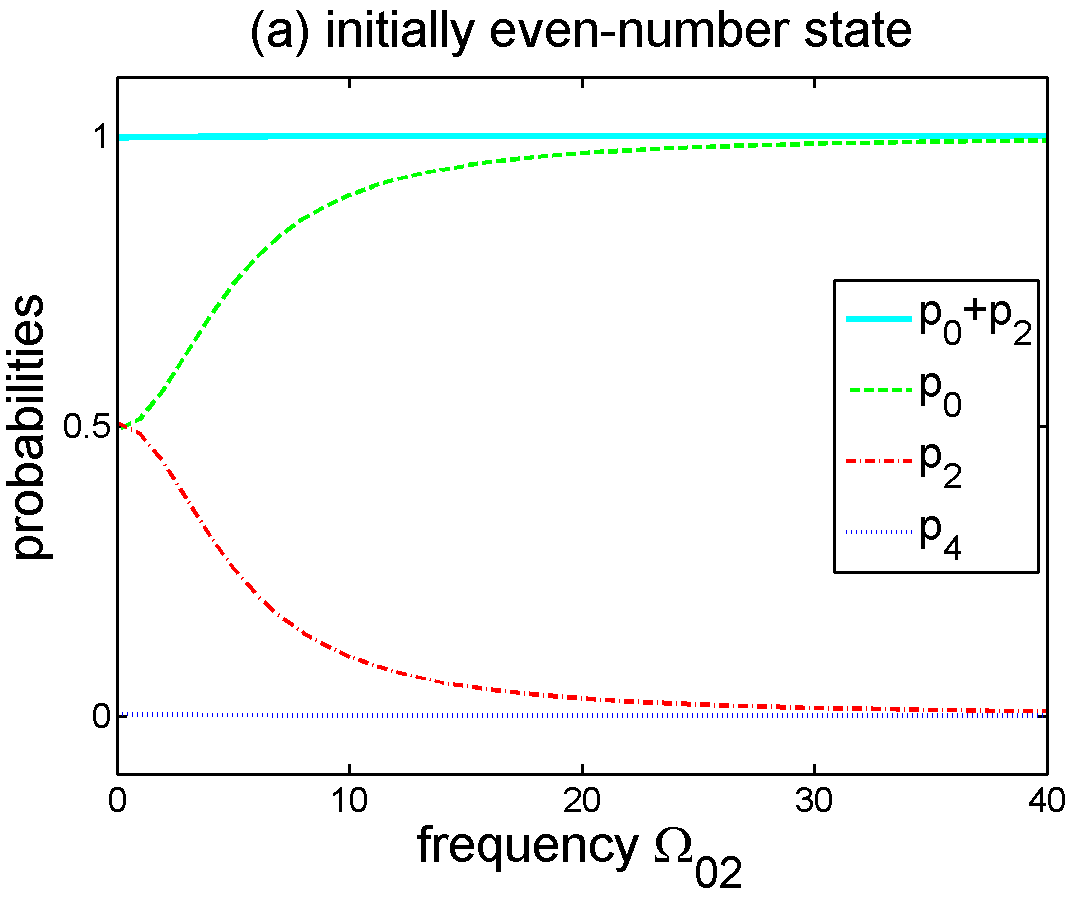}}

\fig{\includegraphics[height=45mm]{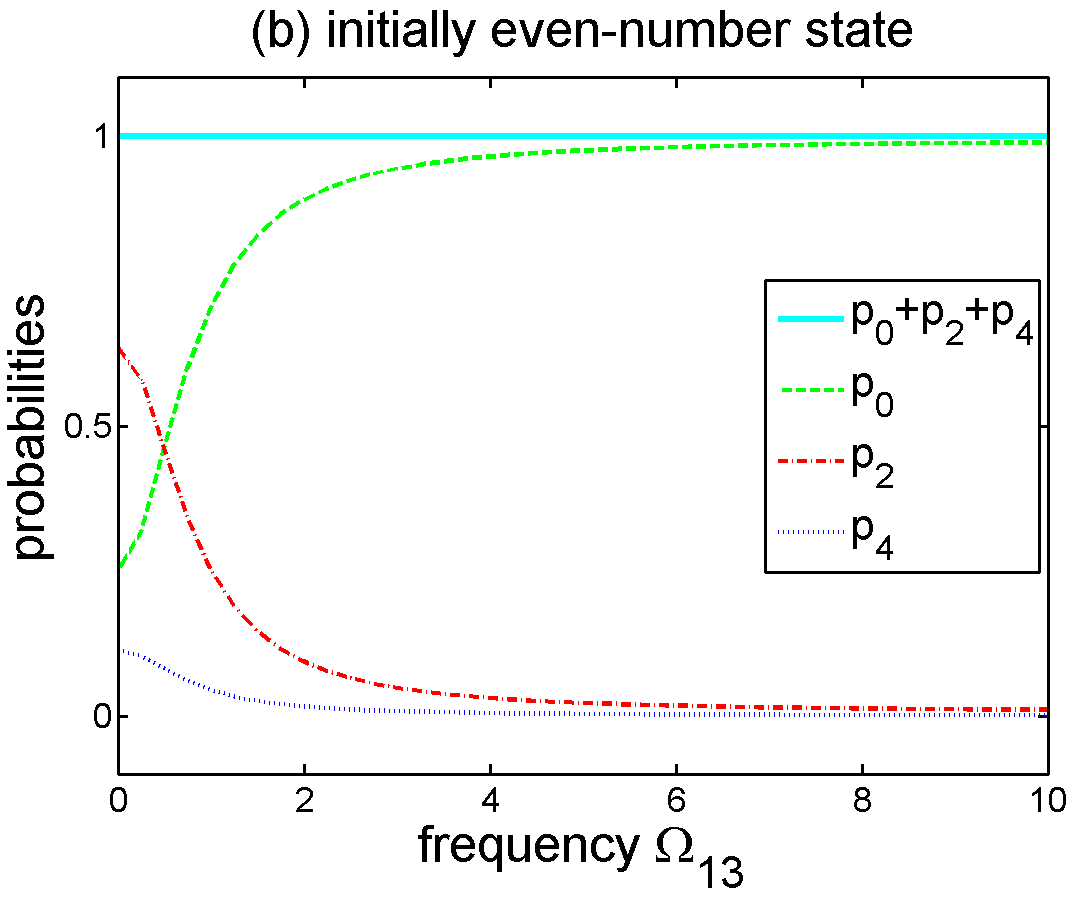}}

\fig{\includegraphics[height=45mm]{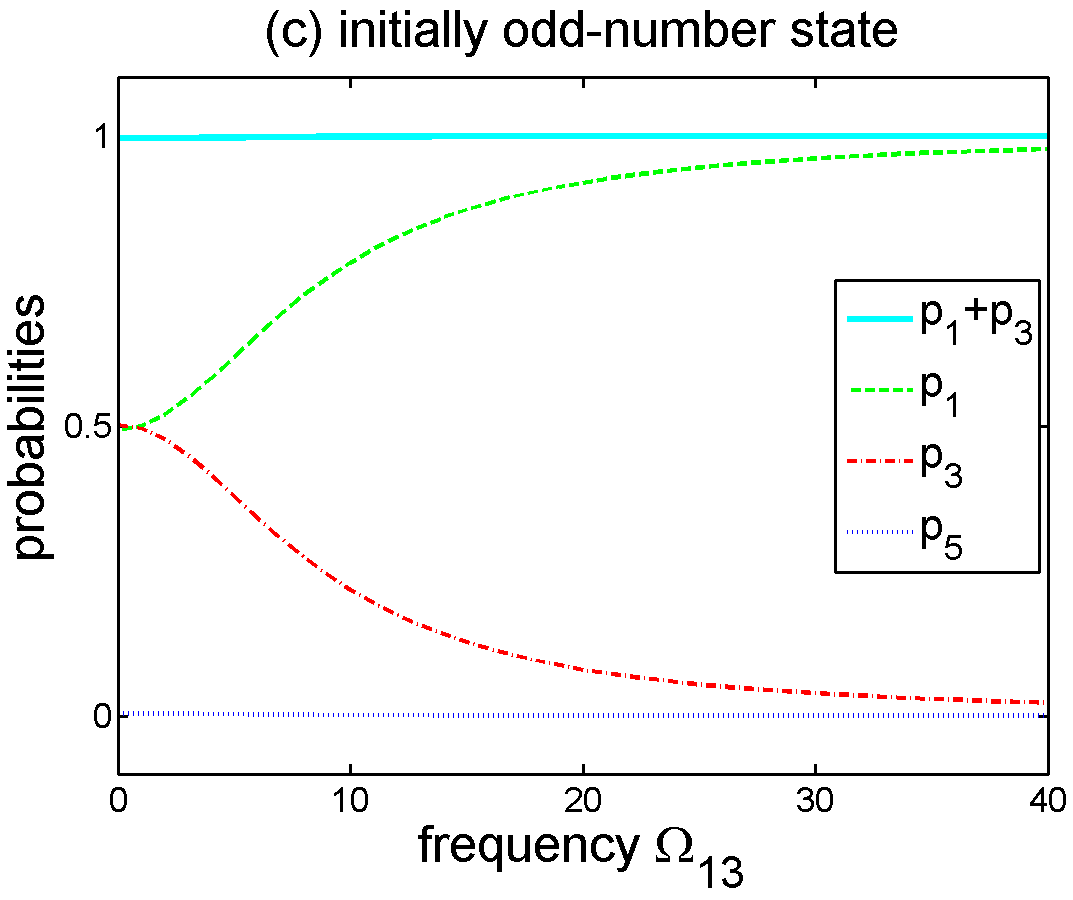}}

\vspace*{-4mm} \caption{(Color online) Model~1 in panel (a) and
Model~2 in (b,c) with two-photon dissipation: The photon-number
probabilities $p_n$ and the fidelity $F=\sum_n p_n$ of the photon
blockade versus the tuning frequency $\Omega_{kl}$ for the
steady-state solutions $\hat \rho_{{\rm ss}}$ of the master
equation~(\ref{ME}), for the Hamiltonian $\hat H_{\rm
s}(\Omega_{kl},\Sigma_{kl})$ with fixed $\Sigma_{kl}=0$, assuming
the initial state to have (a,b) an even or (c) odd number of
photons. Here we set $\delta=\epsilon/\chi=1/6$ and
$\delta'=\gamma/\epsilon=1/25$. The corresponding curves $p_n$
versus $\Omega_{02}$ for $\hat H_{\rm
s}(\Omega_{02},\Sigma_{02}=0)$ and the initial state with odd
number of photons are not presented here because $p_1\approx 1$
and  $p_3\approx 0$ in the whole studied interval, and this is
fully apparent from Fig.~\ref{fig06}(d) as well. It is seen that
even if $\Omega_{kl}\neq 0$, PB can still occur. Nevertheless, for
an initial even (odd) number state, the output steady state
approaches the vacuum (single-photon) state even for a relatively
small $\Omega_{kl}$. Note that plots in panels (a) and (c) look
very similar but they correspond to different probabilities.}
\label{fig09}
\end{figure}

\begin{figure}

\fig{\includegraphics[height=38mm]{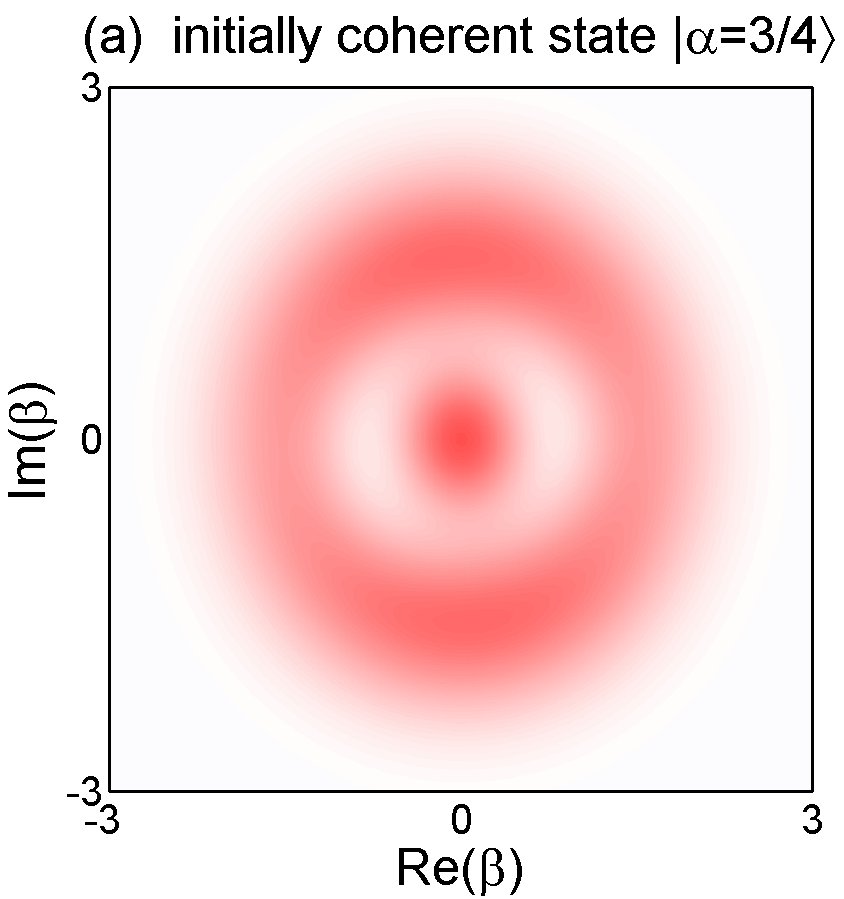}}
\fig{\includegraphics[height=38mm]{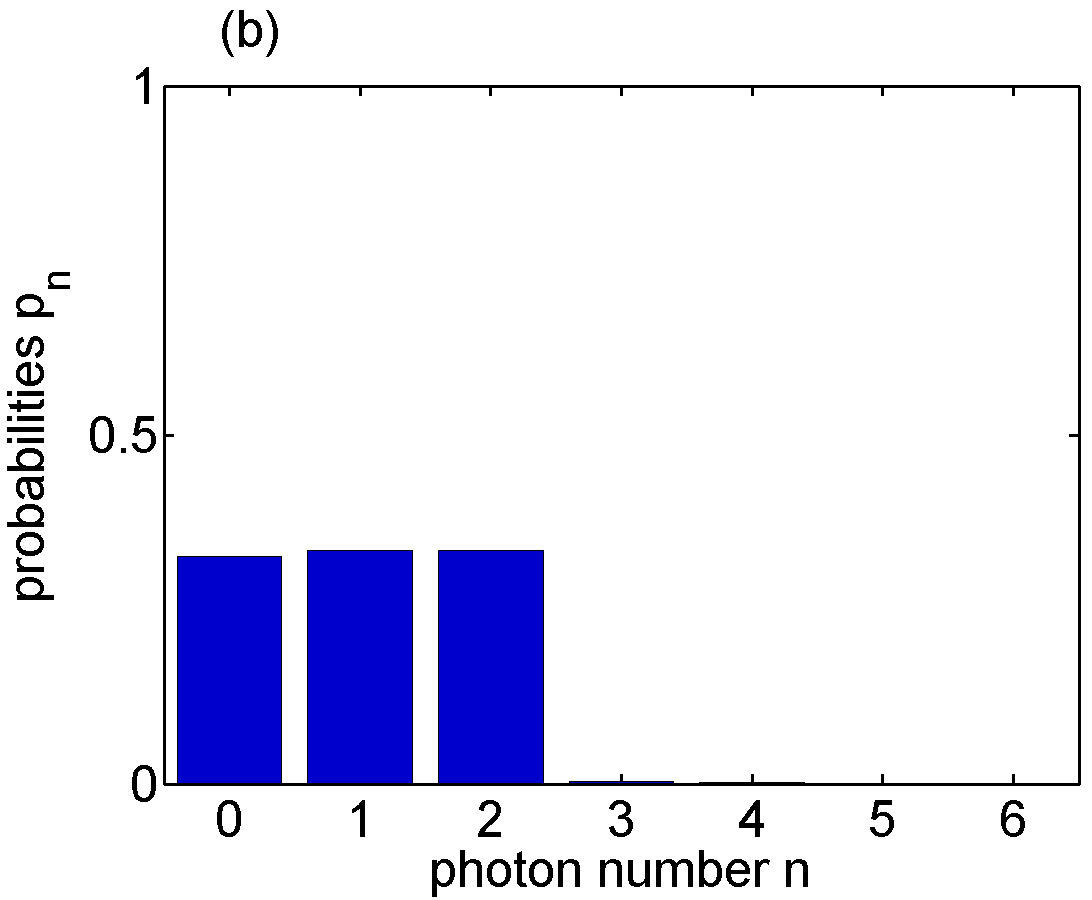}}

\vspace{1mm}

\fig{\includegraphics[height=38mm]{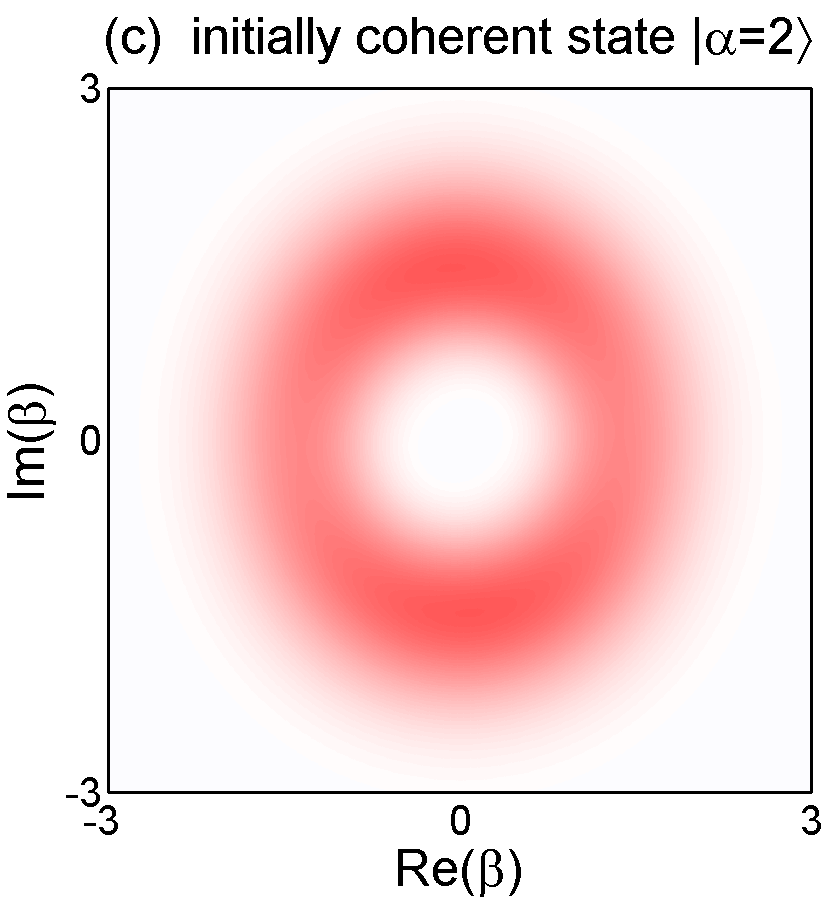}}
\fig{\includegraphics[height=38mm]{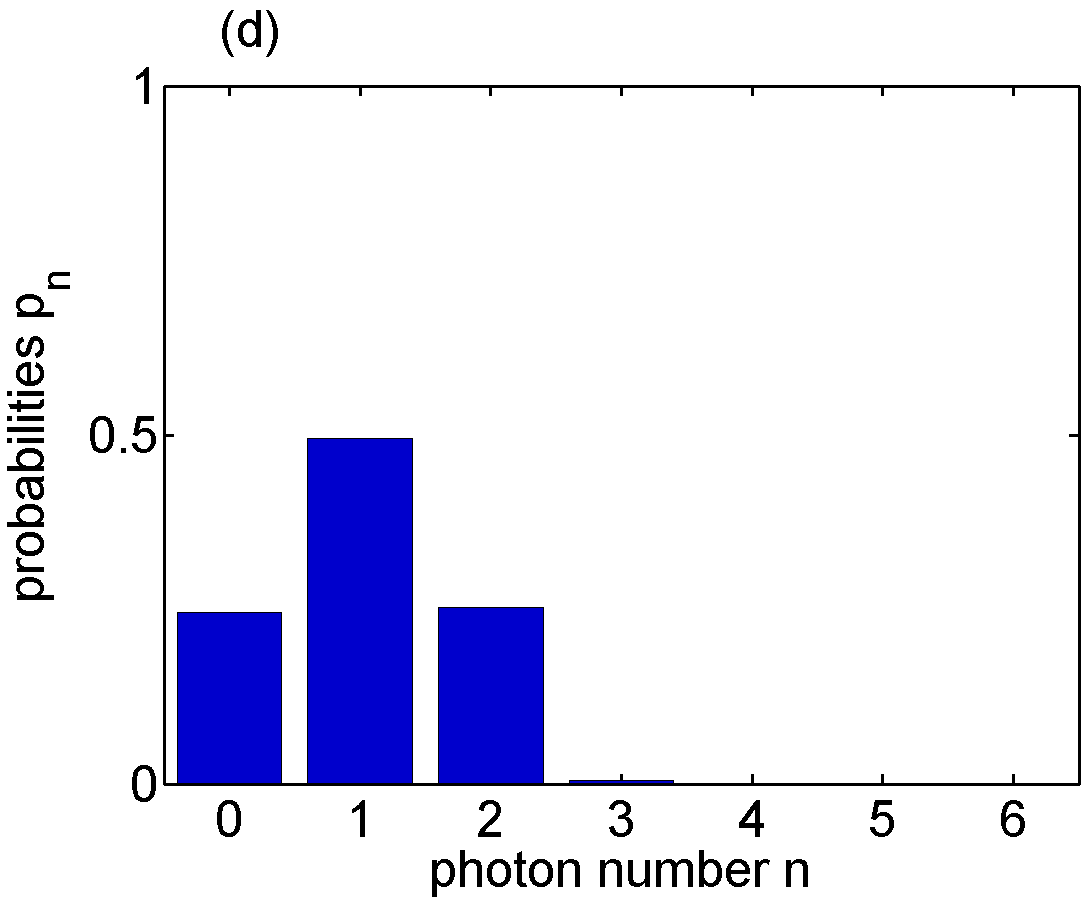}}

\vspace{1mm}

\fig{\includegraphics[height=38mm]{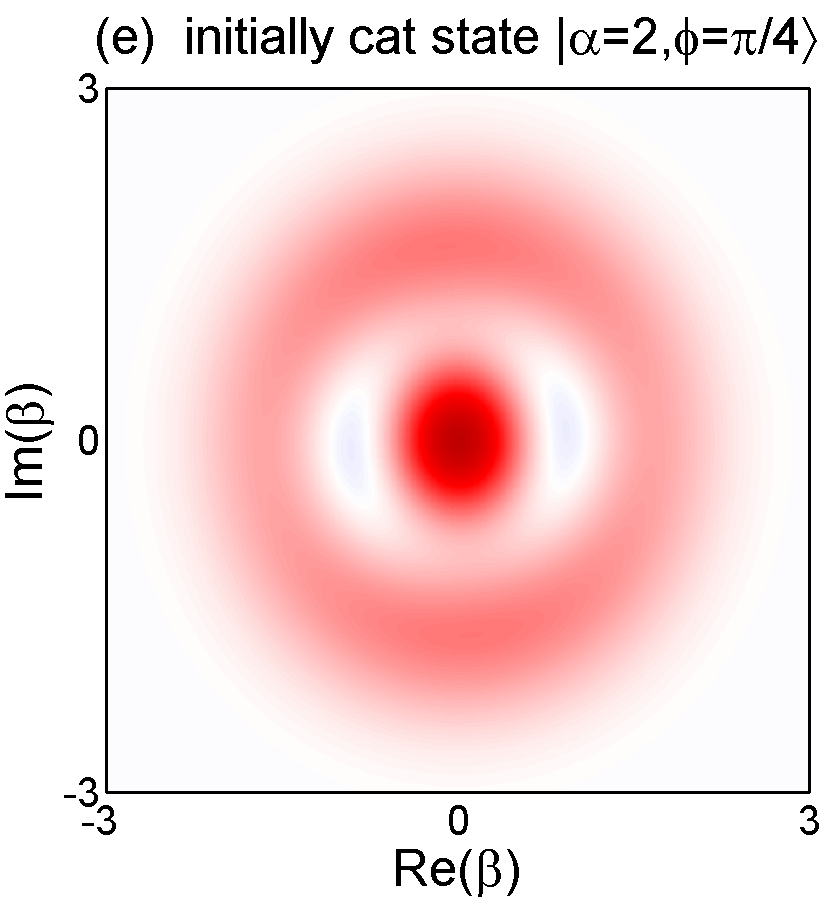}}
\fig{\includegraphics[height=38mm]{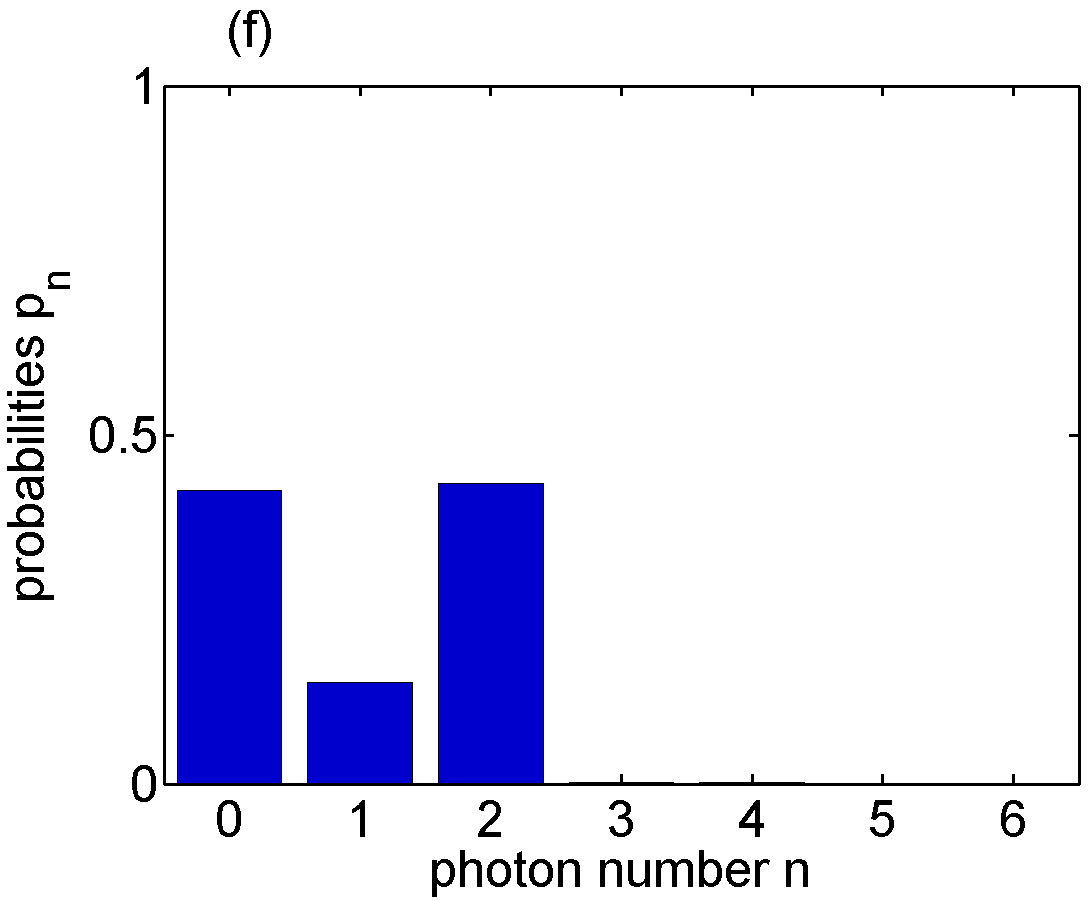}}

\caption{(Color online) Model~1 with two-photon dissipation: Same
as in Fig.~\ref{fig06}, but for different initial states: (a,b)
coherent state $|\alpha=3/4\rangle$, with $p_0\approx p_1\approx
p_2$, (c,d) coherent state $|\alpha=2\rangle$, with $p_1>
p_0\approx p_2$, and (e,f) the cat state
$|\alpha=2,\phi=\pi/4\rangle$, with $p_1< p_0\approx p_2$.}
\label{fig10}
\end{figure}

\begin{figure}

\fig{\includegraphics[height=38mm]{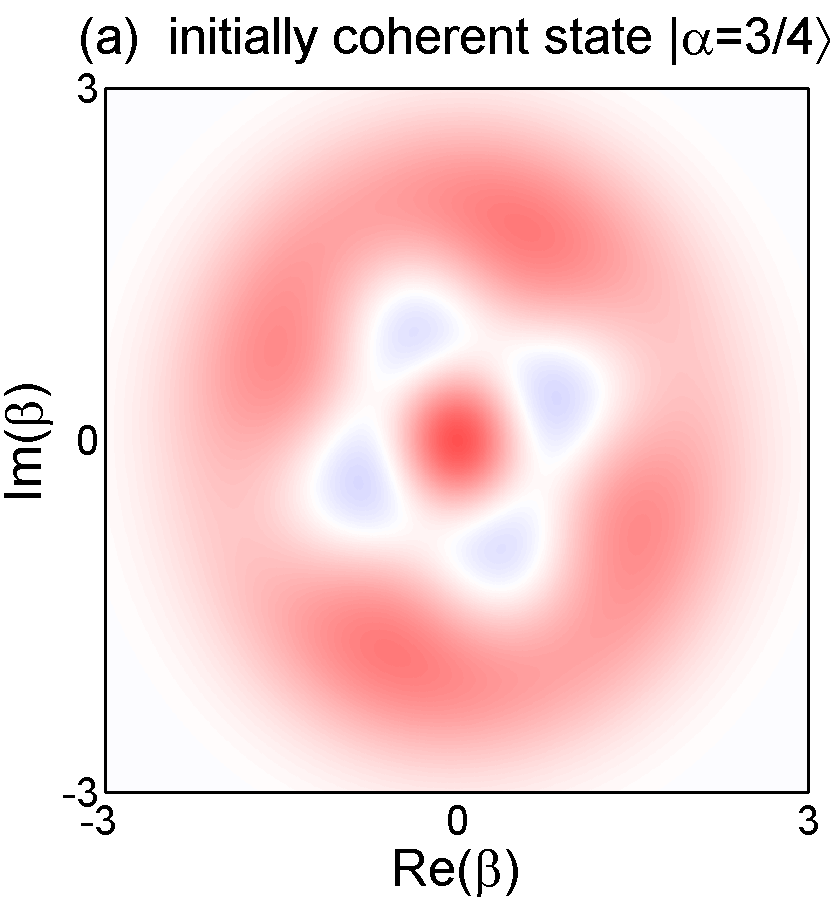}}
\fig{\includegraphics[height=38mm]{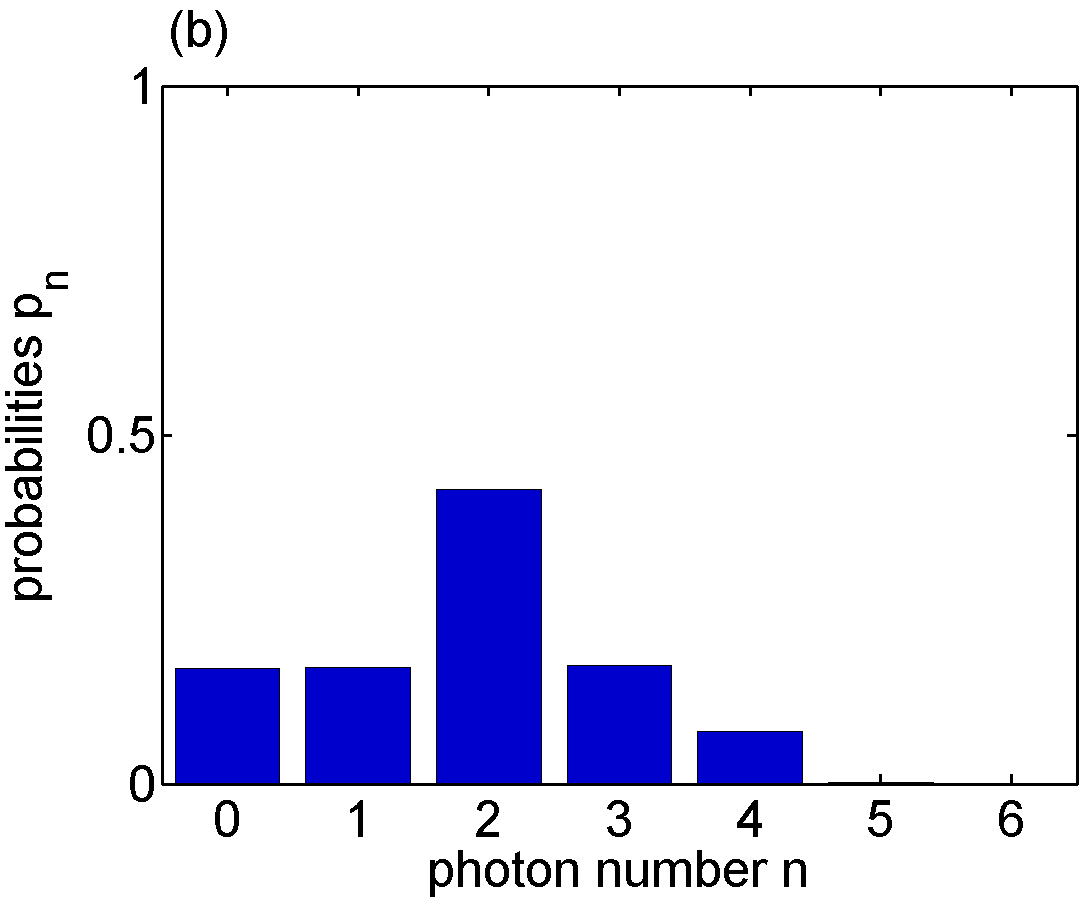}}

\vspace{1mm}

\fig{\includegraphics[height=38mm]{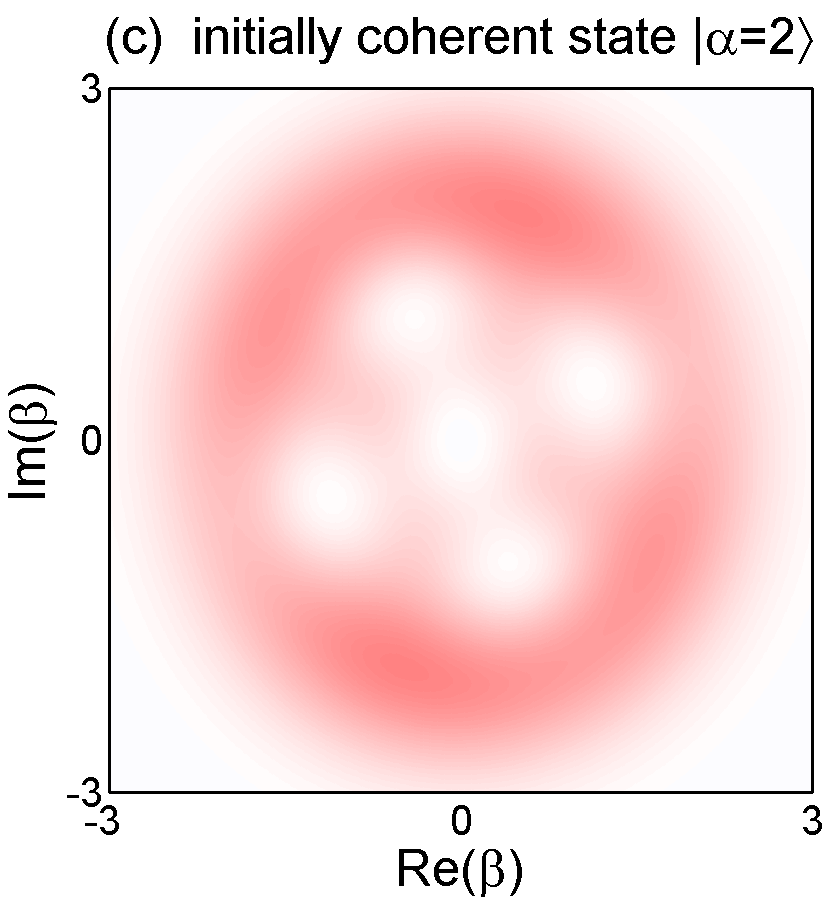}}
\fig{\includegraphics[height=38mm]{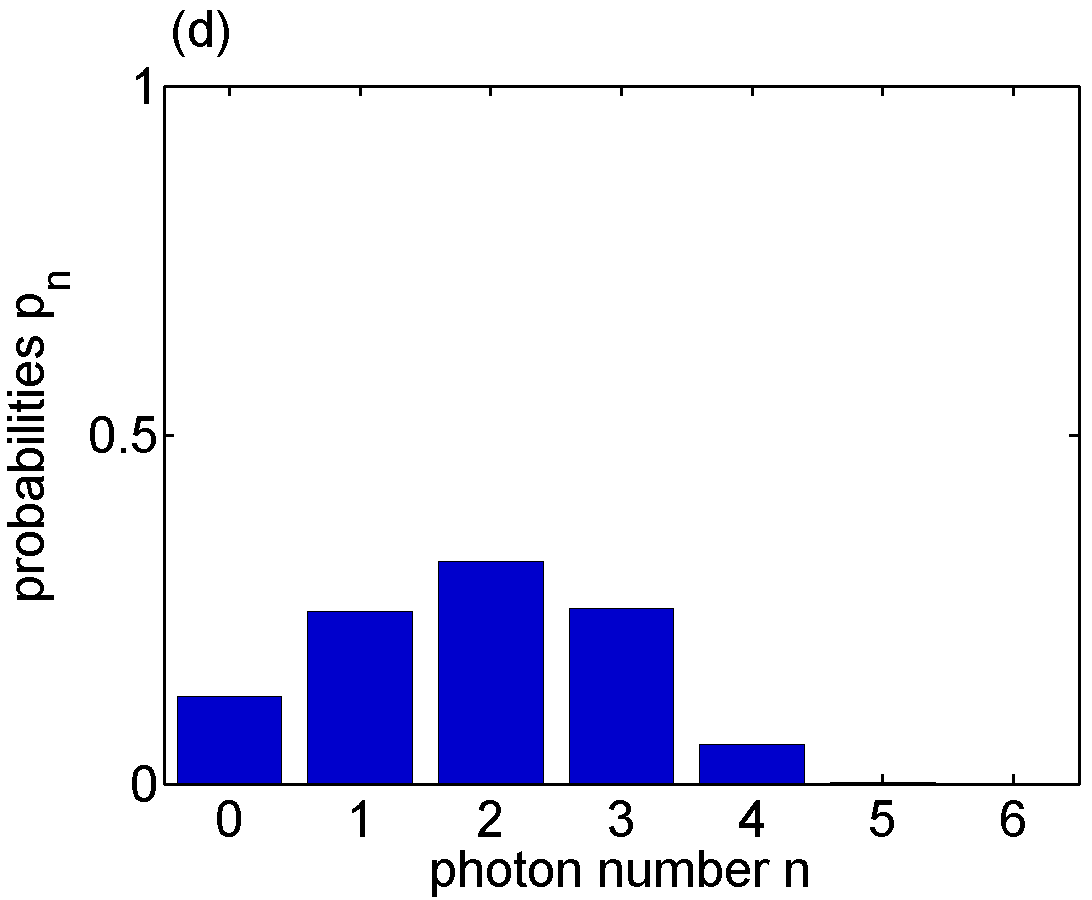}}

\vspace{1mm}

\fig{\includegraphics[height=38mm]{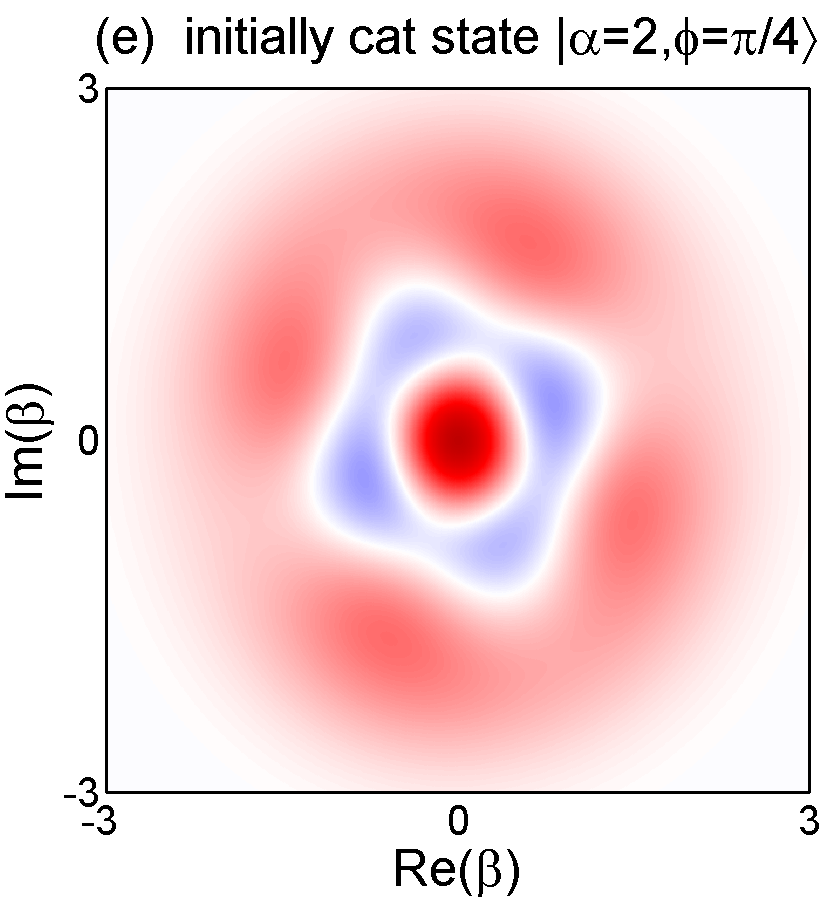}}
\fig{\includegraphics[height=38mm]{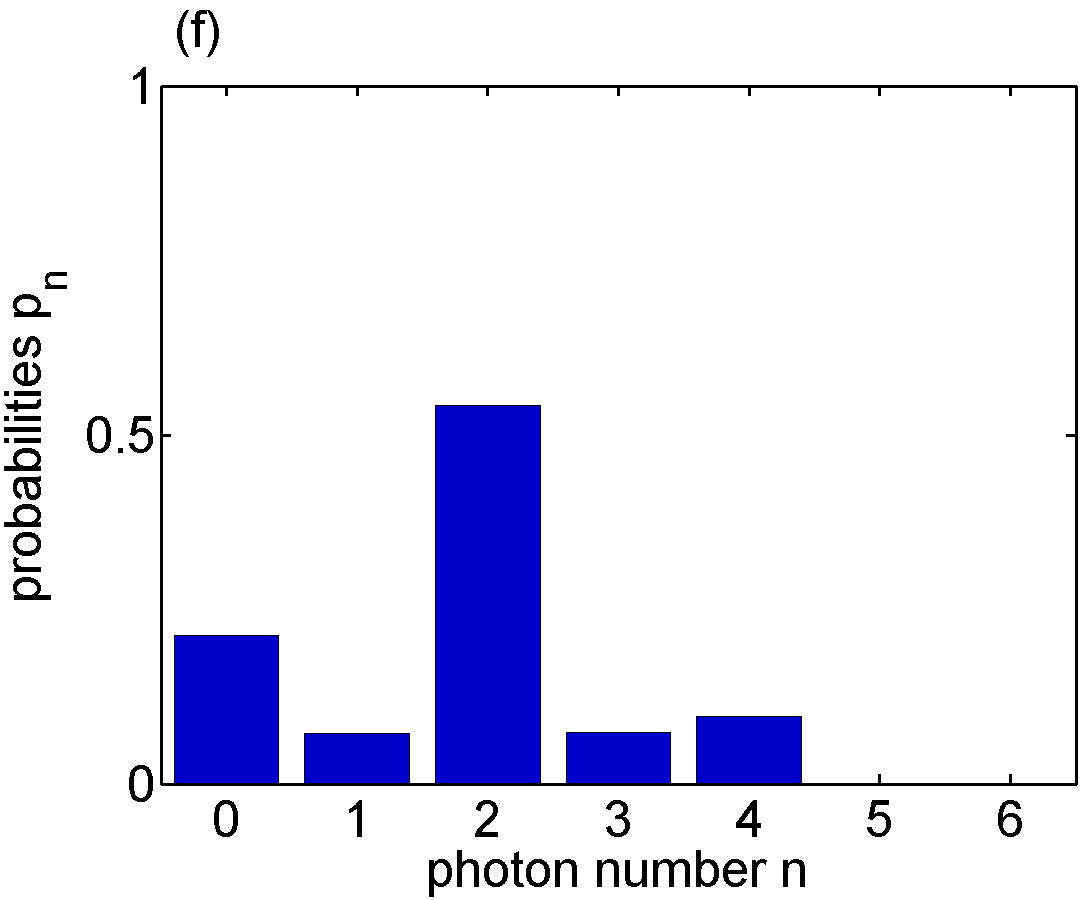}}

\caption{(Color online) Model~2 with two-photon dissipation: Same
as in Fig.~\ref{fig10}, but for the steady-state solutions $\hat
\rho^{13}_{{\rm ss}}$ of the master equation~(\ref{ME}) with the
Hamiltonian $\hat H_{13}$, given by Eq.~(\ref{Hkl}).}
\label{fig11}
\end{figure}

\begin{figure}
 \fig{ \includegraphics[height=35mm]{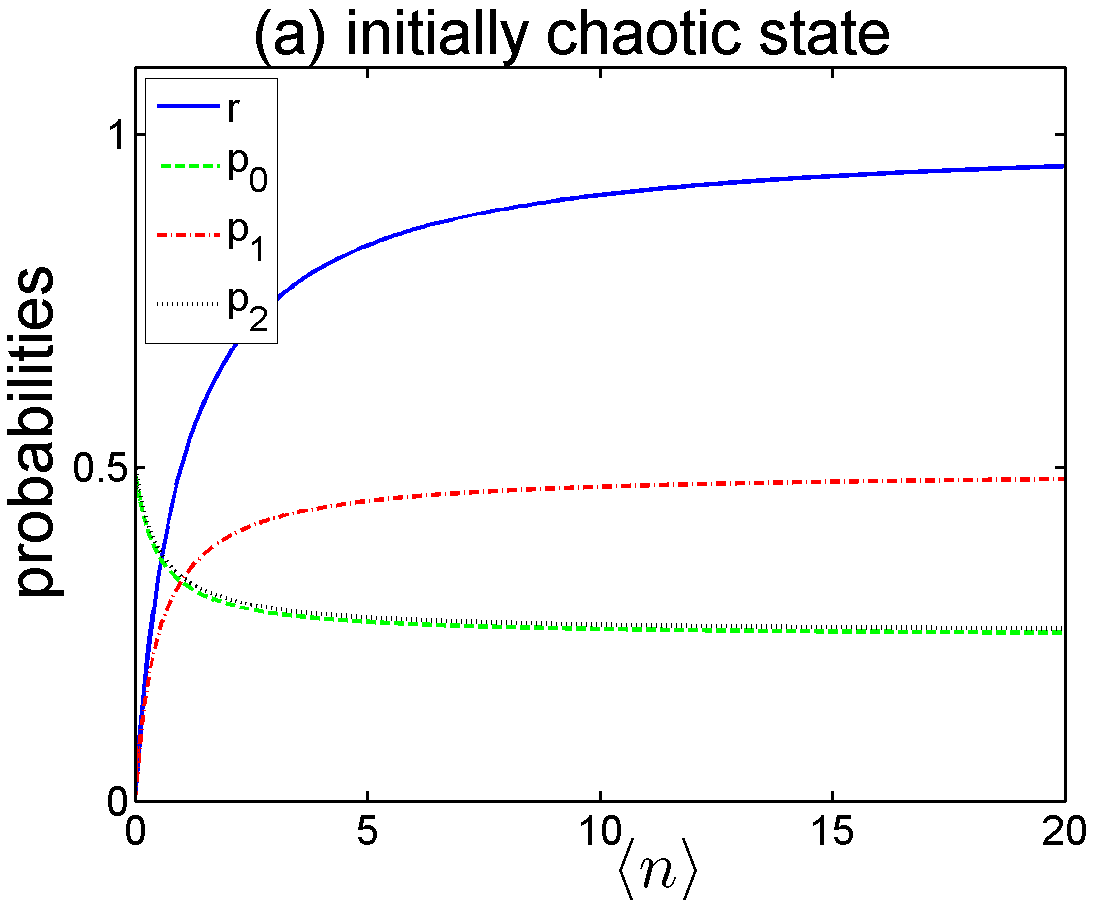}
       \includegraphics[height=35mm]{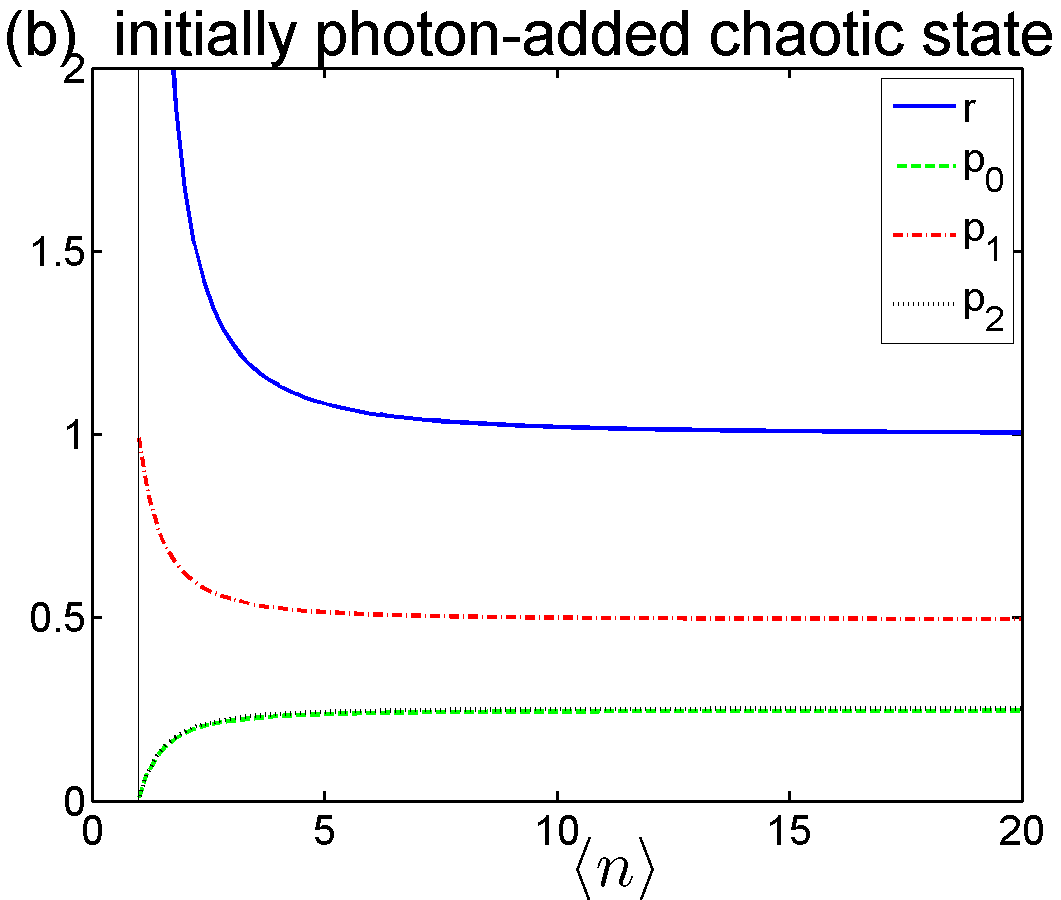}}

\caption{(Color online) Model~1 with two-photon dissipation:
Photon-number probabilities $p_n=\bra{n}\hat \rho^{02}_{{\rm
ss}}\ket{n}$ and ratio $r$ versus mean photon number $\mean{n}$
for different initial cavity fields: (a) chaotic state $\hat
\rho_{{\rm ch}}$ and (b) single-photon-added chaotic state
$\hat{\rho}_{_{\rm AT}}$. } \label{fig12}
\end{figure}

\begin{figure}
 \fig{ \includegraphics[height=34mm]{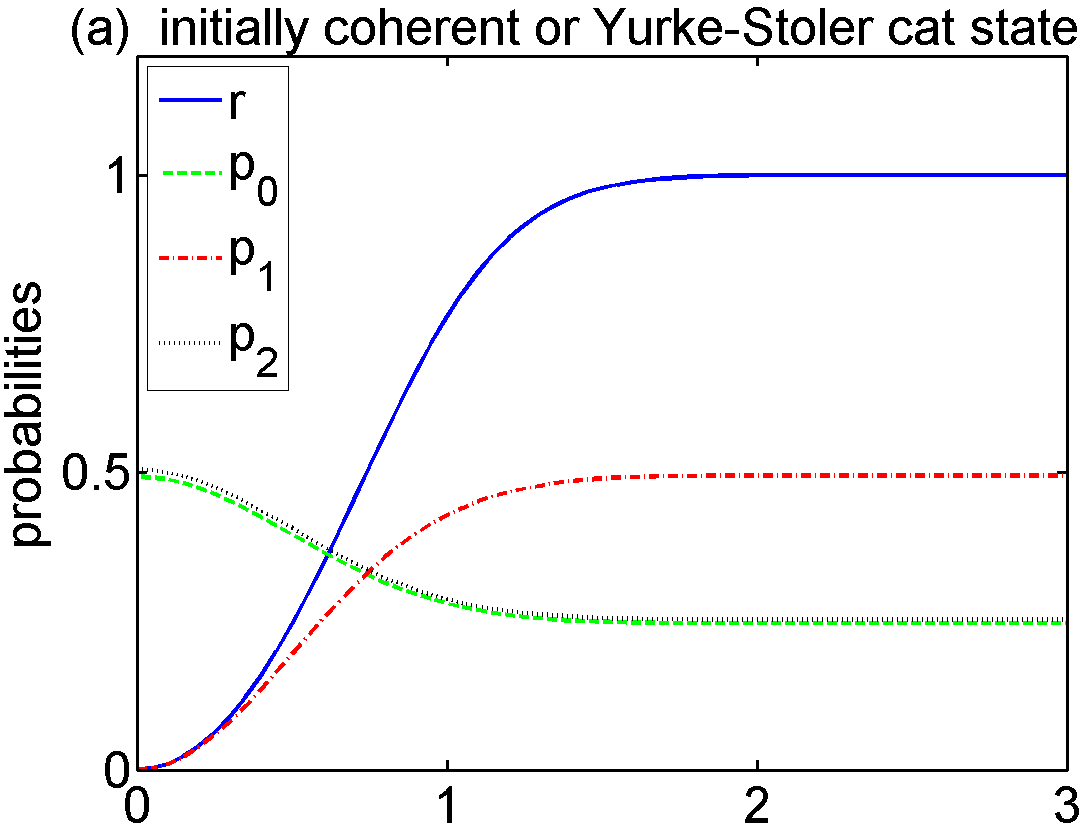} \includegraphics[height=34mm]{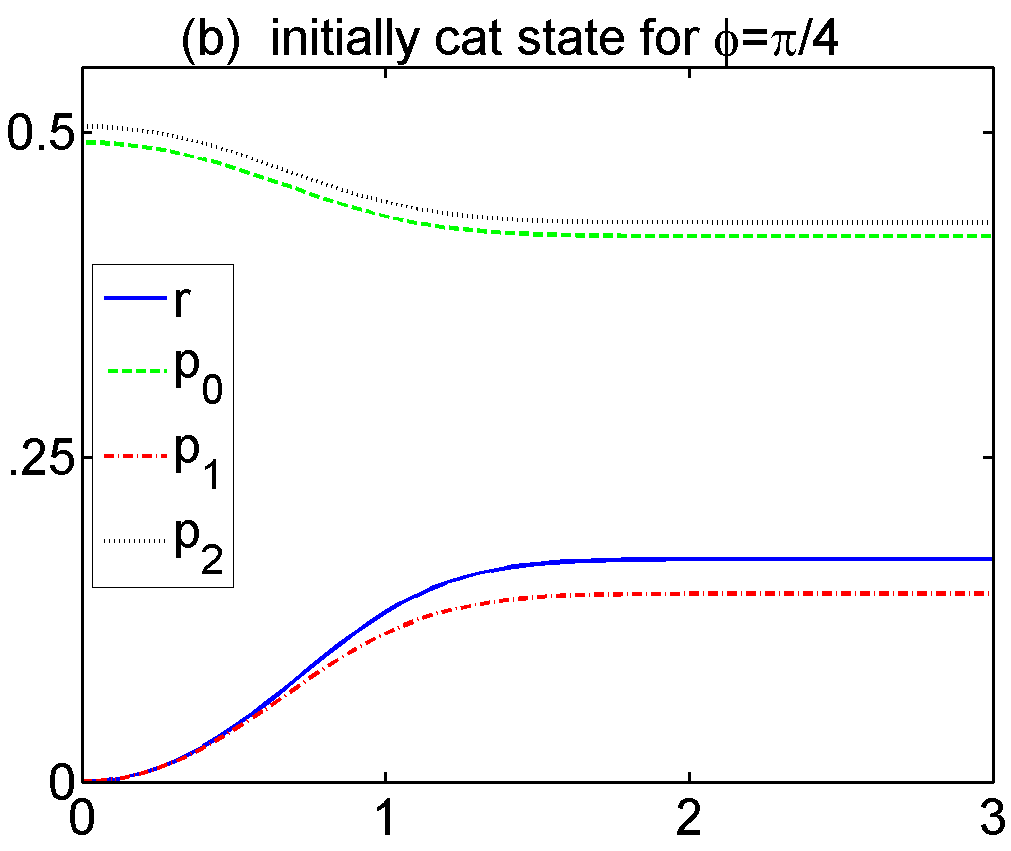}}

 \vspace{1mm}

 \fig{ \includegraphics[height=35mm]{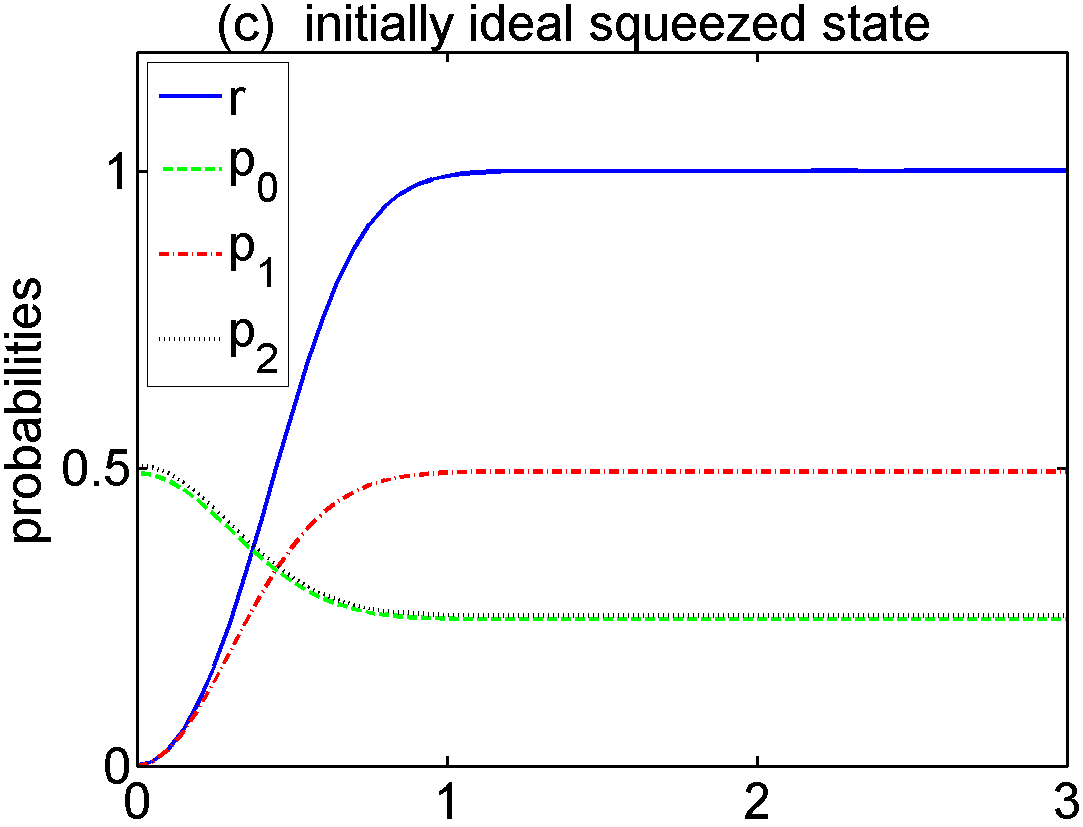}\includegraphics[height=35mm]{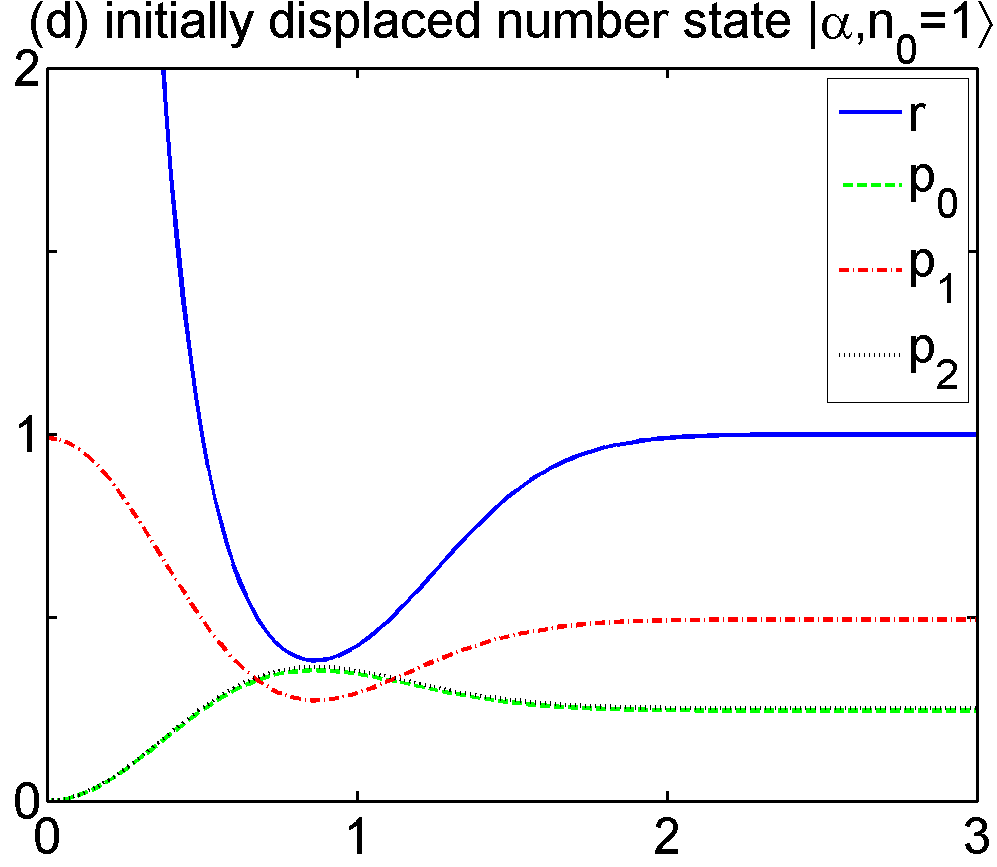}}

 \vspace{1mm}

 \fig{ \includegraphics[height=37mm]{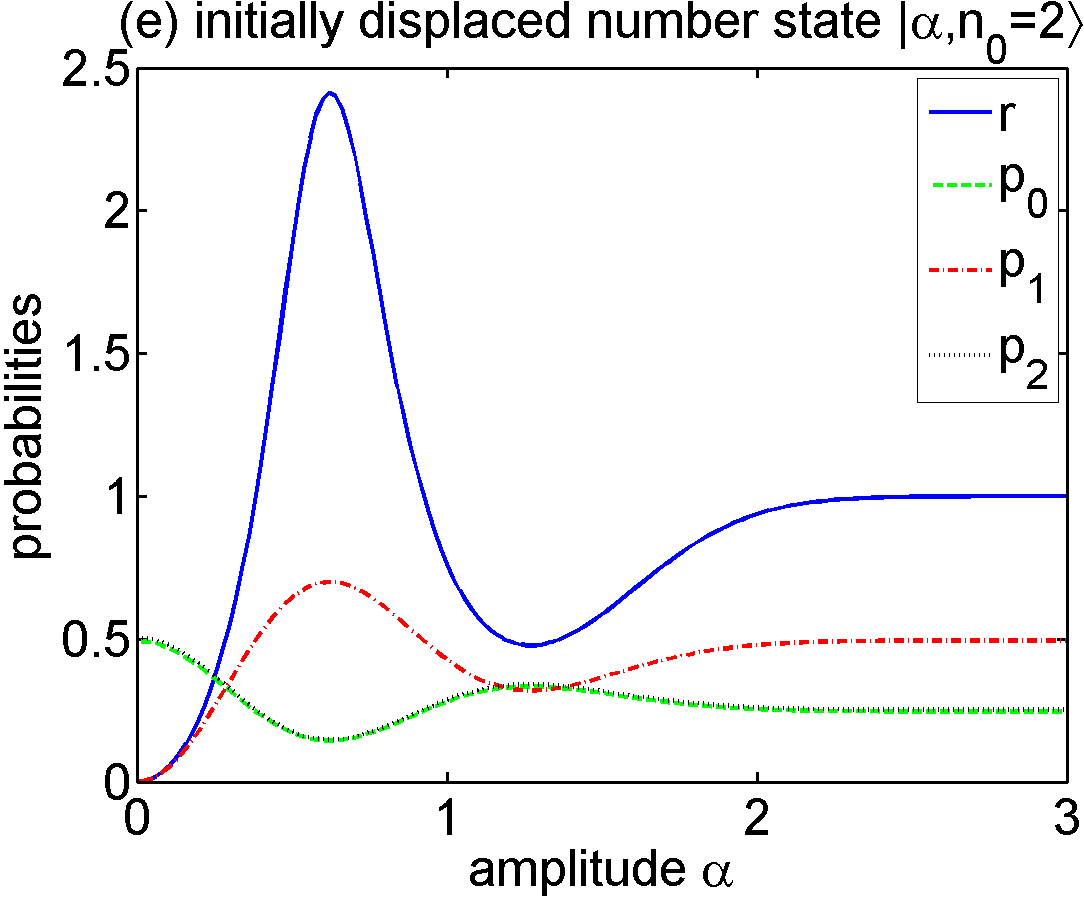}\includegraphics[height=37mm]{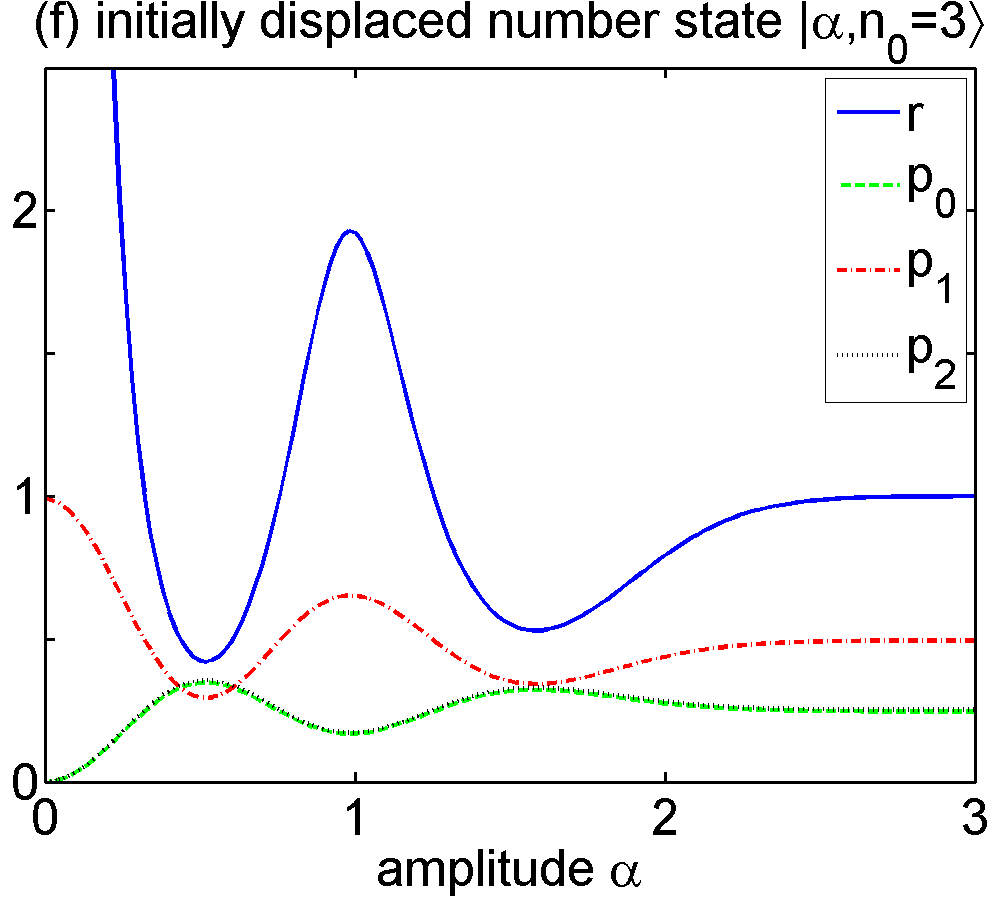}}

\caption{(Color online) Model~1 with two-photon dissipation: The
figure shows how the steady state $\hat \rho^{02}_{{\rm ss}}$ in
the engineered photon blockade depends on initial cavity field
$|\psi_{0}\rangle$. The photon-number probabilities
$p_n=\bra{n}\hat \rho^{02}_{{\rm ss}}\ket{n}$ (for $n=0,1,2$) and
the ratio $r=p_{{\rm odd}} (|\psi_{0}\rangle)/p_{{\rm
even}}(\psi_{0}\rangle)$ versus real amplitude $\alpha$ for
various initial states $|\psi_{0}\rangle$: (a) coherent state
$|\alpha\rangle$ or, equivalently, the Yurke-Stoler cat state
$|\alpha_{_{\rm YS}}\rangle=|\alpha,\phi=\pi/2\rangle$, (b) the
cat state $|\alpha,\phi=\pi/4\rangle$, and (c) the ideal squeezed
state $|\alpha,\xi=1/2\rangle$, and (d,e,f) the displaced number
states $|\alpha,n_{0}\rangle$ with $n_0=1,2,3$. The steady-state
solutions $\hat \rho^{02}_{{\rm ss}}$ of the master
equation~(\ref{ME}) for the Hamiltonian $\hat H_{02}$ are obtained
assuming $\delta=\epsilon/\chi=1/6$ and
$\delta'=\gamma/\epsilon=1/25$.} \label{fig13}
\end{figure}

\begin{figure}

\fig{\includegraphics[height=38mm]{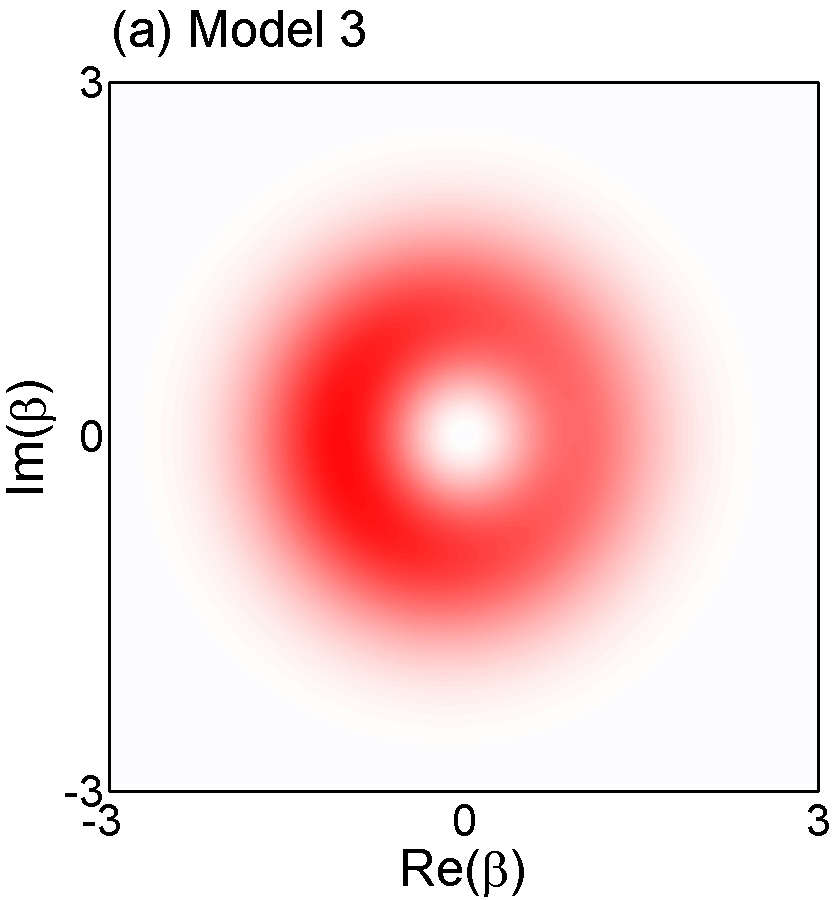}}
\fig{\includegraphics[height=38mm]{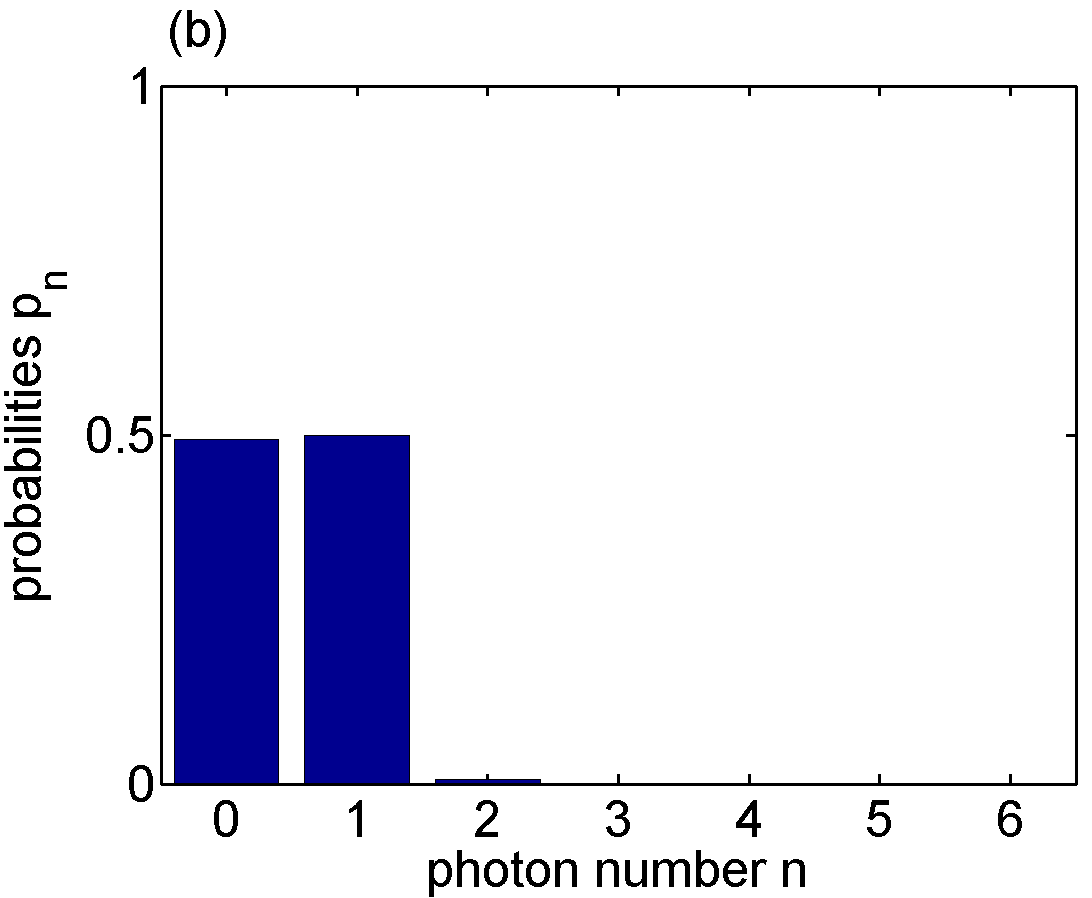}}

\vspace{1mm}

\fig{\includegraphics[height=38mm]{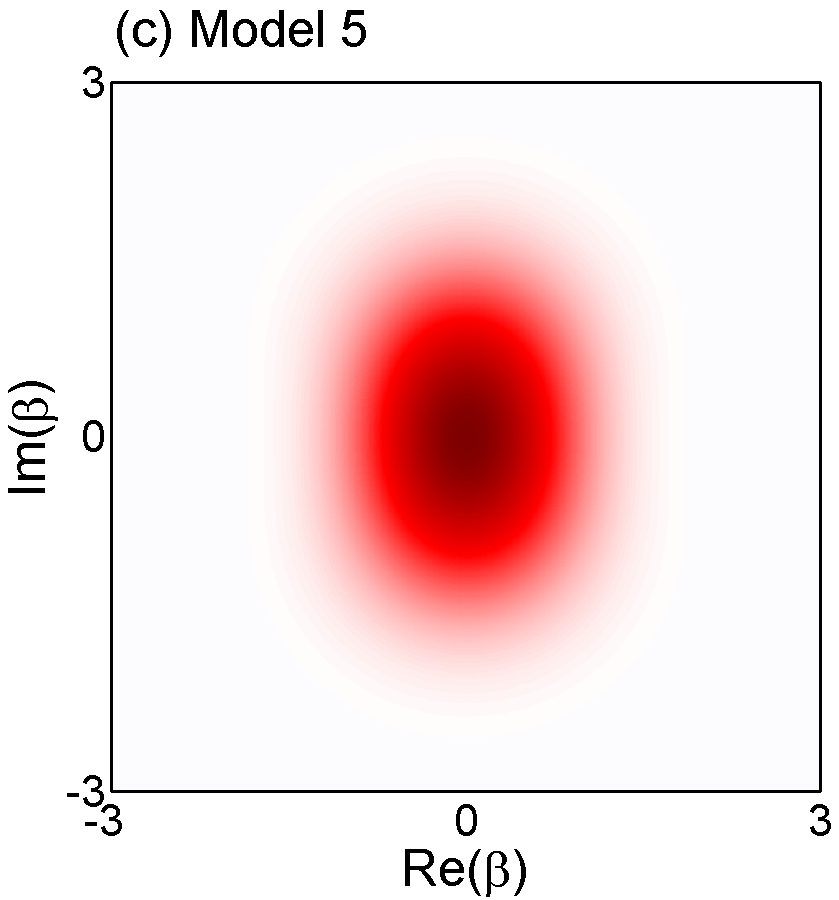}}
\fig{\includegraphics[height=38mm]{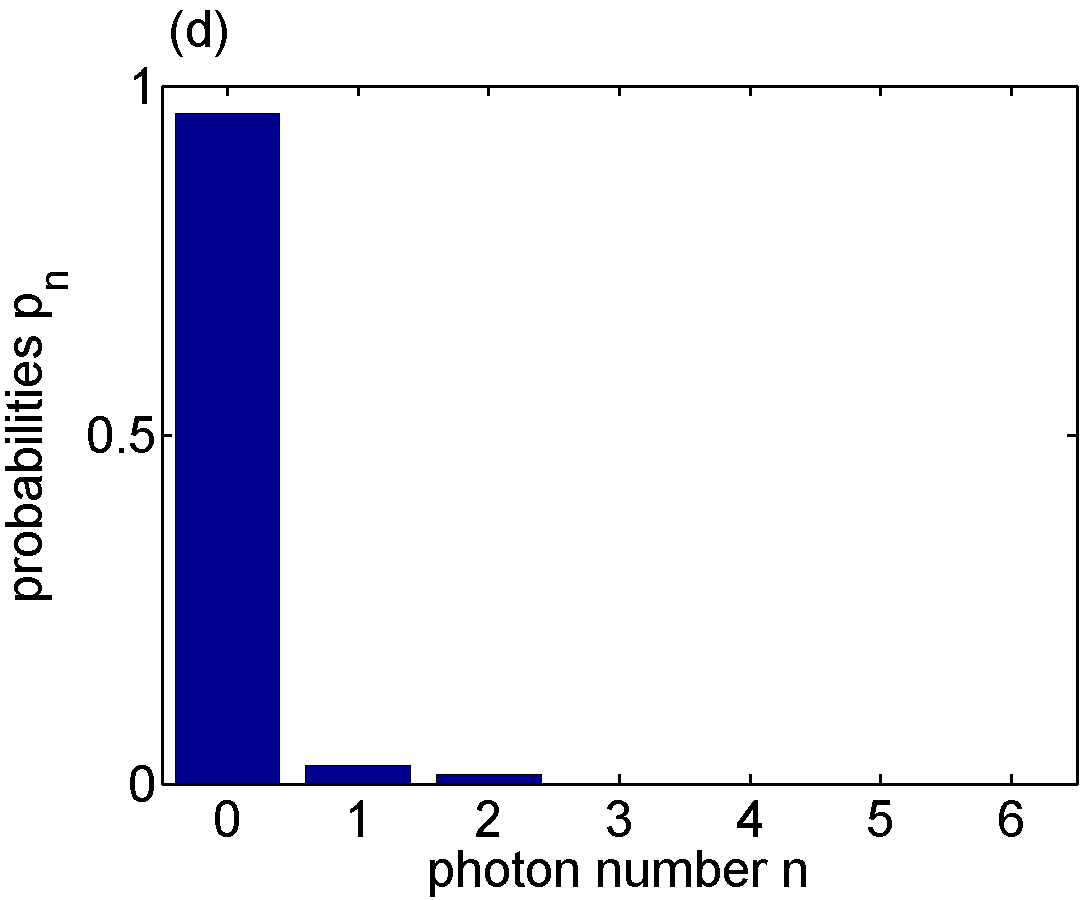}}

\caption{(Color online)  Models~3 in panels (a,b) and 5 in (c,d)
with single-photon dissipation for any initial states: Wigner
functions $W(\beta)$ and photon-number probabilities $p_n$ for the
steady-state solutions of the  master equation~(\ref{ME2}) with
$\gamma_2=\gamma_\perp=0$ for: (a,b) the single-photon driven
Hamiltonian $\hat H_{\rm usual}$, given by Eq.~(\ref{H_usual}),
and (c,d) the two-photon driven Hamiltonian $\hat H_{01}$, given
by Eq.~(\ref{Hkl}) with $k=0,l=1$. We set
$\delta=\epsilon/\chi=1/6$ and $\gamma_{1}/\epsilon=1/25$. Model 3
corresponds to the standard description of photon blockade. It is
worth noting that the same solutions, as shown in panels (a,b),
can be obtained for the two-photon absorption master equation,
given by Eq.~(\ref{ME}) with $\delta'=\gamma/\epsilon=1/25$. This case is referred to as Model~3' in Table~I.}
\label{fig14}
\end{figure}

Steady-state solutions $\hat{\rho}_{{\rm ss}}$ can be obtained by
solving the master equation, given  in Eqs.~(\ref{ME})
and~(\ref{ME2}),  with the condition $\frac{\rm d}{dt}
\hat\rho_{{\rm ss}} \equiv \frac{\rm d}{dt} \hat\rho = 0$, by
using, e.g., the inverse power method (as implemented, e.g., in
Ref.~\cite{Tan99}) or by a direct integration, for long enough
evolution times: $\hat{\rho}_{{\rm ss}}=\hat \rho(t\rightarrow
\infty)$. All our numerical results, shown in
Figs.~\ref{fig06}--\ref{fig14}, are based on these two equivalent
methods. We also applied an analytical approach of finding
approximate solutions of the master equation, given in
Eq.~(\ref{ME}), as described below.

We assume that the ratio of the driving field strength
$\epsilon$ and the Kerr nonlinear coupling $\chi$, and the ratio
of $\chi$ and the damping constant $\gamma$ are small, i.e.,
$\delta=\epsilon/\chi\ll 1$ and $\delta'=\gamma/\epsilon \ll 1$.
Thus, we can analyze the cavity-field Hilbert space of a small
dimension. For example, let us truncate the Hilbert space at the
five-photon Fock state, which corresponds to analyzing a
six-dimensional Hilbert space. We have obtained numerically a very
good agreement between our numerical solutions in the six- and
100-dimensional Hilbert spaces for the parameters chosen in all
figures.

In order to find compact-form analytical solutions, we expanded
our lengthy and complicated solutions (which are not presented
here) in power series of $\delta$ and $\delta'$, and keeping
linear and quadratic terms only.

First, let us assume the initial state of our system is an
even-number state, i.e.,
\begin{equation}
  \hat\rho_{0,\rm even} = \sum_{m=1}^{\infty} p_m \ket{\psi_m}
  \bra{\psi_m}, \quad \ket{\psi_m}= \sum_{n=0}^{\infty} c_{n}^{(m)} \ket{2n},
\label{even_state}
\end{equation}
with arbitrary probabilities $p_m$ and complex amplitudes
$c_{n}^{(m)}$, satisfying the normalization conditions $\sum_m
p_m=\sum_{n} |c_{n}^{(m)} |^2=1$, for $m=1,2,...\,$. Then we find
the steady-state solutions of the master equation, given by
Eq.~(\ref{ME}), for the system described by the Hamiltonians
$\hat{H}_{02}$ and $\hat{H}_{13}$, to be  given, in the standard
Fock basis, by
\begin{equation}
\hat \rho^{{\rm even}}_{{\rm ss}}
\approx\left[\begin{array}{cccccc}
p & 0 & a+ib & 0 & c+id & 0\\
0 & 0 & 0 & 0 & 0 & 0\\
a-ib & 0 & q & 0 & e+if & 0\\
0 & 0 & 0 & 0 & 0 & 0\\
c-id & 0 & e-if & 0 & r & 0\\
0 & 0 & 0 & 0 & 0 & 0
\end{array}\right] \label{rho_even}
\end{equation}
in terms of the coefficients given explicitly in Appendix~A. By
further assuming that $\delta^2\approx \delta'^2\approx
\delta\delta'\approx 0$ then $r\approx c\approx d\approx f=0$ for
Model~1 [see Eqs.~(\ref{A2a}) and (\ref{A2b})]. Thus, it is seen
that the steady state, which can be  generated in this model,
assuming that the cavity field is initially in an even-number
state, is a partially incoherent superposition of effectively only
two number states, $\ket{0}$ and $\ket{2}$, while for Model~2, the
steady state is spanned by  the number states$\ket{0}$, $\ket{2}$,
and $\ket{4}$. See Table~I for comparison.

Now we assume that the initial state of our system is an
odd-number state, i.e.,
\begin{equation}
  \hat\rho_{0,\rm odd} = \sum_{m=1}^{\infty} p_m \ket{\psi_m}
  \bra{\psi_m},\quad  \ket{\psi_m}= \sum_{n=0}^{\infty} c_{n}^{(m)}
  \ket{2n+1}, \
\label{odd_state}
\end{equation}
for any $p_m$ and $c_{n}^{(m)}$, as in Eq.~(\ref{even_state}).
Then the steady-state solution for $\hat{H}_{02}$ and
$\hat{H}_{13}$ can be approximately given by
\begin{equation}
\hat \rho^{{\rm odd}}_{{\rm ss}}\approx
p|1\rangle\langle1|+(1-p)|3\rangle\langle3|+[(a+ib)|1\rangle\langle3|+{\rm
h.c.}]. \label{rho_odd}
\end{equation}
where the coefficients $a$, $b$, and $p$ are given explicitly in
Appendix~A. It is seen that the steady state is spanned by the
number states $\ket{1}$ and $\ket{3}$ only. Actually, by also
ignoring the terms proportional to $\delta^2$, $\delta'^2$, and
$\delta\delta'$, the steady state for Model~1 is just the
single-photon  state, which is not the case for Model~2 (see also
Table~I for comparison).

As an illustration of these results, the Wigner functions and
photon-number probabilities for the numerically calculated
steady-state solutions $\hat \rho^{{\rm even}}_{{\rm ss}}$ and
$\hat \rho^{{\rm odd}}_{{\rm ss}}$ are shown in Figs.~\ref{fig06}
and~\ref{fig07}. On the scale of these plots, there is practically
no difference between our approximate analytical and precise
numerical solutions. Figure~\ref{fig08} shows how the steady-state
number probabilities $p_n$ depend on the driving field strength
$\epsilon$ in units of the damping constant $\gamma$ for an
initial even-number state. Analogous solutions for an initial
odd-number state practically do not depend on $\epsilon/\gamma\in
[0,10]$. Figure~\ref{fig09} shows how the probabilities $p_n$
depend on the tuning frequencies $\Omega_{02}$ (assuming
$\Sigma_{02}=0$) and $\Omega_{13}$ (with $\Sigma_{13}=0$).


We find that the steady-state solution of the master equation,
given by Eq.~(\ref{ME}) assuming $\gamma\ll \epsilon\ll \chi$,
reads
\begin{equation}
\hat \rho_{{\rm ss}}(\hat \rho_{0})=p_{{\rm even}}(\hat
\rho_{0})\hat \rho^{{\rm even}}_{{\rm ss}}+p_{{\rm odd}}(\hat
\rho_{0})\hat \rho^{{\rm odd}}_{{\rm ss}},
\label{general_solution}
\end{equation}
for an arbitrary initial state $\rho_{0}$. This solution is a
weighted sum of the steady-state solutions, given by
Eqs.~(\ref{rho_even}) and~(\ref{rho_odd}), with the weights
\begin{eqnarray}
p_{{\rm even}}(\hat \rho_{0}) & = & \sum_{n=0}^{\infty}
\langle2n|\hat \rho_{0}|2n\rangle,\\
p_{{\rm odd}}(\hat \rho_{0}) & = &
\sum_{n=0}^{\infty}\langle2n+1|\hat \rho_{0}|2n+1\rangle.
\end{eqnarray}
So, it holds $p_{{\rm even}}(\hat \rho_{0})+p_{{\rm odd}}(\hat
\rho_{0})=1$. It is seen that
\begin{eqnarray}
\hat \rho^{{\rm even}}_{{\rm ss}} & \equiv & \hat \rho_{{\rm ss}}
(\sum_{n}c_{n}|2n\rangle)=\hat \rho_{{\rm ss}}(|0\rangle),
 \label{rho_even1}\\
 \hat \rho^{{\rm odd}}_{{\rm ss}}
& \equiv & \hat \rho_{{\rm
ss}}(\sum_{n}c_{n}|2n+1\rangle)=\hat \rho_{{\rm ss}}(|1\rangle).
 \label{rho_odd1}
\end{eqnarray}
for any complex amplitudes $c_n$.

Our result that
\begin{eqnarray}
  \hat \rho_{{\rm ss}}(|0\rangle) &\neq& \hat \rho_{{\rm ss}}(|1\rangle)
\label{even_odd}
\end{eqnarray}
sounds counterintuitive for the following reason: To find the
solution of the master equation, given by Eq.~(\ref{ME}), one can
write separately the equations of motion for all the elements
$\rho_{ij}(t)$ of the density matrix $\hat\rho(t)$. The
steady-state solutions $\bar\rho^{\rm
ss}_{ij}=\lim_{t\rightarrow\infty}\rho_{ij}(t)$ can be obtained by
setting $\partial \bar\rho^{\rm ss}_{ij}/\partial t=0$. Then, it
would appear that the elements $\bar\rho^{\rm ss}_{ij}$ in the
steady state do not depend on the initial conditions. We will show
below that they can depend on the initial states of mixed parity.
Namely, let us create two matrices $\hat\rho'$ and $\hat\rho''$
with the \emph{only} nonzero elements $\rho'_{2i,2j} =
\bar\rho^{\rm ss}_{2i,2j}$ and $\rho''_{2i+1,2j+1} = \bar\rho^{\rm
ss}_{2i+1,2j+1}$ for $i,j=0,1,...$. Then, the even steady-state
density matrix, given by
Eq.~(\ref{rho_even1}), simply reads as $\hat\rho^{\rm ss}_{\rm
even} = \hat\rho'/\tr(\hat\rho')$. Analogously, the odd
steady-state density matrix, given
by Eq.~(\ref{rho_odd1}), is equal to $\hat\rho^{\rm ss}_{\rm
odd} = \hat\rho''/\tr(\hat\rho'')$. The general solution, given in
Eq.~(\ref{general_solution}), is then state-dependent, although in
this limited manner.

These formulas can be confirmed numerically by comparing them with
the solutions for the master equation obtained for sufficiently
long evolution times. Moreover, the steady-state density matrix
elements $\bar\rho^{\rm ss}_{ij}$ can be directly calculated
numerically by finding a vector in the null space of the
Liouvillian superoperator ${\cal L}$~\cite{Tan99}.

We note that the ratio
\begin{equation}
r=\frac{p_{{\rm odd}}(\hat \rho_{0})}{p_{{\rm even}}(\hat
\rho_{0})}=\frac{p_{{\rm odd}}(\hat \rho_{\rm ss})}{p_{{\rm even}}(\hat
\rho_{\rm ss})}
\end{equation}
is  preserved during the system evolution. This is because the
two-photon driving and two-photon dissipation, together with the
photon-number-preserving Kerr interaction, do not mix even and odd
number states. To show how this  engineered photon blockade
depends on the initial states $\hat \rho_{0}$, the ratio $r$ is
plotted in Figs.~\ref{fig10}--\ref{fig13} for a few states $\hat
\rho_{0}$ discussed in the next section.

Finally, we note that these steady-state solutions, as well as our precise
numerical solutions shown in all plots, depend solely on the
ratios $\delta=\epsilon/\chi$ and $\delta'=\gamma/\epsilon$, and
do not depend on the absolute values of $\epsilon$, $\chi$, and
$\gamma$.

\section{Photon blockade for specific initial fields}

Here we analyze how the engineered photon blockade depends on some
typical classical and nonclassical initial states of the cavity
field.

For a coherent state (CS) $|\alpha\rangle$, we have
\begin{eqnarray}
p_{{\rm even}}(|\alpha\rangle) & = &
\frac12[1+\exp(-2|\alpha|^{2})],\nonumber \\
p_{{\rm odd}}(|\alpha\rangle) & = &
\frac12[1-\exp(-2|\alpha|^{2})], \label{P_CS0}
\end{eqnarray}
so their ratio is $r=\tanh(|\alpha|^{2})$. In the limiting cases,
one observes that
\begin{eqnarray}
\lim_{\mean{n}\rightarrow0} p_{{\rm even}}(\hat
\rho_{0})&=&1,\quad
\lim_{\mean{n}\rightarrow0} p_{{\rm odd}}(\hat \rho_{0})=0,\label{P_CS1} \\
\lim_{\mean{n}\rightarrow\infty} p_{{\rm even}}(\hat \rho_{0})&=&
\lim_{\mean{n}\rightarrow\infty} p_{{\rm odd}}(\hat
\rho_{0})=\frac12, \label{P_CS2}
\end{eqnarray}
where $\hat \rho_{0}=|\alpha\rangle\langle\alpha|$ and the
intensity is given by $\mean{n}=|\alpha|^{2}$. A few illustrative
examples of phase-space and photon-number distributions for the
steady-state solutions, for initial coherent states, are shown in
Fig.~\ref{fig10} for Model~1 and in Fig.~\ref{fig11} for Model~2.

For the chaotic (or thermal) state of the cavity field,
\begin{equation}
\hat \rho_{{\rm ch}}=(1-q)\sum_{n=0}^{\infty}q^{n}|n\rangle\langle
n|, \label{chaotic}
\end{equation}
where $\langle n\rangle=q/(1-q)$ is the mean photon number, one
can find that
\begin{eqnarray}
p_{{\rm even}}(\hat \rho_{{\rm ch}})&=&\frac{1}{1+q}=\frac{1+\langle n\rangle}{1+2 \langle n\rangle},\nonumber\\ p_{{\rm
odd}}(\hat \rho_{{\rm ch}})&=&\frac{q}{1+q}=\frac{\langle n\rangle}{1+2 \langle n\rangle},
\end{eqnarray}
so $r(\hat\rho_{{\rm ch}})=q=\langle n\rangle/(\langle
n\rangle+1)$. In the limiting cases of the intensity $\mean{n}$,
one can see that the relations Eqs.~(\ref{P_CS1})
and~(\ref{P_CS2}) hold also for $\hat\rho_{0}=\hat\rho_{{\rm
ch}}$, as for coherent states. Figure~\ref{fig12}(a) shows how the
photon-number probabilities $p_n$ of the steady-state solutions
$\hat \rho_{{\rm ss}}(\hat \rho_{0})$ depend on the mean photon
number $\mean{n}$ of the initial chaotic state $\hat
\rho_{0}=\hat\rho_{{\rm ch}}$.

Now we analyze a few examples of nonclassical initial states.

The single-photon-added chaotic state, introduced by Agarwal and
Tara~\cite{Agarwal92}, can be defined as follows
\begin{equation}
\hat{\rho}_{_{\rm AT}}=\norm\hat{a}^{\dagger}\hat{\rho}_{{\rm
ch}}\hat{a}=\norm\sum_{n=1}^{\infty}nq^{n}|n\rangle\langle
n|,\label{N01}
\end{equation}
where $\hat{\rho}_{{\rm ch}}$ is the chaotic state, given by
Eq.~(\ref{chaotic}), $\hat{a}$ ($\hat{a}^{\dagger}$) is the
annihilation (creation) operator of the field, and
$\norm=(1-q)^{2}/q$ is the normalization constant. The mean photon
number for $\hat{\rho}_{_{\rm AT}}$ is $\langle
n\rangle=(1+q)/(1-q)$. It is interesting to note that this
infinite-dimensional state is nonclassical although diagonal in
the photon-number basis. We find that
\begin{eqnarray}
p_{{\rm even}}(\hat \rho_{_{\rm AT}})&=&\frac{2q}{(1+q)^{2}}=\frac12(1-\mean{n}^{-2}),\nonumber \\
p_{{\rm odd}}(\hat \rho_{_{\rm AT}})&=&\frac{1+q^{2}}{(1+q)^{2}}=\frac12(1+\mean{n}^{-2}),
\end{eqnarray}
so $r(\hat \rho_{_{\rm AT}})=(1+q^{2})/(2q)=
(1+\mean{n}^{2})/(1-\mean{n}^{2})$. In the limit of large number
of photons, $\mean{n}\rightarrow\infty$, again the relation,
given by Eq.~(\ref{P_CS2}), hold as for chaotic states. However,
in the limit of small number of photons, we have
\begin{eqnarray}
\lim_{\mean{n}\rightarrow 1} p_{{\rm even}}(\hat \rho_{_{\rm
AT}})&=&0,\quad \lim_{\mean{n}\rightarrow 1} p_{{\rm odd}}(\hat
\rho_{_{\rm AT}})=1\label{P_AT}, \label{P_chaotic}
\end{eqnarray}
which is the opposite case to the chaotic state, as given by
Eq.~(\ref{P_CS1}). Note that $\mean{n}\ge 1$, because only then
$p_{{\rm odd}}(\hat \rho_{_{\rm AT}})\le 1$. Figure~\ref{fig12}(b)
shows how the photon-number probabilities $p_n=\bra{n}\hat
\rho_{{\rm ss}}(\hat \rho_{0})\ket{n}$ depend on the mean photon
number $\mean{n}$ of the initial single-photon-added chaotic state
$\hat \rho_{0}=\hat\rho_{{\rm AT}}$. This dependence is
fundamentally different from that shown in Fig.~\ref{fig12}(a) for
the initial chaotic state.

Let us also analyze a prototype of Schr\"odinger's cat states
given as a macroscopically distinct superposition of two coherent
states, e.g.,
\begin{equation}
|\alpha,\phi\rangle={\cal {\cal
N}}[|\alpha\rangle+\exp(i\phi)|-\alpha\rangle]
\end{equation}
with the normalization ${\cal {\cal
N}}=\{2[1+\cos\phi\exp(-2|\alpha|^{2})]\}^{-1/2}$ assuming a
complex amplitude $\alpha.$ One can find that
\begin{eqnarray}
  p_{{\rm even}}(|\alpha,\phi\rangle)
  &=&\cos^{2}\Big(\frac{\phi}{2}\Big)\frac{1+\exp(-2|\alpha|^{2})}
  {1+\cos\phi\exp(-2|\alpha|^{2})},\nonumber \\
p_{{\rm odd}}(|\alpha,\phi\rangle)
  &=&\sin^{2}\Big(\frac{\phi}{2}\Big)\frac{1-\exp(-2|\alpha|^{2})}
  {1+\cos\phi\exp(-2|\alpha|^{2})}, \nonumber \\
r&=&\tan ^2(\phi/2)\tanh|\alpha|^{2}.
   \label{P_cat}
\end{eqnarray}
For special choices of $\phi=0,\pi/2,\pi$, the state
$|\alpha,\phi\rangle$ reduces to the renowned cat states. These
include the even CS
\begin{equation}
|\alpha_{+}\rangle\equiv|\alpha,\phi=0\rangle
=\frac{1}{\sqrt{\cosh|\alpha|^{2}}}\sum_{n=0}^{\infty}
\frac{\alpha^{2n}}{\sqrt{(2n)!}}|2n\rangle,
\end{equation}
the Yurke-Stoler cat state $|\alpha_{_{\rm
YS}}\rangle\equiv|\alpha,\phi=\pi/2\rangle$~\cite{Yurke86}, and
the odd CS
\begin{equation}
|\alpha_{-}\rangle\equiv|\alpha,\phi=\pi\rangle
=\frac{1}{\sqrt{\sinh|\alpha|^{2}}}
\sum_{n=0}^{\infty}\frac{\alpha^{2n+1}}{\sqrt{(2n+1)!}}|2n+1\rangle.
\end{equation}
Clearly, $p_{{\rm even}}(|\alpha_{+}\rangle)=1$ and $p_{{\rm
odd}}(|\alpha_{-}\rangle)=1$ for any $\alpha$, this is in contrast
to the states $|\alpha,\phi\rangle$ for other angles $\phi$.
Equation~(\ref{P_cat}) for the Yurke-Stoler cat state
$|\alpha_{_{\rm YS}}\rangle$ implies that
\begin{equation}
p_{{\rm even}}(|\alpha_{_{\rm YS}}\rangle)=p_{{\rm
even}}(|\alpha\rangle).
\end{equation}
Thus, the formulas given by Eqs.~(\ref{P_CS0})--(\ref{P_CS2}),
hold for the state $|\alpha_{_{\rm YS}}\rangle$ too. A few
examples of the Wigner functions and photon-number probabilities
for the steady-state solutions,
obtained for initial cat states, are shown in
Figs.~\ref{fig10}(e,f) and~\ref{fig13}(a,b) for Model~1 and in
Fig.~\ref{fig11}(e,f) for Model~2.

In Fig.~\ref{fig13}, we also analyze the engineered photon
blockade for squeezed and displaced-number initial states. The
ideal squeezed states (also known as the two-photon coherent
states) are defined as
\begin{eqnarray}
|\alpha,\xi\rangle & = & \hat D(\alpha)\hat S(\xi)|0\rangle,
\label{sq_state}
\end{eqnarray}
which are given in terms of the squeeze operator $\hat
S(\xi)=\exp[\tfrac{1}{2}\xi^{*}a^{2}-\tfrac{1}{2}\xi(a^{\dagger})^{2}],$
where $\xi$ is the complex squeezing parameter, and the
displacement operator $\hat D(\alpha)=\exp(\alpha\hat
a^{\dagger}-\alpha^{*}\hat a)$ with a complex amplitude $\alpha$.
The displaced-number states are defined by
\begin{equation}
|\alpha,n_{0}\rangle=\hat D(\alpha)|n_{0}\rangle, \label{DNS}
\end{equation}
which become a coherent state $|\alpha\rangle$ in a special case
of $n_{0}=0$. The photon-number expansions of these  are given in
Appendix~B. These are useful to find the steady-state solutions,
given by Eq.~(\ref{general_solution}). Except for some special
cases, including squeezed vacuum $|\alpha=0,\xi\rangle$ and
coherent state $|\alpha,n_{0}=0\rangle$, the formulas for the
probabilities $p_{{\rm even}}$ and $p_{{\rm odd}}$ are not
compact, thus we present only our numerical results in
Fig.~\ref{fig13}.

\begin{figure}

 \fig{\includegraphics[height=45mm]{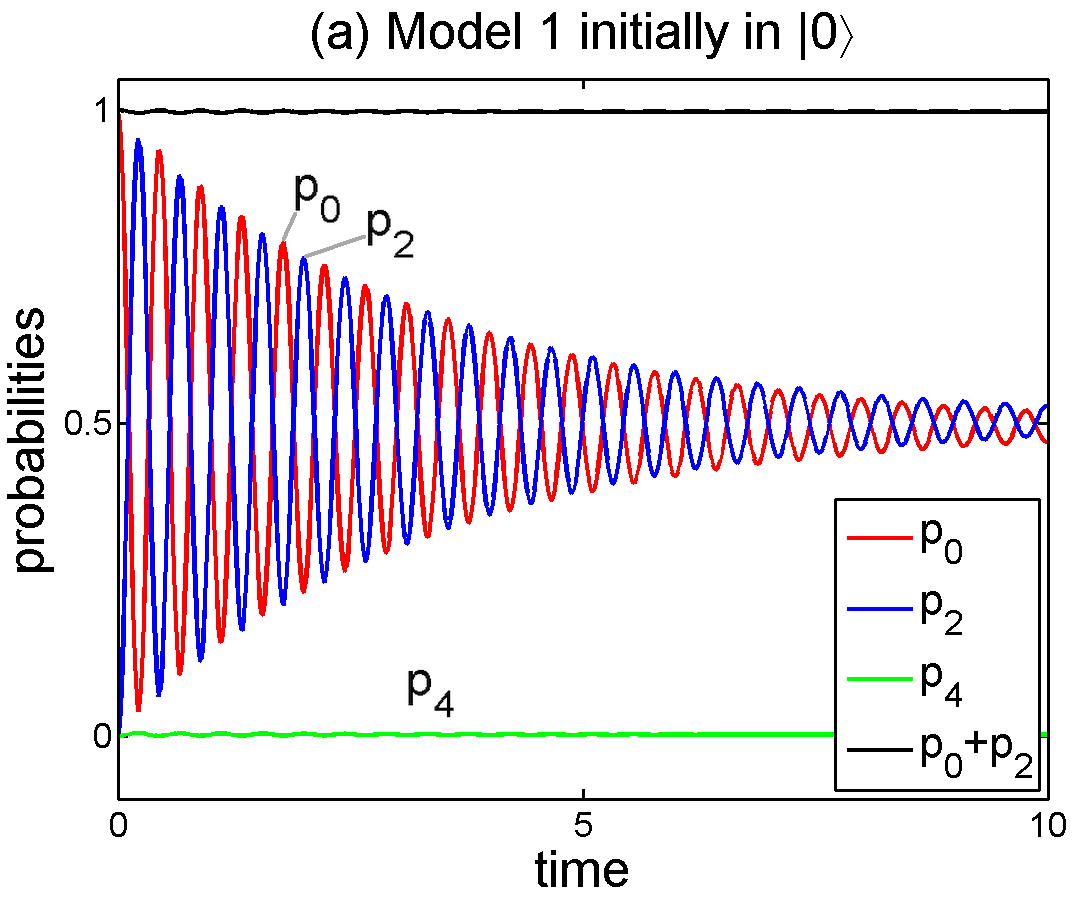}}

 \fig{\includegraphics[height=45mm]{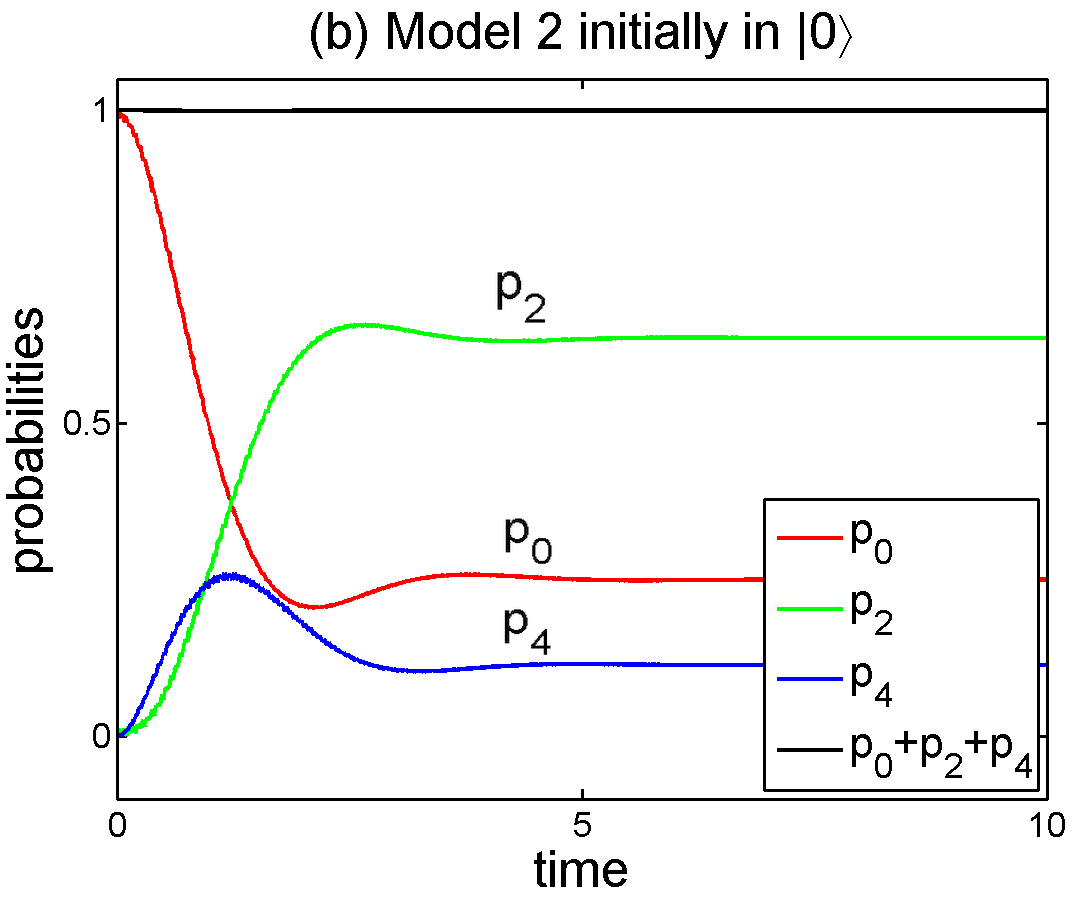}}

 \fig{\includegraphics[height=45mm]{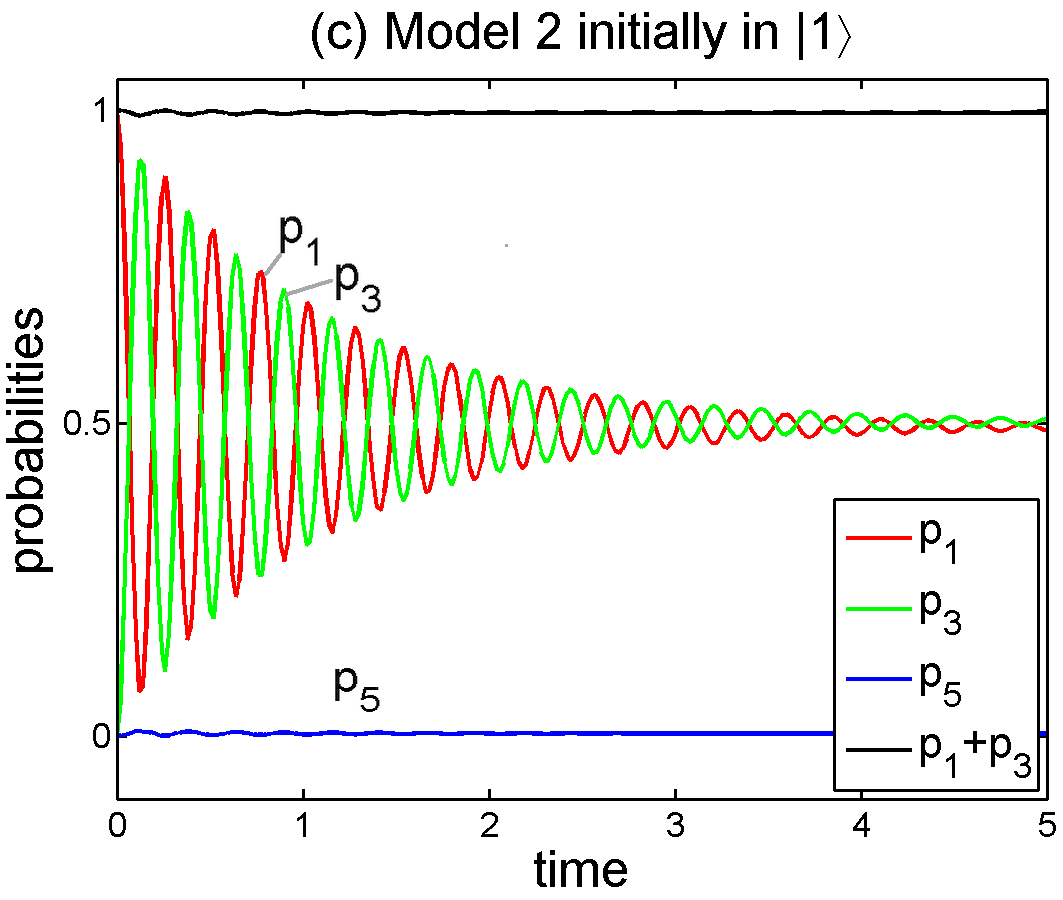}}

 \fig{\includegraphics[height=45mm]{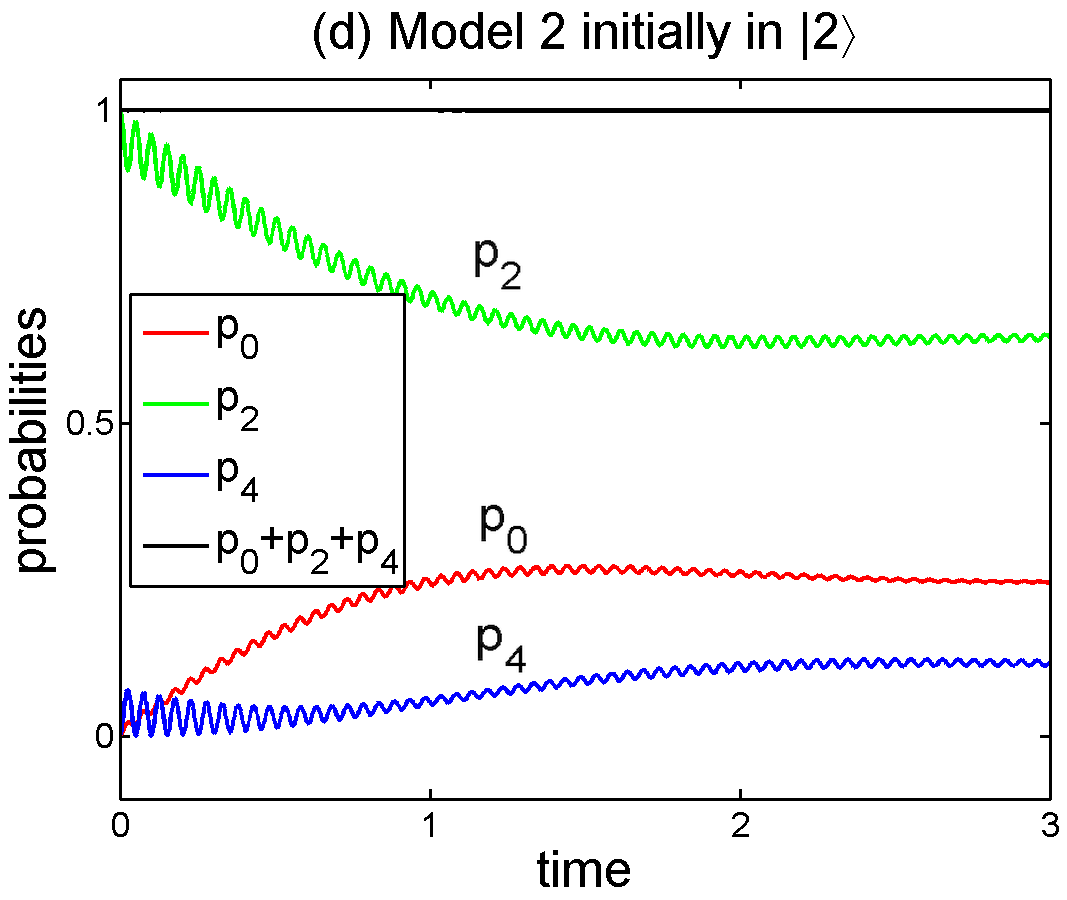}}

\caption{(Color online) Models~1(a) and 2 (b,c,d): Dissipative
evolutions of the photon-number probabilities $p_n(t)$ and the photon-blockade fidelities
$F(t)=\sum_n p_n(t)$ for several initial Fock states $\ket{m}$ (as
indicated in the panel titles) assuming
$\delta=\epsilon/\chi=1/6$, $\delta'=\gamma/\epsilon=1/25$, and
$\epsilon=5$. The decay of Rabi-type oscillations is clearly seen.
Panels (a) and (b,c,d) should be compared with Figs.~1(a)
and~2(a,b,c) demonstrating the corresponding dissipation-free
evolutions in Models 1 and 2, respectively. It is seen that the
decaying states rapidly approach the steady states shown in
Figs.~6 and~7.} \label{fig15}
\end{figure}

\section{Conclusions}

We studied photon blockade (also known as optical-state
truncation) of an optical or microwave light in a Kerr nonlinear
cavity parametrically driven by a two-photon process. The Kerr
nonlinear cavity can be effectively implemented by a standard
linear cavity with a tunable two-level system within the
Jaynes-Cummings model in the dispersive limit. We have also
assumed that the nonlinear system interacts with a nonlinear
reservoir, where two-photon absorption is dominant. We have shown
in Sec.~II how to observe various types of photon blockade effects
(as summarized in Table~I) by properly tuning the frequencies of
the driving field and the two-level system.

Our approach is partially motivated by the observation that
two-photon loss mechanisms often accompany a Kerr
nonlinearity~\cite{Yurke06}. Moreover, quantum reservoirs are a
powerful resource for quantum state engineering (see, e.g.,
Refs.~\cite{Gilles93, Kim93, Gilles94, Guerra97, Ispasoiu00,
Jakob01, Klimov03, Buks06, Yurke06, Karasik08, Kowalewska10,
Boissonneault12, Everitt13, Kumar10, Gevorkyan10, Mogilevtsev13,
Leghtas13, Reiter13, Arenz13, Mirrahimi14, Lu14}). In particular,
the circuit described in Ref.~\cite{Kumar10} for the
implementation of nondemolition measurement using the Kerr effect
seems to be readily applicable not only for generating
Schr\"odinger cat states~\cite{Everitt13}, but also for
implementing our generalized photon blockade via two-photon
dissipation. Anyway, such an implementation would require a
detailed analysis, which is not presented here.

The conditions to observe photon blockade are the following: (i)
the system must exhibit nonlinearity which is much stronger than
the strength of the drive, and (ii) dissipation is weaker than the
drive. (Actually, the second condition can be relaxed, as seen in
Fig.~8.) Thus, in particular, photon blockade can also be observed
even without damping at all. The system evolution is limited to a
few number states, which are determined by the choice of initial
states and the values of the tuning parameters $k,l$ in the
Hamiltonian $H_{kl}$. This effect is referred to as nonstationary
(or time-dependent) photon blockade. As discussed in Sec.~III and
shown in Figs.~1 and 3, Rabi-type oscillations can occur between
some number states, while practically no evolution can be observed
for other states. If the system is damped, then these Rabi-type
oscillations decrease in amplitude, and completely disappear for
longer times, as shown in Fig.~15 both for Models~1 and~2. The
evolutions of these driven and dissipative nonlinear systems
generate nonclassical optically truncated steady states
corresponding to stationary (or steady-state) photon blockade. The
phase-space and photon-number distributions of these states are
shown in Figs.~6--14.

It is well known that a typical steady-state photon blockade does
not depend on the initial state of a Kerr nonlinear system if it
is driven by a single-photon process and coupled to a standard
reservoir, where only a single-photon absorption is allowed, as
illustrated in Figs.~\ref{fig14}(a,b) and listed in Table~I.

In contrast, we have shown that the engineered photon blockade can
depend on the system initial state (see Table~I for comparison of
various photon blockade effects). This state dependence occurs in
a Kerr nonlinear system driven by a two-photon process and
dissipating via an engineered quantum reservoir, where only
two-photon absorption is allowed. These two-photon driving and
dissipation processes result in two completely independent
evolutions of the superpositions of Fock states with either even
or odd numbers of photons. This can be interpreted as two
different evolution-dissipation channels for even and odd-number
states. So, in particular, these states evolve into two different
steady states in the time limit. As the two processes affect only
every second state in the Fock basis and there is no mixing
between the photon numbers of different parity, we can describe
their evolutions in two independent Hilbert spaces. If the initial
state is a superposition of photon-number states of different
parity, then its steady state is a weighted sum of the steady
states achieved independently in the even- and odd-number Hilbert
spaces. The weights in this mixture depend on the photon-number
statistics of the initial states, although in a limited way, as
they depend solely on the ratio of the probabilities of measuring
the odd and even numbers of initial-state photons.

The above results imply that photon blockade phenomena do
\emph{not} depend on the initial states for various other models
listed  in Table~I, for example, if a two-photon driving is
combined with a single-photon dissipation or, vice versa, if a
single-photon driving is accompanied by a two-photon dissipation.
This is because, one of these processes (i.e., the driving or
dissipation) mixes the evolutions of the Hilbert spaces with even
and odd photon numbers.

To confirm these predictions we found approximate analytical
steady-state solutions, given in
Eqs.~(\ref{rho_even})--(\ref{general_solution}), of the master
equation, given by Eq.~(\ref{ME}). We also obtained precise
numerical solutions in a 100-dimensional Hilbert space, as shown
in all our plots. We found a very good agreement between these
numerical and approximate analytical solutions. Moreover, we
observed that they depend solely on the ratios between the driving
field strength $\epsilon$, the Kerr nonlinear coupling $\chi$, and
the damping constant $\gamma$; i.e., $\delta=\epsilon/\chi$ and
$\delta'=\gamma/\epsilon$. So, the absolute values of $\epsilon$,
$\chi$, and $\gamma$ are irrelevant.

We analyzed a few examples of standard infinite-dimensional
quantum optical states including coherent, squeezed, displaced
number, chaotic, photon-added chaotic and Schr\"odinger's cat
states to show how the photon-number statistics of an initial
state influences its steady state. As an illustration of our
results, the Wigner functions and photon-number probabilities for
the steady states are shown in Figs.~\ref{fig06}--\ref{fig14}. We
note that some of these states have nonnegative Wigner functions
(i.e., without regions marked in blue). Nevertheless all them are
nonclassical; i.e., their Glauber-Sudarshan $P$ function is
negative in some regions of phase space.

We hope that our proposal of state-dependent photon blockade via a
two-photon absorbing reservoir is another convincing example
demonstrating how to harness quantum-reservoir engineering for
quantum technology.

\begin{acknowledgments}
This work was supported by the Polish National Science Centre
under Grants No. DEC-2011/03/B/ST2/01903 and
DEC-2012/04/M/ST2/00789.  J.B. was supported by the Palack\'y
University under the Project IGA-P\v{r}F-2014-014. Y.X.L. is
supported by the National Basic Research Program of China Grant
No.~2014CB921401 and the NSFC Grants No.~61025022 and
No.~91321208. F.N. is partially supported by the RIKEN iTHES
Project, MURI Center for Dynamic Magneto-Optics, and a
Grant-in-Aid for Scientific Research (S).
\end{acknowledgments}

\appendix

\section{Approximate steady-state solutions}

Here we give approximate formulas for the coefficients $a,b,...,r$
occurring in the steady-state solutions, given by
Eqs.~(\ref{rho_even}) and~(\ref{rho_odd}), as obtained by
expanding our precise but lengthy solutions (thus, not presented
here) in power series of $\delta=\epsilon/\chi\ll 1$ and
$\delta'=\gamma/\epsilon\ll 1$ and keeping terms only up to
$\delta^2$, $\delta'^2$, and $\delta\delta'$.

The coefficients in Eq.~(\ref{rho_even}), where the initial state
was assumed to have an even number of photons, for the Hamiltonian
$\hat{H}_{02}$ (Model~1) are given by
\begin{align}
 p=\tfrac{1}{2}-\tfrac{9}{32}\delta^{2}+\tfrac{1}{8}\delta'^{2},
 \quad
 q=1-p-r,
 \quad
 r=&\tfrac{3}{32}\delta^{2},
\label{A2a}
\end{align}
for the diagonal terms of $\hat \rho^{{\rm even}}_{{\rm ss}}$ and
\begin{align}
 &a=-\tfrac{3}{8}\sqrt{2}\delta,\quad
 b=\tfrac{1}{4}\sqrt{2}\delta',\quad
 c=\tfrac{5}{64}\sqrt{6}\delta^{2},
 \nonumber\\
 &d=-\tfrac{1}{16}\sqrt{6}\delta \delta',\quad
 e=-\tfrac{1}{8}\sqrt{3}\delta,\quad
 f=0
\label{A2b}
\end{align}
for its off-diagonal terms, while the coefficients for the
Hamiltonian $\hat{H}_{13}$ (Model~2) are found to be
\begin{align}
p =  \tfrac{25}{32}-\tfrac{107}{512}\delta^{2},\;\;\;
q=\tfrac{3}{16}+\tfrac{15}{128}\delta^{2}, \quad
s=\tfrac{5}{768}\delta^{2}, \label{B2a}
\end{align}
and $r =1-p-q-s,$ for the diagonal terms, and
\begin{align}
a = & \tfrac{19}{128}\sqrt{2}\delta,\quad &b=&\tfrac{3}{32}\sqrt{2}\delta,\nonumber\\
c = & -\tfrac{37}{4608}\sqrt{6}\delta^{2},\quad &d=&\tfrac{1}{16}\sqrt{6}-\tfrac{49}{768}\sqrt{6}\delta^{2},\nonumber\\
e = & -\tfrac{5}{64}\sqrt{3}\delta,\quad
&f=&\tfrac{1}{32}\sqrt{3}\delta \label{B2b}
\end{align}
for the off-diagonal terms. For simplicity, we set
$\delta=\delta'$ in Eqs.~(\ref{B2a}) and~(\ref{B2b}).

The coefficients in Eq.~(\ref{rho_odd}), where the initial state
was assumed to have an odd number of photons, for the Hamiltonian
$\hat{H}_{02}$ (Model~1) read
\begin{eqnarray}
 p&=&1-\frac{6\epsilon^{2}}{M}\approx1-\frac{3}{8}\delta^{2}\approx1,\nonumber \\
 a&=&-4\sqrt{6}\frac{\chi\epsilon}{M}\approx-\frac{\sqrt{6}}{4}\delta,\nonumber \\
 b&=&3\sqrt{6}\frac{\epsilon\gamma}{M}\approx\frac{3}{16}\sqrt{6}\delta\delta'\approx0,
\label{A1}
\end{eqnarray}
where $M=16\chi^{2}+12\epsilon^{2}+9\gamma^{2}$, while these
coefficients for the Hamiltonian $\hat{H}_{13}$ (Model~2) are
given by
\begin{align}
 p=1-\frac{2\epsilon^{2}}{M}\approx \frac{1}{2},\quad a=0, \quad
 b=\sqrt{6}\frac{\epsilon\gamma}{M}\approx\frac{\sqrt{6}}{4}\delta',
\label{B1}
\end{align}
where $M=4\epsilon^{2}+3\gamma^{2}$.

\section{Some photon-number expansions}

Here we give some formulas useful for the calculation of the
probabilities $p_{{\rm even}}(\hat \rho_{0})$ and $p_{{\rm
odd}}(\hat \rho_{0})$, and their ratio $r$ for the squeezed and
displaced-number states $\hat \rho_{0}$, shown in
Fig.~\ref{fig13}.

The photon-number expansion of the ideal squeezed states, defined
by Eq.~(\ref{sq_state}), is given by
\begin{equation}
|\alpha,\xi\rangle=\sum_{n}\langle
n|\alpha,\xi\rangle|n\rangle\equiv\sum_{n}c_{n}|n\rangle,
\end{equation}
where
\[
c_{n}=\left(\frac{x}{2}\right)^{n/2}\frac{H_{n}(y)\exp(-z/2)}{\sqrt{n!}\cosh(|\xi|)},
\]
in terms of the Hermite polynomials $H_{n}(y)$,
$x=\tanh(|\xi|)\exp(i{\rm Arg}\xi),$
$y=(2x)^{-1/2}(\alpha+\alpha^{*}x),$ and
$z=|\alpha|^{2}+(\alpha^{*})^{2}x$.

The photon-number expansion of the displaced number states,
defined by Eq.~(\ref{DNS}), is given by~\cite{Oliveira90,Tanas96}:
\begin{eqnarray}
|\alpha,n_{0}\rangle & = & \sum_{n}b_{n}\exp[i(n-n_{0})\,{\rm
Arg}\alpha]|n\rangle,\label{N142}
\end{eqnarray}
where the real amplitudes $b_{n}$ are
\begin{eqnarray}
b_{n}  =  \langle n|\hat{D}(|\alpha|)|n_{0}\rangle  \hspace*{4.5cm}\nonumber \\
  =  C\,\sqrt{\frac{n_{-}!}{n_{+}!}}(-1)^{n_{+}-n}
  |\alpha|^{n_{+}-n_{-}}L_{n_{-}}^{(n_{+}-n_{-})}(|\alpha|^{2}),
\end{eqnarray}
where $L_n^{(m)}(x)$ are the associated Laguerre polynomials,
$C=\exp\left(-\frac{1}{2}|\alpha|^{2}\right)$,
$n_{-}=\min\{n,n_{0}\},$ and
$n_{+}=\max\{n,n_{0}\}=n+n_{0}-n_{-}.$


\end{document}